\documentclass[superscriptaddress,showpacs,nofootinbib,aps,prd,twocolumn]{revtex4}
\usepackage{epsfig}
\usepackage{graphicx}
\usepackage{dcolumn}
\usepackage{bm}

\date{}
\begin{document}


\newcommand{\ds}{\displaystyle}
\newcommand{\mc}{\multicolumn}
\newcommand{\bce}{\begin{center}}
\newcommand{\ece}{\end{center}}
\newcommand{\beq}{\begin{equation}}
\newcommand{\eeq}{\end{equation}}
\newcommand{\bea}{\begin{eqnarray}}

\newcommand{\eea}{\end{eqnarray}}
\newcommand{\cont}{\nonumber\eea\bea}
\newcommand{\cl}[1]{\begin{center} {#1} \end{center}}
\newcommand{\ba}{\begin{array}}
\newcommand{\ea}{\end{array}}

\newcommand{\ab}{{\alpha\beta}}
\newcommand{\cd}{{\gamma\delta}}
\newcommand{\dc}{{\delta\gamma}}
\newcommand{\ac}{{\alpha\gamma}}
\newcommand{\bd}{{\beta\delta}}
\newcommand{\abc}{{\alpha\beta\gamma}}
\newcommand{\eps}{{\epsilon}}
\newcommand{\lam}{{\lambda}}
\newcommand{\mn}{{\mu\nu}}
\newcommand{\mpnp}{{\mu'\nu'}}
\newcommand{\Amuu}{{A_{\mu}}}
\newcommand{\Amuo}{{A^{\mu}}}
\newcommand{\Vmuu}{{V_{\mu}}}
\newcommand{\Vmuo}{{V^{\mu}}}
\newcommand{\Anuu}{{A_{\nu}}}
\newcommand{\Anuo}{{A^{\nu}}}
\newcommand{\Vnuu}{{V_{\nu}}}
\newcommand{\Vnuo}{{V^{\nu}}}
\newcommand{\Fmnu}{{F_{\mu\nu}}}
\newcommand{\Fmno}{{F^{\mu\nu}}}

\newcommand{\abcd}{{\alpha\beta\gamma\delta}}


\newcommand{\bsigma}{\mbox{\boldmath $\sigma$}}
\newcommand{\btau}{\mbox{\boldmath $\tau$}}
\newcommand{\brho}{\mbox{\boldmath $\rho$}}
\newcommand{\bpipi}{\mbox{\boldmath $\pi\pi$}}
\newcommand{\bss}{\bsigma\!\cdot\!\bsigma}
\newcommand{\btt}{\btau\!\cdot\!\btau}
\newcommand{\bnabla}{\mbox{\boldmath $\nabla$}}
\newcommand{\bphi}{\mbox{\boldmath $\tau$}}
\newcommand{\bvarphi}{\mbox{\boldmath $\rho$}}
\newcommand{\bDelta}{\mbox{\boldmath $\Delta$}}
\newcommand{\bpsi}{\mbox{\boldmath $\psi$}}
\newcommand{\bPsi}{\mbox{\boldmath $\Psi$}}
\newcommand{\bPhi}{\mbox{\boldmath $\Phi$}}
\newcommand{\bnab}{\mbox{\boldmath $\nabla$}}
\newcommand{\bpi}{\mbox{\boldmath $\pi$}}
\newcommand{\btheta}{\mbox{\boldmath $\theta$}}
\newcommand{\bkappa}{\mbox{\boldmath $\kappa$}}

\newcommand{\bA}{{\bf A}}
\newcommand{\bB}{\mbox{\boldmath $B$}}
\newcommand{\bC}{\mbox{\boldmath $C$}}
\newcommand{\bp}{\mbox{\boldmath $p$}}
\newcommand{\bk}{\mbox{\boldmath $k$}}
\newcommand{\bq}{\mbox{\boldmath $q$}}
\newcommand{\bfe}{{\bf e}}
\newcommand{\bb}{\mbox{\boldmath $b$}}
\newcommand{\bc}{\mbox{\boldmath $c$}}
\newcommand{\br}{\mbox{\boldmath $r$}}
\newcommand{\bR}{\mbox{\boldmath $R$}}
\newcommand{\bs}{\mbox{\boldmath $s$}}
\newcommand{\bT}{{\bf T}}
\newcommand{\fph}{${\cal F}$}
\newcommand{\aph}{${\cal A}$}
\newcommand{\dph}{${\cal D}$}
\newcommand{\fpi}{f_\pi}
\newcommand{\mpi}{m_\pi}
\newcommand{\Tr}{{\mbox{\rm Tr}}}
\def\Qb{\overline{Q}}
\newcommand{\delu}{\partial_{\mu}}
\newcommand{\delo}{\partial^{\mu}}
%
%
\newcommand{\up}{\!\uparrow}
\newcommand{\upup}{\uparrow\uparrow}
\newcommand{\updo}{\uparrow\downarrow}
\newcommand{\uu}{$\uparrow\uparrow$}
\newcommand{\ud}{$\uparrow\downarrow$}
\newcommand{\auu}{$a^{\uparrow\uparrow}$}
\newcommand{\aud}{$a^{\uparrow\downarrow}$}
\newcommand{\pu}{p\!\uparrow}

\newcommand{\qp}{quasiparticle}
\newcommand{\sa}{scattering amplitude}
\newcommand{\ph}{particle-hole}
\newcommand{\qcd}{{\it QCD}}
\newcommand{\integ}{\int\!d}
\newcommand{\ie}{{\sl i.e.~}}
\newcommand{\etal}{{\sl et al.~}}
\newcommand{\etc}{{\sl etc.~}}
\newcommand{\rhs}{{\sl rhs~}}
\newcommand{\lhs}{{\sl lhs~}}
\newcommand{\eg}{{\sl e.g.~}}
\newcommand{\ef}{\epsilon_F}
\newcommand{\sigt}{d^2\sigma/d\Omega dE}
\newcommand{\sige}{{d^2\sigma\over d\Omega dE}}
\newcommand{\rpaeq}{\beq
\left ( \begin{array}{cc}
A&B\\
-B^*&-A^*\end{array}\right )
\left ( \begin{array}{c}
X^{(\kappa})\\Y^{(\kappa)}\end{array}\right )=E_\kappa
\left ( \begin{array}{c}
X^{(\kappa})\\Y^{(\kappa)}\end{array}\right )
\eeq}
\newcommand{\ket}[1]{| {#1} \rangle}
\newcommand{\bra}[1]{\langle {#1} |}
\newcommand{\ave}[1]{\langle {#1} \rangle}
\newcommand{\half}{{1\over 2}}

\newcommand{\singlespace}{
    \renewcommand{\baselinestretch}{1}\large\normalsize}
\newcommand{\doublespace}{
    \renewcommand{\baselinestretch}{1.6}\large\normalsize}
\newcommand{\bftau}{\mbox{\boldmath $\tau$}}
\newcommand{\bfalpha}{\mbox{\boldmath $\alpha$}}
\newcommand{\bfgamma}{\mbox{\boldmath $\gamma$}}
\newcommand{\bfxi}{\mbox{\boldmath $\xi$}}
\newcommand{\bfbeta}{\mbox{\boldmath $\beta$}}
\newcommand{\bfeta}{\mbox{\boldmath $\eta$}}
\newcommand{\bfpi}{\mbox{\boldmath $\pi$}}
\newcommand{\bfphi}{\mbox{\boldmath $\phi$}}
\newcommand{\bfR}{\mbox{\boldmath ${\cal R}$}}
\newcommand{\bfL}{\mbox{\boldmath ${\cal L}$}}
\newcommand{\bfM}{\mbox{\boldmath ${\cal M}$}}
\def\dblint{\mathop{\rlap{\hbox{$\displaystyle\!\int\!\!\!\!\!\int$}}
    \hbox{$\bigcirc$}}}
\def\ut#1{$\underline{\smash{\vphantom{y}\hbox{#1}}}$}

\def\UNITY{{\bf 1\! |}}
\def\CutPom{{\bf I\!P\!\!\!\!\!\Big{ / }}}
\def\Pom{{\bf I\!P}}
\def\lsim{\mathrel{\rlap{\lower4pt\hbox{\hskip1pt$\sim$}}
    \raise1pt\hbox{$<$}}}         
\def\gsim{\mathrel{\rlap{\lower4pt\hbox{\hskip1pt$\sim$}}
    \raise1pt\hbox{$>$}}}         
\def\beq{\begin{equation}}
\def\eeq{\end{equation}}
\def\bea{\begin{eqnarray}}
\def\eea{\end{eqnarray}}
\def\CutD{{\bf D\!\!\!\!\!\Big{ / }}}

\mbox
{\mbox{}
\hfill{\large FZJ-IKP-TH-2006-12}\\
\mbox{}}
        
\title{ Unitarity 
cutting rules
for the nucleus excitation and topological cross sections
in hard production off nuclei from nonlinear $k_{\perp}$-factorization}

\author{N.N. Nikolaev}%
\email{N.Nikolaev@fz-juelich.de}
\affiliation{Institut f\"ur Kernphysik, Forschungszentrum J\"ulich, D-52425 J\"ulich, Germany}
\affiliation{L.D. Landau Institute for Theoretical Physics, 142432 Chernogolovka, Russia}
\author{W. Sch\"afer}%
\email{Wo.Schaefer@fz-juelich.de}
\affiliation{Institute of Nuclear Physics PAN, PL-31-342 Cracow, Poland}

\date{\today}%

\begin{abstract} 
At the partonic level, a typical final state in small-$x$
deep inelastic scattering off nuclei and hard proton-nucleus 
collisions can be characterized by the 
multiplicity of 
color-excited nucleons. Within the reggeon field theory,
each color-excited nucleon is associated with the unitarity 
cut of the pomeron exchanged between the projectile and nucleus.
In this communication we derive the unitarity
rules for the multiplicity of excited nucleons,
alias cut pomerons, alias topological cross sections,
for typical hard dijet production processes. We demonstrate 
how the coupled-channel non-Abelian intranuclear
evolution of color dipoles, inherent to pQCD,
gives rise to the reggeon field theory diagrams
for final states in terms of the uncut, and two kinds
of cut, pomerons. Upon the proper identification of the
uncut and cut pomeron exchanges, the topological cross
sections for dijet production follow in a straightforward way from the
earlier derived nonlinear $k_\perp$ factorization
quadratures for the inclusive dijet cross sections.
The concept of a coherent (collective) nuclear glue proves
extremely useful for the formulation of the
reggeon field theory vertices of multipomeron --- cut and uncut ---
couplings to particles and between themselves. A departure of our
unitarity cutting rules from the ones suggested by the
pre-QCD Abramovsky-Kancheli-Gribov rules, stems from
the coupled-channel features of non-Abelian intranuclear pQCD. We 
propose a multiplicity re-summation as a tool for the
isolation of topological cross sections for single-jet
production. 
\end{abstract}
\date{\today}
\maketitle




\section{Introduction}

The subject of this communication
is a study of the unitarity cutting rules for
long-range rapidity correlations
 between the forward, and mid-rapidity, jet and
dijet production and multiproduction in the
target nucleus fragmentation region --- one of the important
experimental observables (\cite{FB,Jacobs} and references therein).
In perturbative Quantum Chromo Dynamics (pQCD), hard single-jet or dijet
production in the beam hemisphere of high-energy deep inelastic 
scattering (DIS) or a hard nucleon-nucleon collision can be viewed as 
a hard
photon-gluon fusion, $\gamma^*g_t\to q\bar{q}$, or hard scattering of 
the valence quark of the nucleon, $q g_t\to qg$. At mid-rapidity, 
the dominant gluon-gluon dijets come from $gg_t \to gg$, the 
open charm production
is driven by $gg_t \to c\bar{c}$. Here $g_t$ stands
for the target gluon --- in the target rest frame it is the gluon exchanged 
in the $t$-channel. Hard scattering leaves the target-nucleon debris in 
the color-octet state. In hard scattering 
off a heavy nucleus multiple gluon exchanges are enhanced by a large 
thickness of the target nucleus, and a typical inelastic event leaves behind 
many bound nucleons in the color-excited state. Their multiplicity 
defines the hadronic activity in the nucleus hemisphere, i.e., 
the centrality of 
collisions --- a fundamental  concept in the physics 
of ultrarelativistic nuclear
collisions (\cite{Centrality,PHOBOScentrality}, for reviews and further references 
see \cite{Jacobs}). 

In the reggeon field theory (RFT) 
language, \cite{GribovRFT,AGK}, pQCD offers a perfect definition of
cut pomerons: 
each and every excited nucleon can be associated with the unitarity cut of a 
pomeron exchanged between the beam and target. Then our pQCD 
unitarity rules for color excitation of the nucleus
give a multiplicity of cut pomerons, alias the topological cross
sections --- a subject
of much discussion ever since the pre-QCD unitarity cutting rules
suggested in 1972 by Abramovsky, Gribov and Kancheli (AGK) \cite{AGK}.
A casual reference to the AGK rules appears in numerous discussions
of multiproduction in ultrarelativistic (nuclear) collisions
(\cite{Capella,TreleaniOld,Kaidalov,Treleani,CapellaAGK,KovchegovAGK,Boreskov,BartelsAGK,Braun3P,BraunAGK}, for 
more references see the recent reviews by Bartels and by Kowalski 
in Ref. \cite{AGKsvalka}). 
Here we address the issue of pQCD unitarity rules on an example of 
interaction of small projectiles with large
nuclear targets, where one can take advantage of a
new large parameter -- the thickness (diameter $2R_A$) of a heavy
nucleus of mass number $A$.
We discuss observables of different degrees of
re-summation over this large parameter.

There are two principal novelties in our approach 
to topological cross sections. The first one is a manifest imposition of
the unitarity constraints at each step of the derivation.
This is readily achieved within the
color-dipole $\textsf{S}$-matrix approach to high-energy hard processes
developed in 
Refs.~\cite{SlavaPositronium,NZ91,NZZBFKL,NZ94,NPZcharm,NSSdijet}.
We start from the basic pQCD, follow 
carefully a separation of color-diagonal
and color-excitation interactions of color dipoles, 
diagonalize the non-Abelian intranuclear evolution,
and relate the 
topological cross sections with a fixed multiplicity  
of color-excited
nucleons --- cut pomerons --- to properties of the collective nuclear glue
of spatially overlapping nucleons in the Lorentz-contracted
ultrarelativistic nucleus. A consistent 
treatment of the coupled-channel aspects of the non-Abelian 
intranuclear evolution of color dipoles is behind the
universality classes for nonlinear $k_\perp$-factorization for
hard processes in a nuclear environment 
\cite{Nonlinear,PionDijet,SingleJet,Nonuniversality,Paradigm,QuarkGluonDijet,GluonGluonDijet,VirtualReal}. 
And it is behind the second principal novelty of this paper ---
the finding that in the description
of topological cross sections one needs two kinds of cut pomerons.
One of these two, $\CutPom_r$ (the subscript ``r'' stands for 
the rotation), describes color excitations of the nucleus
by color rotations of the color dipole within the same color
multiplet. The second one, $\CutPom_e$ (the subscript ``e'' stands for 
the excitation), is for the transition between color multiplets 
the dimensions of which differ by the factor ${\cal O}(N_c^2)$,
where $N_c$ is the  number of colors. As a matter of fact, this
distinction between $\CutPom_r$ and  $\CutPom_e$ had to a large
extent been anticipated in the first publication \cite{Nonlinear}
on the nonlinear $k_\perp$-factorization.
Finally, uncut pomerons,
$\Pom$, are associated with the color-diagonal (elastic) intranuclear
interactions of dipoles.

To the leading order of the large-$N_c$ perturbation theory,
we formulate simple RFT diagram rules for the calculation of
topological cross sections for the dijet spectra. To this end,
upon the proper identification of the uncut, $\Pom$, and
the two cut, $\CutPom_r$ and $\CutPom_e$, pomerons our task 
of the derivation
of the topological cross sections boils down to a simple ---
in one stroke --- reinterpretation of the nonlinear 
$k_\perp$ factorization results for the inclusive dijet spectra
\cite{Nonlinear,PionDijet,SingleJet,Nonuniversality,Paradigm,QuarkGluonDijet,GluonGluonDijet,VirtualReal}. 
Here a concept of the collective nuclear glue as a coherent state
of the in-vacuum (reggeized) gluons emerges and proves an extremely useful one.
What we report are only prolegomena to a full fledged
RFT: we confine
ourselves to tree diagrams, the hot issue of 
an extension to pomeron loops
in the spirit of Ref. \cite{LipatovAction} needs  
further scrutiny (also see the reviews \cite{Trianta}). Even at the tree level, and to 
the leading order of large $N_c$ perturbation theory, 
we find a very rich variety of the cut pomeron
multiplicity distributions which vary substantially from
one reaction universality class \cite{Nonuniversality,Paradigm,GluonGluonDijet}
to another, and give rise to a whole family of the $t$-channel
multipomeron vertices for $t$-channel transitions from
one cut pomeron $\CutPom_r$ to  $\CutPom_r$'s and  $\CutPom_e$'s,
and their $s$-channel unitarity cuts. Arguably, this 
distinction between universality classes will persist 
under small-$x$ evolution --- the
unitarity cut content of this evolution is one of the obvious
future applications of the emerging formalism.

As far as the single-jet and still more integrated observables
are concerned, an important, and obvious, feature of 
topological cross sections is the Cheshire Cat grin (CCG)
of the spectator interactions. Specifically,
if one starts with the dijet cross section and integrates over 
the phase space of the spectator jet, then all the spectator 
interaction effects would cancel in the nonlinear
$k_\perp$ factorization quadratures for the single-jet spectrum.
The spectator contribution to the nucleus excitation
and cut pomerons --- the CCG --- stays on. CCG is a
physics observable and has important implications for the
long-range rapidity
correlations between the beam, and mid-rapidity, and the
target nucleus fragmentation regions. In  this respect, our concept 
of the color-excited nucleons, and the associated
cut pomeron, is close to the concept of 
wounded (participant) nucleons \cite{Bialas,Bialas2005}.
The wounded nucleon multiplicity distributions are customarily
evaluated within a certain interpretation of the Glauber
model for the hadron-nucleus interactions 
(\cite{Capella,TreleaniOld,Kaidalov}, for the review see \cite{WoundedNucleons},
for the relation to the collision centrality see \cite{Centrality,PHOBOScentrality,Jacobs}).
Such a Glauber model interpretation is not borne out by 
our pQCD coupled-channel
approach to topological cross sections, and we discuss 
the origin of this
distinction to a great detail. 

On the event-by-event basis, the single mid-rapidity gluon 
spectra are infested by the CCG of the spectator, and comover, 
parton interactions.
What is remarkable is that the effect of spectator parton interactions
can be eliminated entirely, and meaningful topological cross
sections for single-jet spectra can be defined, upon
certain multiplicity re-summations. Upon these re-summations,
the single mid-rapidity gluon spectra serve as a definition
of the $t$-channel transitions of the in-vacuum $\CutPom_r$ to
multiple $\CutPom_e$'s and  $\CutPom_r$'s. The emerging
multipomeron vertices are universal and do not depend on the
projectile. We strongly advocate an analysis of the experimental
data using this multiplicity re-summation technique.

\begin{widetext}
\mbox{} 
\begin{figure}[!h]
\begin{center}
\includegraphics[width = 3cm, height= 12.0cm,angle=270]{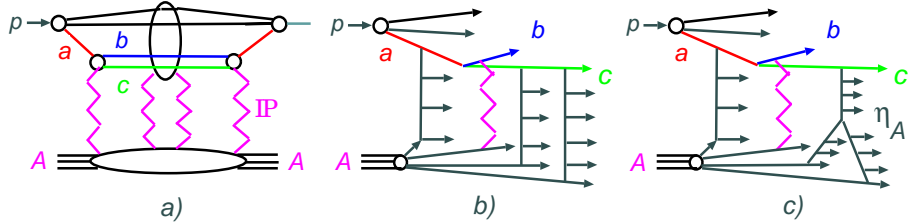}
\caption{ a) A contribution of multipomeron exchanges to the forward elastic proton-nucleus
scattering amplitude from final states containing the dijet $bc$,
(b) an example of the unitarity cut with independent hadronization
of color strings over all rapidities from the target to a projectile,
(c)  an example of the unitarity cut with the fan structure of 
hadronizing strings with string junction at the rapidity $\eta_A$.}
\label{fig:CutPomerons}
\end{center}
\end{figure}
\mbox{}
\end{widetext}
One important application of cutting rules for these long-range rapidity
correlations is a nonperturbative quenching (stopping) of leading
jets and the dependence of this quenching on the
nucleus excitation (for reviews  on quenching see
\cite{Ledoux,HIJING,Jacobs,Berndt}).
One customarily associates with  
cut pomerons color strings stretched between the color-excited
nucleons of the nucleus and the projectile system.
Hadronization of the color string into small-$p_\perp$ hadrons
--- the underlying minimal-bias event for hard dijets --- slows down the
projectile system. The energy flow from 
the projectile system to the nucleus,
and the accompanying nonperturbative quenching (stopping) 
of leading jets, depends on the hadronization model.
In one extreme
scenario, which goes back to the ITEP-Orsay model
of mid-70's \cite{Capella,Kaidalov}, strings hadronize 
independently of each other
in the whole rapidity span from the nucleus up to the projectile,
see Fig. \ref{fig:CutPomerons}b. In such a scenario the projectile loses a finite 
fraction of its energy per color string, so that the nonperturbative
quenching of forward jets and particles, and
the related breaking of the limiting fragmentation, would persist at high energy.
In still another extreme scenario, suggested by the formation
length considerations, hadronization of color strings
attached to a high-$p_\perp$ parton in the final state, is
independent only up to rapidities 
\beq
\eta_A \approx \log{1\over 2R_A m_p},
\label{eq:1.1}
\eeq
from the target nucleus rapidity \cite{Kancheli,NZfusion,NikDav,HIJING}
(here $m_p$ is the proton mass). 
When viewed in the anti-laboratory frame with a Lorentz-contracted
ultrarelativistic nucleus, at  rapidities $\eta >\eta_A$
the strings overlap
spatially. One would argue that the overlapping strings  
fragment as one string, see Fig. \ref{fig:CutPomerons}c,
for the recent string phenomenology see \cite{Pajares}. In such a scenario, 
the projectile loses a finite energy per color string. In a dilute nucleus,
strings of length $\sim R_A$ are stretched also between different
color centers in the nucleus, and the energy loss will be larger 
than the collisional loss for the recoil of struck quarks 
in the nucleus \cite{Recoil}. In any case, it is imperative to understand the relevant
topological cross sections and identify the quenching of the
parton parent to the specific observed jet
and  the much discussed breaking of the
limiting fragmentation \cite{Busza,PHOBOSpseudorap,ArsenePseudoRap}. This is our major
task in this study.

Our treatment is applicable when the beam and final state
partons interact coherently over the whole longitudinal
extension of the nucleus,
\beq
x={(Q^*)^2+M_{\perp}^2 \over 2m_p\nu}
\lsim x_A ={1\over 2R_A m_p} \approx 0.1 A^{-1/3}\,, 
\label{eq:1.2}
\eeq
where $M_{\perp}$ is the transverse mass of the dijet, 
$\nu$ is the energy of the projectile
parton in the target rest frame 
and  $(Q^*)^2$ is its virtuality: $ (Q^*)^2=Q^2$ in DIS
and  $ (Q^*)^2=\bp_q^2$ in $qA$ collisions, where $\bp_q$ is the
transverse momentum of the incident valence quark in the beam proton
\cite{NZfusion,NZ91}. At the 
Relativistic Heavy Ion Collider (RHIC) this amounts to precisely 
the leading dijets in the proton fragmentation region 
(for the discussion of the possible upgrade of detectors at RHIC II 
for the improved coverage of the proton fragmentation region
see \cite{RHIC_II}).

The presentation of the main material is organized as follows. In 
Secs. II and III we introduce the basics of the color-dipole approach,
formulate the master formula for the 
dijet spectrum and present the unitarity based derivation of
the topological cross sections --- alias the multiplicity distribution
of cut pomerons ---  in terms of the multiplicity
of color excited nucleons. Here the major novelty is an integral 
representation for the corresponding nuclear $\cal{S}$-matrices in
terms of elastic and color-excitation free-nucleon interactions,
which has a built-in unitarity property. Sec. IV is an 
introduction to the correspondence between the color dipole and RFT
approaches. We define the collective nuclear glue based on
the color-dipole $\textsf{S}$-matrix, comment on its coherent
state property, and establish its connection to quasielastic scattering
of quarks. The subject of Sec. V is a nonlinear $k_\perp$-factorization
for topological cross sections for the universality classes 
of coherent diffraction (CD), inelastic and pseudo-diffractive DIS 
off nuclei and their RFT interpretation in terms of the cut
and uncut pomeron exchanges. Here we 
identify two kinds of cut pomerons, $\CutPom_e$ and $\CutPom_r$, 
as an indispensable feature of pQCD.  Based on this
new finding and the identification of the r\^ole of $\CutPom_e$ 
and $\CutPom_r$, we show how the topological cross sections follow
in one stroke from the nonlinear $k_\perp$ factorization
quadratures for inclusive dijet cross sections. Inspired
by this success, in Sec. VI we
address an important issue of whether the topological cross
sections can be guessed from the Glauber model results
for the fully integrated total cross section. We show that within pQCD 
this is 
not the case, and derive new results for the multiplicity
distribution of cut pomerons in the fully integrated cross
sections. The possible phenomenological applications of our
results to the multiplicity distributions in the
backward (nucleus) hemisphere and long-range rapidity
correlations between leading jets and rapidity spectra of
backward particles in DIS off nuclei are commented on in Sec. VII.
Here we introduce the multiplicity re-summations which enable one
to eliminate the spectator quark interaction effects in the
nucleus excitation.
In Secs. VIII, IX and X we extend our treatment of DIS to $qg$ 
final states in $qA$ collisions, $gg$ final states in $gA$ collisions
and open charm $c\bar{c}$ states in $gA$ collisions, respectively. 
We formulate the RFT diagram rules for each universality class
and offer a fresh look at slight variations from one projectile
to another.
In Sec. XI we briefly revisit topological cross sections in 
single mid-rapidity gluon  production. Here the CCG precludes the
determination of the relevant multiplicity of cut pomerons 
on an event-by-event basis. Still upon certain
multiplicity re-summations, one can define single-jet 
topological cross sections 
which exhibit a remarkable universality for the quark and
gluon projectiles modulo to the Casimirs in the 
Regge-factorizable coupling of
the projectile to a pomeron $\CutPom_r$. We comment on the
variety of triple-pomeron couplings which appear in the RFT
description of the re-summed topological cross sections.
In Sec. XII we summarize our principal conclusions. The
technicalities of the diagonalization of the non-Abelian
evolution of color dipoles which is a basis for
the derivation of topological cross sections are presented
in Appendices A,B,C.


\section{The master formula for hard dijet production off 
free nucleons and nuclei}

To the lowest order in pQCD, the underlying hard subprocess are of the
general form $ag_t \to bc$. In the photon fragmentation region of
DIS the projectile is the virtual photon,
$a=\gamma^*$, in the proton fragmentation region of 
proton-nucleus collisions one would focus
on interactions of valence quarks, $a=q$, the production 
of open charm and of mid-rapidity
jets in proton-nucleus collisions will be dominated by gluons
of the projectile hadron, $a=g$. From the laboratory frame standpoint, 
the dijet production is an excitation of the perturbative $|bc\rangle$ Fock state
of the physical projectile $|a\rangle_{phys}$ by one-gluon $t$-channel
exchange with the
target nucleon. In our discussion we 
follow a general treatment of
multiple gluon exchanges in nuclear targets developed 
in, and use the principal notations from, our earlier publications 
\cite{NPZcharm,Nonlinear,SingleJet,QuarkGluonDijet,GluonGluonDijet,VirtualReal}.

To the lowest order in the perturbative transition $a\to bc$ the Fock 
state expansion for the physical state $|a\rangle_{phys}$ reads
\beq
 \ket{a}_{phys} = \sqrt{Z_a}\ket{a}_0 + \Psi(z_b,\br) \ket{bc}_0,
\label{eq:2.1}
\eeq
where $\Psi(z_b,\br)$ is the probability amplitude to find the $bc$ system
with the separation (color dipole) $\br$ in the two-dimensional impact parameter space,
the subscript $"0"$ refers to bare partons. The perturbative
coupling of the $a\to bc$ transition is re-absorbed into the light-cone
wave function $\Psi(z_b,\br)$ which
depends on the virtuality of the incident parton $a$, i.e., $Q_a^2=Q^2$ for 
the virtual photon in DIS or $Q_a^2=\bp_a^2$ in $qA$ collisions, where
$\bp_a$ is the transverse momentum of the parton $a$ in the proton.
Explicit expressions for various pQCD subprocess in terms of the
familiar parton splitting functions are found in \cite{SingleJet}.
In this paper we focus on the lowest order (Born) excitation processes $a\to bc$ 
without production of more secondary partons in the rapidity span between 
$\eta_a$ and $\eta_c$. 

\begin{figure}[!h]
\begin{center}
\includegraphics[width = 2.8cm,height=7.3cm, angle = 270]{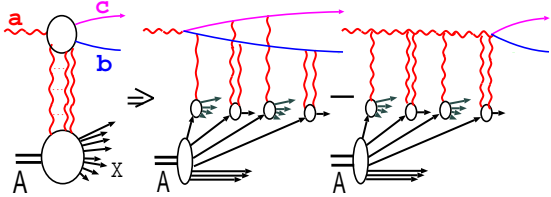}
\caption{Typical contribution to the excitation amplitude for $a A \to b c X$,
with multiple color excitations of the nucleus. 
The amplitude receives contributions from processes that involve interactions 
with the nucleus after and before the virtual 
decay which interfere destructively. Notice a distinction between
color-excitation (single-gluon, i.e., cut pomeron, exchange) and
color-diagonal 
(elastic two-gluon, uncut pomeron, exchange) interactions between the target nucleons and
the projectile and produced parton systems.}
\label{fig:single-jet_a_to_bc}
\end{center}
\end{figure}

We use the light-cone $\textsf{S}$-matrix approach. For the sake of simplicity
we work in the $aA$ collision frame. If $\bb \equiv \bb_a$ 
is the impact parameter of the  projectile $a$, then 
\beq
\bb_{b}=\bb+z_c\br, \quad\quad \bb_{c}=\bb-z_b\br\, ,
\label{eq:2.2}
\eeq
where  $z\equiv z_b \,,\,z_c= (1-z_b)$ is the partition of
the beam momentum between the final state partons.
Participating partons propagate along a straight path and the interaction is
coherent over the whole nucleus in the high-energy limit of $x \lsim x_A$.
The contribution from transitions
$a\to bc$ inside the target nucleus vanishes in the high energy limit.

Evidently,  the $\textsf{S}$-matrix for interaction of the $bc$ state
equals $\textsf{S}_{bc}(\bb_b,\bb_c)= \textsf{S}_b(\bb_b) \textsf{S}_c(\bb_c)$
and the action of the $S$-matrix 
on $\ket{a}_{phys}$ takes a simple form
\bea
&& \textsf{S}\ket{a}_{phys} =
\sqrt{Z_a}S_a(\bb) \ket{a}_0 \nonumber\\
&+&
S_b(\bb_b) S_c(\bb_c)\Psi(z,\br) \ket{bc}_0 =S_a(\bb) \ket{a}_{phys}
\nonumber \\
& +&  [ S_b(\bb_b) S_c(\bb_c) - S_a(\bb) ]
\Psi(z,\br) \ket{bc}_0 \, . 
\label{eq:2.3} 
\eea 
Here we
explicitly decomposed the final state into the elastically
scattered $\ket{a}_{phys}$ and the excited state $\ket{bc}_{0}$.
The wave function renormalization $\sqrt{Z_a}$ in (\ref{eq:2.1})
does not enter the excitation amplitude.
The two terms in the latter describe a scattering on the target of
the $bc$ system formed way in front of the target and the
transition $a\to bc$ after the interaction of the state
$\ket{a}_0$ with the target, as illustrated in
Fig. \ref{fig:single-jet_a_to_bc}. Whenever it wouldn't cause any confusion, we shall
use the short hand notations, $\textsf{S}(\bb)\equiv \textsf{S}_a(\bb)$
for the beam-target $\textsf{S}$-matrix and $\textsf{S}(\bB)
\equiv\textsf{S}_{bc}(\bb_b,\bb_c)$ for the $bc$-target $\textsf{S}$-matrix 
where $\bB=\{\bb_b,\bb_c\}$ is a short hand notation for the impact parameters
of 
the two-parton system.

Usually one discusses fully inclusive nuclear cross 
sections summed over all nuclear -- and nucleon -- excitations. 
The specific problem of interest in this
communication --- the properties of final states tagged by the degree of
excitation of the target nucleus and their RFT
interpretation --- calls for a somewhat more detailed
treatment of color properties of final states.
The probability amplitude for the two-jet spectrum is given by the
Fourier transform 
\bea 
&&\int d^2\bb_b d^2\bb_c \exp[-i(\bp_b\bb_b +
\bp_c\bb_c)]\nonumber\\
&\times& \bra{A_f}[ \textsf{S}(\bB) - \textsf{S}(\bb) ] \ket{A_{in}}_{c_f c_i}
\Psi(z,\br)\,,
\label{eq:2.4} 
\eea 
where we show explicitly the matrix element over the configuration
and color space wave functions of the target $\ket{A_{in}}$ and
final state nucleus $\bra{A_f}$. 
We average the cross section over the
unobserved color states $c_i$ of the incident parton. We sum over the unobserved
color states $c_f$ of the dijet. Then,
the differential cross section for a specific final state $\bra{A_f}$ of
the target is proportional
to the modulus squared of (\ref{eq:2.4}),
\bea 
&&{d\sigma(aA\to (bc)A_f)\over dz d^2\bp_b d^2\bp_c}={1\over (2\pi)^4}
\int d^2\bb_b' d^2\bb_c' \nonumber\\
&\times& \int d^2\bb_b d^2\bb_c \exp[-i \bp_b(\bb_b -\bb_b')-
i\bp_c(\bb_c-\bb_c')]\nonumber\\
&\times& \Psi^{*}(z,\br')\Psi(z,\br)\nonumber\\
&\times&{1\over {\textsf{dim[in]}}} \sum_{c_f c_i}
\bra{A_{in}}[ \textsf{S}^{\dagger}(\bB')- \textsf{S}^{\dagger}(\bb') ]
\ket{A_f}_{c_i c_f}
\nonumber\\
&\times&
\bra{A_f}[ \textsf{S}(\bB)  - \textsf{S}(\bb) ] \ket{A_{in}}_{c_f c_i}, 
\label{eq:2.5} 
\eea 
where $\textsf{dim[in]}$ is the number of color states in the $SU(N_c)$ multiplet
the incident parton belongs to.

Hereafter we describe the final state in terms of $\bp \equiv \bp_b,\, z\equiv
z_b$ 
and the dijet acoplanarity momentum $\bDelta=\bp_b+\bp_c$. We also
introduce $\bs=\bb_b-\bb_b'$, in terms of which $\bb_c-\bb_c'=
\bs-\br+\br'$. Notice that $\bs$ is conjugate to the acoplanarity momentum
$\bDelta$:
\bea
&&\exp[-i \bp_b(\bb_b -\bb_b')-
i\bp_c(\bb_c-\bb_c')]=\nonumber\\
&=&\exp[-i\bDelta\bs -i(\bp-\bDelta)(\br-\br')] 
\label{eq:2.6} 
\eea
Our main task will be an evaluation of the
color trace which emerged in (\ref{eq:2.5}), 
\begin{widetext}
\bea
{\cal T}(\bB,\bB',\bb,\bb')& =& {\rm Tr}\Bigl\{\bra{A_{in}}
[ \textsf{S}^{\dagger}(\bB')- \textsf{S}^{\dagger}(\bb') ] \ket{A_f}
\bra{A_f}[ \textsf{S}(\bB)  - \textsf{S}(\bb) ]\ket{A_{in}} \Bigr\}\nonumber\\
 &=&{\rm Tr}\Bigl\{\bra{A_{in}}\textsf{S}^{\dagger}(\bB')\ket{A_f}
\bra{A_f}\textsf{S}(\bB)\ket{A_{in}} \Bigr\} 
+ 
{\rm Tr}\Bigl\{\bra{A_{in}}\textsf{S}^{\dagger}(\bb')\ket{A_f}
\bra{A_f}\textsf{S}(\bb)\ket{A_{in}} \Bigr\}\nonumber\\
& -&{\rm Tr}\Bigl\{\bra{A_{in}}\textsf{S}^{\dagger}(\bB')\ket{A_f}
\bra{A_f}\textsf{S}(\bb)\ket{A_{in}} \Bigr\}
-
 {\rm Tr}\Bigl\{\bra{A_{in}}\textsf{S}^{\dagger}(\bb')\ket{A_f}
\bra{A_f}\textsf{S}(\bB)\ket{A_{in}} \Bigr\}\nonumber\\
&=&{\cal S}_A^{(4)}(\bB',\bB)+{\cal S}_A^{(2)}(\bb',\bb)-
{\cal S}_A^{(3)}(\bB',\bb)-{\cal S}_A^{(3)}(\bb',\bB),
\label{eq:2.7} 
\eea
\end{widetext}
(wherever it would not cause a confusion, we suppress the 
superscript $(n)$ for the $n$-parton states).
The crucial point is that $\textsf{S}^{\dagger}$ can be interpreted 
as an $\textsf{S}$-matrix for the scattering of antiquarks
\cite{SlavaPositronium,NPZcharm,Nonlinear,SingleJet,QuarkGluonDijet}. 
Then 
four terms in the last line of (\ref{eq:2.7}) are contributions 
from the intermediate nuclear state with a certain
set of color-excited and ground-state nucleons, $\ket{A_f}$, to
the $\textsf{S}$-matrices of the interaction of color-singlet multiparton states 
with the ground state of the nucleus $\ket{A_{in}}$. This
point will be explained to a more detail below in Sec. III.


\section{The master formula for topological nuclear cross sections}


\subsection{Basics of the color dipole approach}

 To the lowest
order in pQCD, the inelastic color triplet-antitriplet 
dipole-nucleon interaction is driven 
by one-gluon exchange and the  total inelastic 
cross section is described by the two-gluon 
exchange approximation \cite{NZ91,Nonlinear}.
Correspondingly, the $\textsf{S}$-matrices of the
quark-nucleon and antiquark-nucleon interaction must be evaluated
to the second order in pQCD. They equal, respectively, 
\bea 
&&\textsf{S}(\bb_q)  = \openone +i\hat{\chi}(\bb_q) -{1\over
  2}\hat{\chi}^2(\bb_q)
\nonumber\\
&=&\openone + iT^a V_a\chi(\bb_q)- {1\over 2}
T^aT^a \chi^2(\bb_q)\, , \nonumber\\
&&\textsf{S}^\dagger(\bb_{\bar{q}})  =\openone-i\hat{\chi}(\bb_{\bar{q}}) -{1\over
  2}\hat{\chi}^2(\bb_{\bar{q}})\nonumber\\
&=& 
\openone
- iT^aV_a \chi(\bb_{\bar{q}}) - {1\over 2}T^aT^a \chi^2(\bb_{\bar{q}})\,,
\label{eq:3.A.1}
\eea
were $\hat{\chi}(\bb)=T^a V_a\chi(\bb)$, proportional to the strong coupling
$\alpha_S$, is the eikonal for the quark-nucleon single-gluon 
exchange, the summation goes over the color-octet indices $a$.
The terms linear in $\chi(\bb)$ describe the scattering with
color excitation of the target nucleon. 
To the desired pQCD accuracy, in the second order terms in (\ref{eq:3.A.1})
we only keep the contribution which is diagonal
in the target color state. The distinction between the 
color-excitation and color-diagonal -- elastic -- interaction is
explicit in Fig. \ref{fig:single-jet_a_to_bc}. 
The vertex $V_a$ for excitation of the nucleon $g^a N \to N^*_a$ 
into color-octet state is so normalized that after application of closure over
the final state excitations $N^*$ the vertex $g^a g^b
NN$ equals $\bra{ N} V_a^\dagger V_b \ket{ N}=\delta_{ab}$. 
The second order terms in (\ref{eq:3.A.1})
do already use this normalization. The  $\textsf{S}$-matrix (\ref{eq:3.A.1})
satisfies the unitarity condition $\textsf{S}(\bb_q)\textsf{S}^\dagger(\bb_q)=\openone$.

The $\textsf{S}$-matrix of the
$(q\bar{q})$-nucleon interaction equals
\beq
\textsf{S}_{q\bar{q}}(\bb_q,\bb_{\bar{q}})=
{\bra{N}{\rm Tr}[\textsf{S}(\bb_q)\textsf{S}^\dagger(\bb_{\bar{q}})] 
\ket{N} \over \bra{N}  {\rm Tr}\openone \ket{N}} \,.
\label{eq:3.A.2}
\eeq
 The corresponding profile function is 
${\Gamma}_{q\bar{q}}(\bb_q,\bb_{\bar{q}})= 1 - \textsf{S}_{q\bar{q}}(\bb_q,\bb_{\bar{q}})$.
The dipole cross section for interaction of the color-singlet
$q\bar{q}$ dipole, $\br=\bb_q-\bb_{\bar{q}}$,  with the free
nucleon
is obtained
upon the integration over the overall impact parameter, 
\bea
\sigma(x,\br) =
C_F \int
d^2\bb_q [\chi(\bb_q)-\chi(\bb_{q}-\br)]^2\, , 
\label{eq:3.A.3} 
\eea
whereas for a dipole made of a 
parton $R$ and antiparton $\overline{R}$ in the color representation $R$
\bea
&&\sigma_{R\overline{R}}(x,\br) = 2\int d^2\bb_{q} 
\Gamma_{R\overline{R}}(\bb_q,\bb_q-\br) \nonumber \\
&=&\int
d^2\bb_q{\bra{N} {\rm Tr}[\hat{\chi}(\bb_q)-\hat{\chi}(\bb_{q}-\br)]^2\ket{N} 
 \over \textsf{dim[R]}}\nonumber\\
&=& {C_R \over C_F} \sigma(x,\br)\, .
\label{eq:3.A.4} 
\eea
Here $C_R$ is the quadratic Casimir operator for, and $\textsf{dim[R]}$
is the dimension of, the multiplet $R$.

The $k_{\perp}$-factorization formula in terms of the 
glue in the target reads \cite{NZ94,NZglue}
\bea 
\sigma(x,\br) &=& \int d^2\bkappa f (x,\bkappa)
[1-\exp(i\bkappa\br)]\nonumber\\
&=&{1\over 2}  \int d^2\bkappa f (x,\bkappa)\nonumber\\
&\times& [1-\exp(i\bkappa\br)][1-\exp(-i\bkappa\br)],\nonumber\\
 \label{eq:3.A.5} 
\eea
where
\bea
f (x,\bkappa)&=& {4\pi\alpha_S(r)\over N_c}
\cdot {1\over \kappa^4} \cdot {\cal
F}(x,\kappa^2)
\label{eq:3.A.6} 
\eea
and
\beq
{\cal F}(x,\kappa^2) = {\partial G(x, \kappa^2) \over \partial \log \kappa^2}
 \label{eq:3.A.7}
\eeq 
is  the unintegrated glue in the
target nucleon. The second integral form of the dipole
cross section is the preferred one from the viewpoint of
the summation of four Feynman diagrams with  
$t$-channel two-gluon exchange. 
We shall also encounter the dipole cross 
section for large $q\bar{q}$ dipoles,
\bea
&&\sigma_0(x) \equiv \sigma(x,\infty)\nonumber\\
&=&2 C_F \int
d^2\bb_q \chi^2(\bb_q)
=
\int d^2\bkappa f(x,\bkappa) .
\label{eq:3.A.8} 
\eea

The leading Log${1\over x}$ evolution of the dipole
cross section is governed by the color-dipole
Balitsky-Fadin-Kuraev-Lipatov (BFKL) evolution
\cite{NZZBFKL,NZ94}, the same evolution for the
unintegrated gluon density is governed by the familiar
momentum-space BFKL equation \cite{BFKL}. 
Hereafter, unless it may cause a confusion, we
suppress the variable $x$ in the gluon densities,
dipole cross sections and $\textsf{S}$-matrices.


\subsection{Basics of the Glauber-Gribov approach}

 We follow the standard treatment of a 
nucleus as {\it an uncorrelated dilute gas of 
color-singlet nucleons} with the total
wave function
\beq
\ket{A_{in}} = \Psi_{in}(\br_A,\dots ,\br_i)\prod_{i=1}^{A}
\ket{N_i}\,,
\label{eq:3.B.1}
\eeq
where $\ket{N_i}$ is the intrinsic wave function of the
nucleon $N_i$ with the coordinate $\br_i=(\bb_i,z_i)$ and 
$\Psi_{in}(\br_A,\dots ,\br_i)$ is the configuration-space wave
function. Relativistic partons propagate along straight-path
trajectories, see the frozen configuration of
the nucleus, and the $\textsf{S}$-matrix for interaction with the nucleus
equals (here $\bB=\bb_a,\{\bb_b,\bb_c\}$) \cite{Glauber,Gribov}
\beq
\textsf{S}_A(\bB-\bb_{\{A\}})=
\prod_{i=1}^{A}\textsf{S}(\bB-\bb_{i})\, .
\label{eq:3.B.2}
\eeq
The matrix product in the color space and the longitudinal 
ordering  $z_A \geq z_{A-1}\geq \dots\geq z_2 \geq z_1$
are understood.

For the transition from the ground state
of the target nucleus to the final state with the nucleon $N_i$ produced
in the state $\ket{N_{if}}$ one needs 
\begin{widetext}
\bea
\bra{A_f}\textsf{S}_A(\bB)\ket{A_{in}} &=& \int d^3\br_1\dots d^3\br_A 
\Psi_{f}^*(\br_A,\dots ,\br_1) \Psi_{in}(\br_A,\dots ,\br_1)\nonumber\\
&\times&\Biggl(\bra{N_{Af}}\textsf{S}(\bB-\bb_{A})\ket{N_A} \cdot
\bra{N_{(A-1)f}}\textsf{S}(\bB-\bb_{A-1})\ket{N_{(A-1)}} \nonumber\\
&\times& \dots
\times 
\bra{N_{2f}}\textsf{S}(\bB-\bb_{2})\ket{N_2} \cdot
\bra{N_{1f}}\textsf{S}(\bB-\bb_1)\ket{N_1}\Biggr)_{c_f c_i}\,.
\label{eq:3.B.3}
\eea
The next step is an evaluation of the multiparton ${\cal S}_A(\bB',\bB)$ in (\ref{eq:2.7}):
\bea
{\cal S}_A(\bB',\bB)&=& A!\int d^3\br_1'\dots d^3\br_A' d^3\br_1\dots d^3\br_A
\nonumber\\
&\times&
\theta(z_A'-z_{A-1}') \dots \theta(z_2'-z_1')\
\theta(z_A-z_{A-1})\dots\theta(z_2-z_1)\nonumber\\
&\times&  \Psi_{in}^*(\br_A',\dots,\br_1')\Psi_{f}(\br_A',\dots,\br_1')
\Psi_{f}^*(\br_A,\dots,\br_1) \Psi_{in}(\br_A,\dots,\br_1)\nonumber\\
&\times&{\rm Tr}\Biggl\{
\bra{N_1}\textsf{S}^{\dagger} (\bB'-\bb_1')\ket{N_{1f}}\cdot
\bra{N_2}\textsf{S}^{\dagger}(\bB'-\bb_{2}')\ket{N_{2f}}\nonumber\\
&\times& \dots \times\bra{N_A}\textsf{S}^{\dagger}(\bB'-\bb'_{A})\ket{N_{Af}} \cdot
\bra{N_{Af}}\textsf{S}(\bB-\bb_{A})\ket{N_A} \nonumber\\
&\times& \dots \times
\bra{N_{2f}}\textsf{S}(\bB-\bb_{2})\ket{N_2} \cdot
\bra{N_{1f}}\textsf{S}(\bB-\bb_{A-1})\ket{N_1}\Biggr\}\,.
\label{eq:3.B.4}
\eea
\begin{figure}[!h]
\begin{center}
\includegraphics[width = 5cm,angle=270]{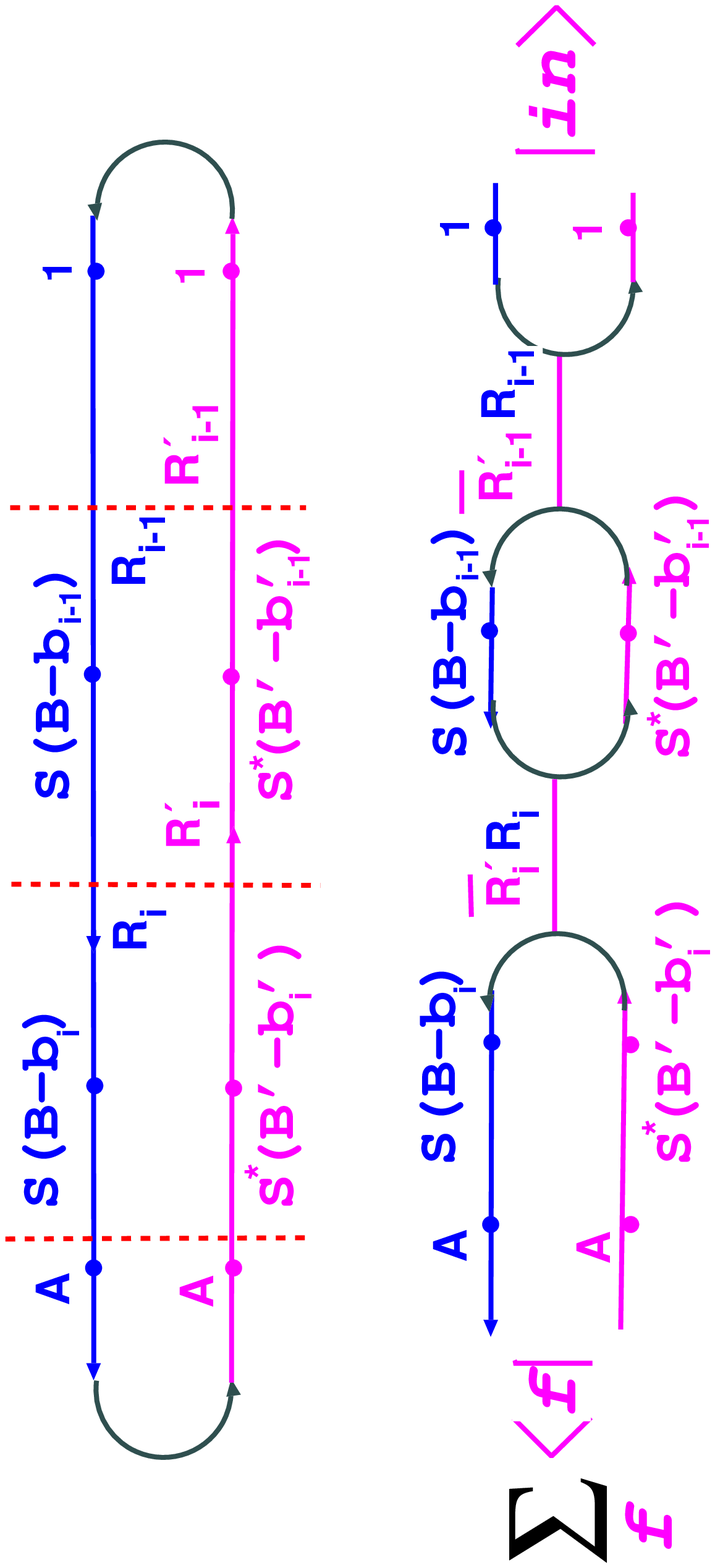}
\caption{The top diagram shows a 
trace ${\cal S}_A(\bB',\bB)$ of the product of strings of nucleonic matrix
elements of $\textsf{S}$-matrices and $\textsf{S}^\dagger$-matrices 
which enters the calculation of the dijet spectrum. The
bottom diagram shows the result for ${\cal S}_A(\bB',\bB)$ after the Fierz transformation to
the basis of multiparton Fock states $R_i\overline{R}_i'$. The color representation
structure of the sum over final states is specified in Appendices A-C. }
\label{fig:NuclearColorLoop}
\end{center}
\end{figure}

\end{widetext}
The color trace in the integrand of (\ref{eq:3.B.4}) contains the product
of the string of $\textsf{S}$-matrices and the string of 
$\textsf{S}^\dagger$-matrices
 and graphically can be represented
by the color loop diagram in the top part of Fig. \ref{fig:NuclearColorLoop}.  
Notice the inverted $z$-ordering in the string for $\text{S}_A^\dagger$,
by which, after
the Fierz transformation, the
color loop diagram reduces to the product of matrix elements
\beq
\bra{N_i}\textsf{S}^{\dagger}(\bB'-\bb'_{i})\ket{N_{if}}\cdot
\bra{N_{if}}\textsf{S}(\bB-\bb_{i})\ket{N_i}. 
\label{eq:3.B.5}
\eeq
for transitions $R_{i-1}\overline{R}_{i-1}´ \to R_i\overline{R}_i´ $
with the intermediate state nucleon in the color state $\ket{N_{if}}$,
see below Sec. III.C. 
One sums over all allowed 
multiparton color-singlet 
states
$R_i\overline{R}_i´=a\bar{a},a\{\bar{b}\bar{c}\},\bar{a}\{bc\},\{bc\}\{\bar{b}\bar{c}\}$, 
where  $R_i=\overline{R}_i´$ are color representations 
of the parton and antiparton systems.
This is shown schematically in the bottom part of
Fig. \ref{fig:NuclearColorLoop}. 


\subsection{Fully inclusive nuclear cross section}

First we recall the relevant Glauber-Gribov 
formalism for the fully inclusive case, when one sums over nuclear final
states and all color excitations of nucleons. The application of 
the configuration-space closure relation over nuclear final states
gives
\bea
&&\sum_f \Psi_{f}(\br_A',\dots ,\br_1')
\Psi_{f}^*(\br_A,\dots ,\br_1)\nonumber\\
&=& 
 \prod_{i=1}^A \delta(\br_i'-\br_i) 
=   \prod_{i=1}^A \delta(z_i'-z_i)
\delta(\bb_i'-\bb_i).\nonumber\\
 \label{eq:3.C.1}
\eea
Then, upon the application of the
closure relation, $\sum_f\ket{N_{if}}\bra{N_{if}}=1$, Eq. (\ref{eq:3.B.5})
reduces to the single-nucleon matrix element
\bea
&&\sum_{f} \bra{N}\textsf{S}^{\dagger}(\bB'-\bb'_{i})\ket{N_{if}} \cdot
\bra{N_{if}}\textsf{S}(\bB-\bb_{i})\ket{N} =\nonumber\\
&&\bra{N}\textsf{S}^{\dagger}(\bB'-\bb'_{i})\textsf{S}(\bB-\bb_{i})\ket{N}
= {\cal S}(\bC-\bb_{i}),
\label{eq:3.C.2} 
\eea
 where ${\cal S}(\bC-\bb_{i})$ is the $\textsf{S}$-matrix in
the basis of  multiparton
states $R_i\bar{R}_i´$ and $\bC$ is a short hand notation for  sets of
impact parameters $\bC=\{\bB',\bB\},\{\bB',\bb\},\{\bb',\bB\},\{\bb',\bb\}$ . 
The coupled-channel multiparton 
${\cal S}(\bC)$ defines the coupled-channel profile function,
${ \hat \Gamma}(\bC) = \openone - {\cal S}(\bC)$,
and the coupled-channel operator of dipole cross sections, which 
is obtained upon the integration over the overall
impact parameter $\bc$:
\beq
\hat{\Sigma}(\bC)=2\int d^2\bc\hat{\Gamma}(\bC)\, .
 \label{eq:3.C.3}
\eeq

Eq. (\ref{eq:3.B.4}) becomes a target-nucleus expectation value  
\bea
{\cal S}_A(\bC)&=& A!\int  d^3\br_1\dots d^3\br_A
\nonumber\\
&\times&\theta(z_A-z_{A-1})\dots \theta(z_2-z_1)\nonumber\\
&\times&  \Psi_{in}^*(\br_A',\dots ,\br_1') \Psi_{in}(\br_A,\dots ,\br_1)\nonumber\\
&\times&
\prod_{i=1}^A \Bigl[1-\hat{\Gamma}(\bC-\bb_i)\Bigr] 
\label{eq:3.C.4}
\eea
In the dilute nucleonic gas approximation
\beq
|\Psi_{in}(\br_A,\dots ,\br_1)|^2 =\left({1\over A}\right)^A\prod_{i=1}^A
n_A(\br_i)\, ,
 \label{eq:3.C.5}
\eeq
and 
\bea
&&{\cal S}_A(\bC)= A! \int  dz_1\dots dz_A
\nonumber\\
&\times&\theta(z_A-z_{A-1})\dots \theta(z_2-z_1)\nonumber\\
&\times& \prod_{i=1}^A \Biggl\{ {1\over A} \int d^2\bb_i n_A(z_i,\bb_i)
\Bigl[1-\hat{\Gamma}(\bC-\bb_i)\Bigr] \Biggr\}\,, \nonumber\\
\label{eq:3.C.6}
\eea
where the nuclear matter  density is
normalized according to $\int d^3\vec{r} \, n_A(\br) = A$.
The size of the multiparton system is much smaller than the radius 
of the nucleus and 
\bea
&&\int d^2\bb_i n_A(z_i,\bb_i)\hat{\Gamma}(\bC-\bb_i)=\nonumber\\
&=&n_A(z_i,\bb)
\int d^2\bc \hat{\Gamma}(\bC)={1\over 2}\hat{\Sigma}(\bC) n_A(z_i,\bb)\,,\nonumber\\
\label{eq:3.C.7} 
\eea 
where $\bb$ is the impact parameter of the multiparton system with 
respect to the center of mass of the target nucleus.
The $z$-integrations
introduce the optical thickness of the nucleus
\bea
T(\bb)= \ds \int_{-\infty}^\infty dz \, n_A(\bb,z).
\label{eq:3.C.8} 
\eea 
Then  the coupled-channel Glauber-Gribov formula
\cite{Glauber,Gribov}
for the $n$-parton states is 
\bea
&&{\cal S}_A^{(n)}(\bC)=
\Bigl[ 1-{1\over 2A} \hat{\Sigma}^{(n)}(\bC)T(\bb)\Bigr]^A\nonumber\\
&=&
\exp\Bigl[-{1\over 2}\hat{\Sigma}^{(n)}(\bC)T(\bb)\Bigr]= 
{\textsf S}[\bb,\Sigma^{(n)}(\bC)].\nonumber\\
\label{eq:3.C.9}
\eea
The exponentiation holds for medium to heavy nuclei.


\subsection{Separation of elastic and color-excitation scattering}

The generalization of the Glauber-Gribov representation to final
states with fixed number of color excited nucleons proceeds
as follows.
Let the propagating system - either incident parton $a$ or the
produced state $bc$ undergo color-excitation interactions with nucleons
$k_1,\dots ,k_\nu$ at impact parameters $\bc_1,\dots ,\bc_\nu$, 
and color-diagonal, elastic, interactions with
all other nucleons $\{1,\dots ,k_1-1\}$, $\{k_1+1,\dots ,k_2-1\}$,\dots ,$
\{k_\nu+1,\dots,A\}$ at corresponding sets of impact parameters $\{\bb_i\}$. A configuration-space
closure over nuclear final states is understood, we are only after the
color algebra.
The case of $\nu=0$, i.e., diffractive interaction without color-excitations,
must be included too.  
The operator of interest, summed over fixed configurations of scatterers,  is
\bea
&& {\cal G}_{\nu}(\bC,\{\bb_j\},\{\bc_i\})= 
\nonumber\\
&=&\sum^A_{k_\nu > k_\nu-1>\dots >k_1}{\rm Tr}\Biggl\{
\Bigl[\prod_{i=1}^{k_1-1}\bra{N_i}
\textsf{S}^{\dagger} (\bB'-\bb_i)\ket{N_i} \Bigr]\nonumber\\
&\times& \bra{N_1}\textsf{S}^{\dagger}(\bB'-\bc_1)\ket{N^*_1}\nonumber\\
&\times& \Bigl[\prod_{i=k_1+1}^{k_2-1}\bra{N_i}
\textsf{S}^{\dagger} (\bB'-\bb_i)\ket{N_i}\Bigr] \nonumber\\ 
&\times& \dots \times\bra{N_{\nu}}\textsf{S}^{\dagger}(\bB'-\bc_{\nu})\ket{N^*_{\nu}}\nonumber\\ 
&\times& 
\Bigl[\prod_{i=k_\nu+1}^{A}\bra{N_i}
\textsf{S}^{\dagger} (\bB'-\bb_i)\ket{N_i}\Bigr] \nonumber\\ 
&\times&
\Bigl[\prod_{i=k_\nu+1}^{A}\bra{N_i}\textsf{S}(\bB-\bb_i)\ket{N_i}\Bigr] 
\bra{N^*_{\nu}}\textsf{S}(\bB-\bc_{\nu})\ket{N_{\nu}} \nonumber\\ 
&\times& \dots 
\times \Bigl[\prod_{i=k_1+1}^{k_2-1}\bra{N_i}
\textsf{S}(\bB - \bb_i)\ket{N_i} \Bigr]\nonumber\\
&\times& \bra{N^*_1}\textsf{S}(\bB-\bc_1)\ket{N_1}
\prod_{i=1}^{k_1-1}\Bigl[\bra{N_i}\textsf{S} (\bB-\bb_i)\ket{N_i} \Bigr]\Biggr\}\nonumber\\
&+&\delta_{\nu 0} {\rm Tr}\Biggl\{
\Bigl[\prod_{i=A}^{1}\bra{N_i}
\textsf{S}^{\dagger} (\bB'-\bb_i)\ket{N_i} \Bigr]\nonumber\\
&\times&
\prod_{i=1}^{A}\Bigl[\bra{N}\textsf{S} (\bB-\bb_i)\ket{N} \Bigr]\Biggr\}\,.
\label{eq:3.D.1}
\eea
Here $\ket{N^*_i}$ denotes the color-excited nucleon.
Based on the above described Fierz transformation technique, 
we reduce (\ref{eq:3.D.1}) to the product of multiparton operators.
\begin{figure}[!h]
\begin{center}
\includegraphics[width = 5cm]{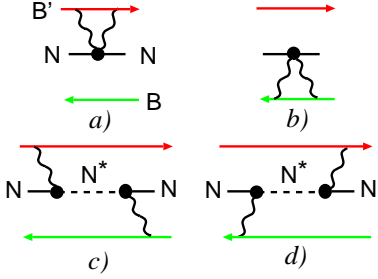}
\caption{Color-diagonal elastic $(a,b)$ and color-excitation $(c,d)$ 
scattering of multiparton states. In elastic scattering both gluons couple
to either the partonic subsystem at impact parameters $\{\bB\}$
or the antipartonic subsystem at impact parameters $\{\bB'\}$. In
color-excitation scattering one gluon couples to the partonic subsystem
and the another to the antipartonic subsystem.}
\label{fig:SigmaElEx}
\end{center}
\end{figure}

Let $\hat{\chi}(\bB)$ 
and $\hat{\chi}(\bB')$ be the eikonals for the interaction
with the nucleon of the partonic and antipartonic system, respectively. 
The terms linear in $\hat{\chi}(\bB)$ and and $\hat{\chi}(\bB')$
in the expansion (\ref{eq:3.A.1}) do not contribute to color-diagonal
interactions, where one encounters 
\bea
&&\openone-\hat{\Gamma}_{el}(\bC-\bb)=\nonumber\\
&=&\bra{N}\textsf{S}^{\dagger} (\bB'-\bb)\ket{N}
\bra{N}\textsf{S} (\bB-\bb)\ket{N}= \nonumber\\
&=& \bra{N}\textsf{S}^{\dagger} (\bB'-\bb)\ket{N}
\bra{N}\openone\ket{N}\nonumber \\
&+&\bra{N}\openone\ket{N}
\bra{N}\textsf{S} (\bB-\bb)\ket{N}-\openone\nonumber\\
&=& \openone-{1\over 2} \Bigl[
\bra{N}\hat{\chi}^2(\bB-\bb)\ket{N}
\nonumber\\
&+&
\bra{N}\hat{\chi}^2(\bB'-\bb)\ket{N}\Bigr].
\label{eq:3.D.2}
\eea

It sums the Feynman diagrams of Fig. \ref{fig:SigmaElEx}a,b 
when both gluons are exchanged with either partonic or antipartonic
subsystem. The profile function 
$\hat{\Gamma}_{el}(\bC)$ is still an operator 
in the space of multiparton color-singlet
states. 
The color-excitation scattering is described by diagrams of 
Fig. \ref{fig:SigmaElEx}c,d in which one of the $t$-channel 
gluons is exchanged with the 
partonic subsystem while the second is exchanged with the 
antipartonic subsystem. Only the terms linear in 
$\hat{\chi}(\bB)$ and and $\hat{\chi}(\bB')$
in the expansion (\ref{eq:3.A.1}) do contribute. The 
diagonal matrix elements vanish, $\bra{N}\hat{\chi}(\bB' -\bc)\ket{N}=0$,
and 
one can sum over unobservable color states 
of excited nucleons $N^*$ making use of the closure:
\bea
&&\hat{\Gamma}_{ex}(\bC)=
\nonumber\\
&-&\sum_{N^*}
\bra{N}\textsf{S}^{\dagger}(\bB'-\bc)\ket{N^*}\bra{N^*}\textsf{S}(\bB-\bc)\ket{N}
\nonumber\\
&=&- \sum_{N^*}\bra{N}\hat{\chi}(\bB' -\bc)\ket{N^*}\bra{N^*}\hat{\chi}(\bB-\bc)\ket{N}
\nonumber\\
&=& -\bra{N}\hat{\chi}(\bB' -\bc)\hat{\chi}(\bB-\bc)\ket{N}.
\label{eq:3.D.3}
\eea
Now we notice that $
\hat{\Gamma}(\bC)=\hat{\Gamma}_{ex}(\bC)+\hat{\Gamma}_{el}(\bC)$
and \bea
 \hat{\Sigma}_{ex}(\bC)+\hat{\Sigma}_{el}(\bC)=\hat{\Sigma}(\bC)
\label{eq:3.D.4}
\eea   
 holds for the corresponding cross-section operators, $
\displaystyle \hat{\Sigma}_{el,ex}(\bC)= 2\int d^2\bc \Gamma_{el,ex}(\bC)$.


\subsection{The unitarity relation for color excitation of the nucleus}

In terms of $\Gamma_{ex}(\bC)$ and $\Gamma_{el}(\bC)$ we obtain
\begin{widetext}
\bea
&&{\cal G}_{\nu}(\bC,\{\bb_j\},\{\bc_i\})=  \sum^A_{k_\nu > k_\nu-1>\dots >k_1} 
\Bigl\{\prod_{i=k_\nu+1}^{A}[1-\hat{\Gamma}_{el}(\bC-\bb_i)] \Bigr\}
\cdot \Bigl\{ -\hat{\Gamma}_{ex}(\bC-\bc_\nu) \Bigr\}\nonumber\\
&\times&\Bigl\{\prod_{i=k_{\nu-1}+1}^{k_{\nu}-1}
[1-\hat{\Gamma}_{el}(\bC-\bb_i)] \Bigr\} 
\dots
\Bigl\{\prod_{i=k_1+1}^{k_2-1}[1-\hat{\Gamma}_{el}(\bC-\bb_i)] \Bigr\}\nonumber\\
&\times& 
\Bigl\{ -\hat{\Gamma}_{ex}(\bC-\bc_1) \Bigr\}
\Bigl\{\prod_{i=1}^{k_1-1}[1-\hat{\Gamma}_{el}(\bC-\bb_i)] \Bigr\}
+ \delta_{\nu 0}
\Bigl\{\prod_{i=1}^{A}[1-\hat{\Gamma}_{el}(\bC-\bb_i)] \Bigr\}\,,
\label{eq:3.E.1}
\eea
where the last term, $\propto  \delta_{\nu 0}$, describes CD.
Now,  consider an expansion of
\beq
\prod_{i=1}^{A}[1-\hat{\Gamma}(\bC-\bb_i)]=
\prod_{i=1}^{A}[1-\hat{\Gamma}_{el}(\bC-\bb_i)-\hat{\Gamma}_{ex}(\bC-\bb_i)]
\label{eq:eq:3.E.2}
\eeq
in powers
of $\hat{\Gamma}_{ex}$.  One must
be careful about the ordering of the not-commuting   
$\hat{\Gamma}_{ex}(\bC)$ and $\hat{\Gamma}_{el}(\bC)$:
\bea
&&\prod_{i=1}^{A}[1-\hat{\Gamma}_{el}(\bC-\bb_i)-\hat{\Gamma}_{ex}(\bC-\bb_i)]=
\prod_{i=1}^{A}[1-\hat{\Gamma}_{el}(\bC-\bb_i)]\nonumber\\
&+&
\sum_{k_1=1}^{A}\Bigl\{\prod_{i=k+1}^{A}[1-\hat{\Gamma}_{el}(\bC-\bb_i)] \Bigr\}
\Bigl\{ -\hat{\Gamma}_{ex}(\bC-\bb_{k_1}) \Bigr\}
\Bigl\{\prod_{i=1}^{k-1}[1-\hat{\Gamma}_{el}(\bC-\bb_i)] \Bigr\} \nonumber\\
&+&
\sum_{k_2>k_1=1}^A
\Bigl\{\prod_{i=k_2+1}^{A}[1-\hat{\Gamma}_{el}(\bC-\bb_i)] \Bigr\}
\Bigl\{ -\hat{\Gamma}_{ex}(\bC-\bb_{k_2}) \Bigr\}
\Bigl\{\prod_{i=1}^{k_2-1}[1-\hat{\Gamma}_{el}(\bC-\bb_i)] \Bigr\}\nonumber\\
&\times& 
\Bigl\{ -\hat{\Gamma}_{ex}(\bC-\bb_{k_1}) \Bigr\}
\Bigl\{\prod_{i=1}^{k_1-1}[1-\hat{\Gamma}_{el}(\bC-\bb_i)] \Bigr\}
+\dots =\sum_{\nu=0}^A {\cal G}_{\nu}(\bC,\{\bb_j\},\{\bc_i\}),
\label{eq:3.E.3}
\eea
\end{widetext}
where $\{\bc_i\} \equiv \{\bb_{k_i}\}$.
This identification of ${\cal G}_{\nu}(\bC,\{\bb_j\},\{\bc_i\})$ 
furnishes a proof of the unitarity relation: 
the reaction operators
for topological cross
sections with color excitation of $\nu$ nucleons
sum up to exactly the  reaction operator for the total cross section.

The operators $\hat{\Gamma}_{ex}(\bC)$ and $\hat{\Gamma}_{el}(\bC)$ 
separate specific color transitions and, in the general case, 
will have certain infrared sensitivity. The above proof of the 
unitarity shows that this infrared sensitivity exactly cancels 
out in the total cross section described by the infrared-safe 
$\hat{\Gamma}(\bC)$.


\subsection{Color excitation of a nucleus:
cut and uncut pomerons}

The above proof of the unitarity relation spares the re-derivation 
of the nuclear matrix element
\bea
{\cal S}_{A,\nu}(\bC)= \bra{A_{in}} {\cal
  G}_{\nu}(\bC,\{\bb_j\},\{\bc_i\})
\ket{A_{in}}.
\label{eq:3.F.1}
\eea
It is obtained by a direct expansion of 
the Glauber-Gribov result (\ref{eq:3.C.9}) in powers of 
$\hat{\Sigma}_{ex}$. Going from  $z_i$ to a depth in the nucleus in
units of $T(\bb)$,
\bea
\beta_i T(\bb)= \int_{-\infty}^{z_i} dz n_A(\bb,z),
\label{eq:3.F.2}
\eea
we find
\bea
&&{\cal S}_{A,\nu}(\bC)
= \int_{0}^1  d\beta_{\nu} \dots d\beta_{1} \nonumber\\
&\times&  
\theta(1 -\beta_\nu)\theta(\beta_\nu -\beta_{\nu-1}) \dots \theta(\beta_2-\beta_1)\theta(\beta_1)\nonumber\\
&\times&
\exp\Bigl[-{1\over 2}(1 -\beta_\nu)\hat{\Sigma}_{el}(\bC)T(\bb)\Bigr] \nonumber\\
&\times&
\Bigl\{ -{1\over 2}\hat{\Sigma}_{ex}(\bC)T(\bb)\Bigr\}\nonumber\\
&\times&
\exp\Bigl[-{1\over
  2}(\beta_\nu-\beta_{\nu-1})\hat{\Sigma}_{el}(\bC)T(\bb)\Bigr] \dots \nonumber\\
&\times& 
\Bigl\{ -{1\over 2}\hat{\Sigma}_{ex}(\bC)T(\bb) \Bigr\}
\exp\Bigl[-{1\over 2}\beta_1 \hat{\Sigma}_{el}(\bC)T(\bb)\Bigr]
\nonumber\\
&+& \delta_{\nu 0} \exp\Bigl[-{1\over 2}\hat{\Sigma}_{el}(\bC)T(\bb)\Bigr]\, .
\label{eq:3.F.3}
\eea

An alternative derivation of (\ref{eq:3.F.3}) is useful. The heavy-nucleus 
Glauber-Gribov result (\ref{eq:3.C.9}) can be viewed as a solution of
the differential equation
\bea
{d\over dz} {\cal S}_A(\bC,z)= -{1\over 2}n_A(\bb,z)\hat{\Sigma}(\bC)
{\cal S}_A(\bC,z)\,
\label{eq:3.F.4}
\eea
i.e., 
\bea
&&{d\over d\beta} {\cal S}_A(\bC,\beta)= -{1\over 2}T(\bb)\hat{\Sigma}(\bC)
{\cal S}_A(\bC,\beta)\nonumber\\
&=&-{1\over 2}T(\bb)
[\hat{\Sigma}_{el}(\bC)+\hat{\Sigma}_{ex}(\bC)]{\cal S}_A(\bC,\beta)
\label{eq:3.F.5}
\eea
subject to the boundary condition $ {\cal S}_A(\bC,0)=\openone$.
A full thickness of the nucleus corresponds to $\beta=1$.
Solve this equation treating $\hat{\Sigma}_{ex}$ as a perturbation. The $\nu$-th
iteration will give precisely (\ref{eq:3.F.3}).

Finally, the counterpart of ${\cal T}(\bB,\bB',\bb,\bb')$
 of Eq. (\ref{eq:2.7}) takes the form
\bea
&&{\cal T}_{\nu}(\bB,\bB',\bb,\bb')={\cal S}_{A,\nu}^{(4)}(\bB',\bB)+{\cal
 S}_{A,\nu}^{(2)}(\bb',\bb)
\nonumber\\
&-&
{\cal S}_{A,\nu}^{(3)}(\bB',\bb)-{\cal S}_{A,\nu}^{(3)}(\bb',\bB) \, .
\label{eq:3.F.6} 
\eea
Equations (\ref{eq:3.F.3}) and (\ref{eq:3.F.6}) are our 
manifestly unitary master formulas for  
the hard scattering accompanied by color excitation of  $\nu$-nucleons.
They allow us to classify the color-excitation final states according to
the universality classes for nonlinear $k_\perp$ factorization 
introduced in \cite{Nonuniversality,QuarkGluonDijet}. A contact with the
reggeon field theory is obvious: {\it{ each and every $\hat{\Sigma}_{ex}$ will
be associated with a cut pomeron, while the exponentials of 
$\hat{\Sigma}_{el}$ re-sum absorption corrections for multiple exchanges 
by the uncut pomerons}}.


\section{Collective nuclear glue: cut and uncut nuclear pomerons and 
sequential quasielastic rescattering
of quarks}


\subsection{Collective glue from the color-dipole $\textsf{S}$-matrix}

In the master formula (\ref{eq:2.5}) for the dijet cross section one 
encounters the Fourier transform of the nuclear
${\textsf S}$-matrix  ${\textsf S}[\bb,\hat{\Sigma}(\bC)]$.
The matrix elements of $\hat{\Sigma}(\bC)$  
are superpositions of the elementary dipole cross sections
and, consequently, ${\textsf S}[\bb,\Sigma(\bC)]$ will be a 
product of ${\textsf S}$ matrices for elementary dipoles.
The Fourier transform of  ${\textsf S}[\bb,\sigma(x,\br)]$
for the elementary $q\bar{q}$ dipole gives the amplitude
for coherent excitation of hard dijets in $\pi A$
collisions - the reference process for the definition of
the collective nuclear unintegrated glue 
\cite{NSSdijet,Nonlinear,VirtualReal}. 

Specifically, 
for the color triplet-antitriplet dipoles we
define
\bea
&&\Phi(\bb,x,\bkappa) ={1\over (2\pi)^2} 
\int d^2\br \textsf{S}[\bb,\sigma(x,\br)]\exp[-i \bkappa \br]\nonumber\\
&=& \textsf{S}[\bb,\sigma_0(x)] \delta^{(2)}(\bkappa)   + \phi(\bb,x,\bkappa),
\label{eq:4.A.1}
\eea
\bea
1- \textsf{S}[\bb,\sigma(x,\br)]&=& \int d^2\bkappa \phi(\bb,x,\bkappa)
[1-\exp(i\bkappa\br)]\nonumber\\
&=&{1\over 2}  \int d^2\bkappa \phi(\bb,x,\bkappa)\nonumber\\
&\times&[1-\exp(i\bkappa\br)] [1-\exp(-i\bkappa\br)].\nonumber\\
\label{eq:4.A.2}
\eea
where $\phi(\bb,x,\bkappa)$ is a collective unintegrated glue per unit
area in the impact parameter plane. The second form of (\ref{eq:4.A.2}) 
is the preferred one from the Feynman diagram viewpoint.
The collective nuclear glue admits a simple expansion
in terms of the collective glue, $f^{(j)}(x,\bkappa)$,
for $j$ spatially overlapping nucleons of the Lorentz-contracted
ultrarelativistic nucleus:
\bea
\phi(\bb,x,\bkappa)= {1\over \sigma_0(x)}\sum_{j=1} w_j\Big(\nu_A(\bb)\Big)
f^{(j)}(x,\bkappa).
\label{eq:4.A.3}
\eea
Here 
\bea
&&f^{(j)}(x,\bkappa) = {1\over \sigma_0(x)}\nonumber\\
&\times&
\int d^2\bkappa_1
\cdot  f(x,\bkappa_1) f^{(j-1)}(x,\bkappa-\bkappa_1), 
\label{eq:4.A.4}
\eea
where $ f^{(1)}(x,\bkappa)= f(x,\bkappa)$, an extension to $j=0$ is
$f^{(0)}(x,\bkappa)=\sigma_0 (x) \delta^{(2)}(\bkappa)$,  
\bea
w_j\Big(\nu_A(\bb)\Big) ={1\over j!} \nu_A^j(\bb)
\textsf{S}[\bb,\sigma_0(x)]
\label{eq:4.A.5}
\eea
is a probability 
to find $j$ spatially overlapping nucleons in the tube of cross 
section ${1\over 2}\sigma_0(x)$ in the Lorentz-contracted
ultrarelativistic nucleus and   
\beq
\nu_A(\bb)={1\over 2}\sigma_0(x)  T(\bb)
\label{eq:4.A.6}
\eeq
is the thickness of the nucleus
in units of the interaction length for large dipoles.
All $f^{(j)}(x,\bkappa)$ have the same normalization 
$\int d^2\bkappa f^{(j)}(x,\bkappa)=\sigma_0(x)$.


\subsection{Collective glue: from the optical theorem perspective
to a coherent state of the in-vacuum gluons}
\begin{figure}[!h]
\begin{center}
\includegraphics[width = 7.4cm,angle=270]{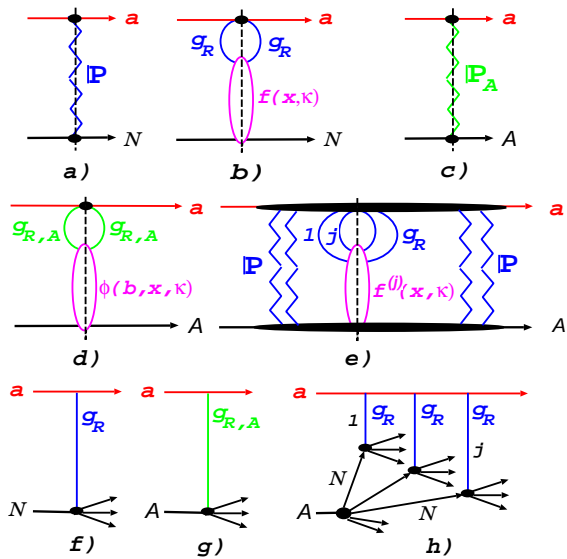}
\caption{ The cut pomeron interpretation of the unintegrated glue:
(a) the optical theorem unitarity cut for $aN$ scattering; 
(b) the unitarity cut in terms 
of the exchange by two in-vacuum reggeized gluons in
the $t$-channel and the 
free-nucleon  
glue $f(x,\bkappa)$; (c) the optical theorem for a nuclear target 
and (d) its tentative interpretation in terms of the unitarity
cut of the exchange by two coherent nuclear
gluons $g_{R,A}$  in
the $t$-channel, and the collective nuclear gluon density 
$\phi(\bb,x,\bkappa)$; (e) the contribution to the 
cut nuclear pomeron from the $j$ in-vacuum gluon component 
of the coherent nuclear
gluon $g_{R,A}$ and the related collective gluon density
$f^{(j)}(x,\bkappa)$, the absorption corrections for the exchange
by uncut pomerons on either side of the unitarity cut
are indicated; (f) the final state in the unitarity diagram
(b) in terms of the quasielastic scattering of the quark
$a$ by the $t$-channel exchange by the in-vacuum gluon
$g_R$, (g) quasielastic scattering of a quark $a$  
off a nucleus by exchange by the coherent nuclear gluon $g_{R,A}$;
(h) $t$-channel exchange by a $jg_R$ component of
the coherent 
nuclear glue in terms of the sequential $j$-fold 
quasielastic $aN$ scattering in the $aA$ interaction,
the absorption by uncut pomerons explicit in diagram
(e) is suppressed here.}
\label{fig:CoherentGlue}
\end{center}
\end{figure}

The expansion (\ref{eq:4.A.3}) admits two interpretations. 
On the one hand, from the optical theorem perspective,
$f(x,\bkappa)$ describes the unitarity cut  $\CutPom$
of the $t$-channel pomeron  $\Pom$ -- a color-singlet composite
state of two (reggeized) in-vacuum 
gluons $g_R$ in the $t$-channel, Figs.
\ref{fig:CoherentGlue}a,b. Likewise,
the $q\bar{q}$ dipole-nucleus total cross section
equals \cite{NZ91}
\bea
{d\sigma_A(x,\br)\over d^2\bb}= 2\int d^2\bkappa\phi(\bb,x,\bkappa)[1-\exp(i\bkappa \br)].
\label{eq:4.B.1}
\eea 
For very large dipoles we tentatively identify the (inelastic) 
quark-nucleus cross
section 
\bea
{d\sigma_{qA}(x) \over d^2\bb}= \int d^2\bkappa\phi(\bb,x,\bkappa)
\label{eq:4.B.2}
\eea 
and, in the optical theorem perspective, view it as a unitarity cut
of the exchange by a nuclear pomeron $\Pom_A$, Fig. \ref{fig:CoherentGlue}c.
In the analogy to the in-vacuum pomeron, we view  $\Pom_A$
as a composite state of two nuclear (reggeized) gluons $g_{R,A}$. 
The cut pomeron  $\Pom_A$ in terms of  $g_{R,A}$ on two sides
of the unitarity cut, and the r\^ole of the collective $\phi(\bb,x,\bkappa)$,
are shown in Fig. \ref{fig:CoherentGlue}d.

Furthermore, the Poisson distribution (\ref{eq:4.A.5})
suggests that such an operationally defined $g_{R,A}$
can be viewed as a coherent state \cite{CoherentState}
of the in-vacuum gluons $g_R$
with the average multiplicity $\langle j \rangle =\nu_A(\bb)$.
The $j$-gluon composite component of $g_{R,A}$ describes
the $j$-gluon unitarity cut, Fig. \ref{fig:CoherentGlue}e,
of the  $2jg_R$ component of the pomeron  $\Pom_A$.
This composite state $\Pom_A^{(2j)}$ 
will be associated with the collective glue $f^{(j)}(x,\bkappa)$
and is universal for all targets. What changes from one 
nuclear target to another is the probability 
$w_j\Big(\nu_A(\bb)\Big)$ in the expansion (\ref{eq:4.A.3}).
Finally, the $k$-th order term in
\beq
\textsf{S}[\bb,\sigma_0(x)]= \sum_{j=0} {1\over k!}
(-1)^k [{1\over 2} T(\bb)]^k\Big[\int d^2\bkappa f(x,\bkappa)\Big]^k
\label{eq:4.B.3}
\eeq
can be viewed as an absorption correction for $k$ uncut pomeron
exchanges as indicated in 
Fig. \ref{fig:CoherentGlue}e.
These highly tentative operational definitions 
for uncut and cut pomeron
exchanges need a qualification which can only be achieved
by consideration of specific final states.

\subsection{Collective glue and quasielastic quark-nucleus scattering}

On the other hand, according to Ref. \cite{VirtualReal},
the differential cross section of inclusive quasielastic quark-nucleon
scattering, $qN\to q'N^*$, summed over all excitations of the
target nucleon, is linear $k_\perp$-factorizable,
\beq
{d\sigma_{Qel}\over d^2\bq}= {1\over 2} f(x,\bq)\,,
\label{eq:4.C.1}
\eeq
where $\bq$ is the transverse momentum of the scattered
quark.
One would identify the unintegrated glue with the
cut pomeron contribution to the leading quark spectrum,
cf. diagrams. \ref{fig:CoherentGlue}b and \ref{fig:CoherentGlue}f.
This $k_\perp$-factorization is exact to the Born approximation,
to higher orders it holds to the leading log${1\over x}$
(LL${1\over x}$) approximation, when the radiation
energy loss is neglected.

The total cross section of the quasielastic scattering integrates to
\beq
\sigma_{Qel} = {1\over 2} \sigma_0(x)
\label{eq:4.C.2}
\eeq
and the differential cross section of the $\nu$-fold quasielastic
scattering equals
\bea
{d\sigma_{Qel}^{(\nu)}(\bq) \over d^2\bq}&=& {1\over \sigma_{Qel}}
\int d^2\bkappa {d\sigma_{Qel}^{(\nu)}(\bkappa) \over d^2\bkappa}
{d\sigma_{Qel}^{(\nu-1)}(\bq-\bkappa) \over d^2\bq}
\nonumber\\
&=& {1\over 2} f^{(\nu)}(x,\bq).
\label{eq:4.C.3}
\eea
To the Born approximation, quasielastic scattering
exhausts the inelastic quark-nucleus interaction.
According to Ref. \cite{VirtualReal}, the linear  $k_\perp$-factorization
(\ref{eq:4.C.1}) extends to nuclear targets too,
\beq
{d\sigma_{Qel,A}\over d^2\bb d^2\bq}=  \phi(\bb,x,\bq),
\label{eq:4.C.4}
\eeq
and here we invoke a similarity between the sets of diagrams
\ref{fig:CoherentGlue}b,f  and \ref{fig:CoherentGlue}d,g.
Making use of (\ref{eq:4.A.3}) and (\ref{eq:4.C.3}), it
can be cast in the form of
the Glauber multiple-scattering expansion for quasielastic
scattering \cite{Glauber}:
\beq
{d\sigma_{Qel,A}\over d^2\bb d^2\bq}
=
\sum_{j=1} w_j\Big(\nu_A(\bb)\Big){d\sigma_{Qel}^{(j)}(\bq) \over \sigma_{Qel}d^2\bq}
.
\label{eq:4.C.5}
\eeq
Correspondingly, the probability $w_{\nu}(\nu_A(\bb))$ to
find $\nu$ spatially overlapping nucleons in the Lorentz-contracted
nucleus amounts, in the target-nucleus frame, to the
probability of the $\nu$-fold quasielastic
scattering of the quark. In the AGK language,
$d\sigma_{Qel}^{(\nu)}(\bq)/d^2\bq$
describes the contribution from $\nu$ cut pomerons to the
leading quark spectrum. Here we emphasize
that the final state in Fig. \ref{fig:CoherentGlue}h
correspond to precisely the Mandelstam cut structure \cite{Mandelstam}
of Fig. \ref{fig:CoherentGlue}e. Screening by uncut pomerons is encoded
in $\textsf{S}[\bb,\sigma_0(x)]$ --- it gives the alternating
sign series in terms of the uncut multipomeron exchanges, familiar from models
for absorption corrections.


\subsection{Intranuclear distortion of color dipoles and
multipomeron exchanges}

In the calculation of 
topological cross sections we shall often
encounter the Glauber-Gribov ${\textsf S}$-matrix   
$ \textsf{S}[\bb,{\half}\beta\sigma(x,\br)]$  for
a finite slice $[0,\beta]$ of a nucleus. It defines the
corresponding collective glue for a slice of a nucleus
\bea
 \textsf{S}[\bb, \beta \sigma(x,\br)] = 
\int d^2 \bkappa \, \Phi(\beta; \bb, x,\bkappa) \, \exp[i \bkappa \br] .\nonumber\\
\label{eq:4.D.1}
\eea
Evidently, $\Phi(\beta; \bb, x,\bkappa)$ will expand in terms
of the same $f^{(j)}(x,\bkappa)$, only the expansion coefficients
would change:
\bea
w_j\Big(\beta \nu_A(\bb)\Big) ={1\over j!} \beta^j \nu_A^j(\bb)
\textsf{S}[\bb,\beta\sigma_0(x)].
\label{eq:4.D.2}
\eea

A pertinent quantity in our derivations will be the coherently distorted wave functions
of color dipoles
\bea
\Psi(\beta;z,\br) &\equiv&  \textsf{S}[\bb,\beta \sigma(x,\br)] \, 
\Psi(z,\br).
\label{eq:4.D.3}
\eea
In the practical evaluations in the momentum space it is convenient
to use
\bea
\Psi(\beta;z,\bp) &=& \int d^2\br 
\exp[-i \bp\br] 
\Psi(\beta;z,\br) \nonumber\\
&=& 
\int d^2\bkappa 
\Phi(\beta,\bb,x,\bkappa) 
\Psi(z, \bp-\bkappa)\nonumber\\
&=&\textsf{S}[\bb,\beta\sigma_0(x)]
\Psi(z,\bp)\nonumber\\&+&
\int d^2\bkappa \phi(\beta;\bb,x,\bkappa) 
\Psi(z, \bp-\bkappa),\nonumber\\
\label{eq:4.D.4}
\eea
where $\phi(\beta,\bb,x,\bkappa)$ is positive defined one. The 
wave function distortions shall always 
be described by the uncut  multipomeron exchanges. In terms of 
an exchange by the uncut collective nuclear pomeron $\Pom_A(\beta)$ 
\bea
\Psi(\beta;z,\bp) &=& \Psi(z,\bp) -
{1\over 2}\int d^2\bp_1 d^2\bkappa \nonumber\\
&\times&[2\delta(\bp-\bp_1)-\delta (\bp-\bkappa-\bp_1) \nonumber\\
&-&\delta (\bp+\bkappa-\bp_1)]
\phi(\beta;\bb,x,\bkappa)\Psi(z,\bp_1) \nonumber\\
&\equiv&
\Psi(z,\bp) - \Biggl(\Pom_A(\beta) \otimes \Psi\Biggr)(z, \bp).\nonumber\\
\label{eq:4.D.5}
\eea
The Feynman diagram interpretation of the four terms in 
the convolution in (\ref{eq:4.D.5}) 
is the same as for the amplitude of diffractive DIS \cite{NZ91,NSSdijet}. 
The kernel $\Pom_A(\beta)$ can further
be expanded in terms of the in-vacuum pomeron exchanges. 
If we define
\bea
\Biggl(\Pom \otimes \Psi\Biggr)(z,\bp) &\equiv& 
{1\over 2}\int d^2\bp_1 d^2\bkappa \nonumber\\
&\times&[2\delta(\bp-\bp_1)-\delta (\bp-\bkappa-\bp_1) \nonumber\\
&-&
\delta (\bp+\bkappa-\bp_1)]
f(x,\bkappa)
\Psi(z,\bp_1),\nonumber\\
\label{eq:4.D.6}
\eea
then the multipomeron exchange expansion for (\ref{eq:4.D.5})
 would read
\begin{widetext}
\bea
\Psi(\beta;z,\bp)&=& \Psi(z,\bp) 
- \sum_{\nu=1} {1\over \nu!} (-1)^{\nu-1} 
\Big[{\beta \over 2} T(\bb)\Big]^\nu
\Biggl(\underbrace{\Pom \otimes\dots\otimes \Pom}_{\nu}\otimes
\Psi\Biggr)(z,\bp)\nonumber\\
&=&\Psi(z,\bp) -\sum_{\nu=1} {1\over \nu!} (-1)^{\nu-1} \Big[{\beta \over 2}
T(\bb)\Big]^\nu\nonumber\\
&\times&
\left({1\over 2}\right)^\nu \int d^2 \bp_1 d^2\bkappa_1\dots  
d^2\bkappa_\nu
f(x,\bkappa_1)\dots f(x,\bkappa_\nu) \Psi(\beta;z,\bp_1)\nonumber\\
&\times& \sum_{i,j,k,l=0}  {\nu! \over i! j! k! l!}
(-1)^{k+l} \delta(\nu-i-j-k-l)
\delta (\bp-\sum\limits_1^k\bkappa_m+\sum\limits_{k+1}^{k+l}\bkappa_n-\bp_1).
\label{eq:4.D.7}
\eea
\end{widetext}
Although only a special subset of the momenta
$\bkappa_{m,n}$ appears in the last delta-function, the
 combinatorial structure of Eq. (\ref{eq:4.D.7}) corresponds
to summing over all the relevant permutations  of 
$\bkappa_1,\dots,\bkappa_\nu$.

Now we proceed to applications to DIS and other hard processes.
The derivation of non-Abelian coupled-channel evolution
for multiparton states is found in 
\cite{Nonlinear,SingleJet,QuarkGluonDijet,GluonGluonDijet,VirtualReal}.
The relevant results for the elastic and 
color-excitation operators are reported in Appendices A-C. 


\section{Nonlinear $k_{\perp}$-factorization for topological
cross sections in DIS off nuclei}

The incident photon, $\gamma^*$, is a color-singlet parton.
In the pQCD expansion, DIS starts with the excitation of the 
$q\bar{q}$ pair which can be in either color-singlet
or color-octet state. Speaking of the octet at
arbitrary $N_c$ should not cause any confusion.  


\subsection{DIS: RFT for the universality class of coherent diffraction}

In CD excitation of the color-singlet $\{q\bar{q}\}_0$ 
the target nucleus is
retained in the ground state and $\nu=0$. 
Combining 
\beq 
{\cal
  S}_{A,0}^{(4)}(\bB´,\bB)=\textsf{S}[\bb,\Sigma_{11}^{el}(\bC)T(\bb) ]
\label{eq:5.A.1}
\eeq
with the results for the two- and three-body states, Eq. (\ref{eq:AppA.1.4}) 
from Appendix A,  we obtain
\bea
{\cal T}_{0}(\bB,\bB',\bb,\bb)&=&
\Bigr\{1-\exp\Bigl[-{1\over 2}\sigma(x,\br)T(\bb)\Bigr]\Bigl\}
\nonumber\\
&\times&
\Bigr\{1-\exp\Bigl[-{1\over 2}\sigma(x,\br')T(\bb)\Bigr]\Bigl\}.\nonumber\\
\label{eq:5.A.2}
\eea
The CD quark-antiquark dijet spectrum
equals \cite{NSSdijet,Nonlinear,QuarkGluonDijet}
\bea
{d\sigma_{\nu}(\gamma^*A\to \{q\bar{q}\}_0A)\over dz d^2\bp d^2\bDelta }
&=&
{1\over (2\pi)^2}\delta_{\nu 0}\delta(\bDelta) \nonumber\\
&\times&\left| \Psi(1;z,\bp)- \Psi(z,\bp)\right|^2,\nonumber\\
\label{eq:5.A.3}
\eea
where $\delta(\bDelta)$ approximates a sharp diffractive peak
of width $\bDelta^2 \lsim 1/R_A^2$.   

Now we proceed to the RFT  
reinterpretation of this simple result. Making use 
of Eq. (\ref{eq:4.D.5}), we first represent (\ref{eq:5.A.3})
in terms of the collective nuclear gluon diagrams, Fig. 
\ref{fig:KMdiffractiveDIS},
\begin{widetext}
\bea
{d\sigma_{\nu}(\gamma^*A\to \{q\bar{q}\}_0 A)\over d^2\bb dz d^2\bp d^2\bDelta }
&=&
{1\over 4 (2\pi)^2}\delta_{\nu 0}\delta(\bDelta)
\int d^2\bp_1 d^2\bp_2 d^2\bkappa_1 d^2\bkappa_2\nonumber\\
&\times& 
[2\delta(\bp-\bp_1) - \delta (\bp-\bkappa_1-\bp_1) -
\delta (\bp+\bkappa_1-\bp_1)]
\nonumber\\
&\times& 
[2\delta(\bp-\bp_2)-\delta (\bp-\bkappa_2-\bp_2) -
\delta (\bp+\bkappa_2-\bp_2)]\nonumber\\
&\times&\phi(\bb,x,\bkappa_1)\phi(\bb,x,\bkappa_2)
\Psi^*(z,\bp_2)\Psi(z,\bp_1)\nonumber\\
&=&{1\over  (2\pi)^2}\delta_{\nu 0}
\int d^2\bp_1 d^2\bp_2 d^2\bkappa_1 d^2\bkappa_2
\phi(\bb,x,\bkappa_2)\phi(\bb,x,\bkappa_1)\nonumber\\
&\times& 
\CutD_{CD}(\gamma^*\to \{q\bar{q}\}_0;\Pom_A,\Pom_A;\bp,\bp_1,\bkappa_1,\bp_2,\bkappa_2),
\label{eq:5.A.4}
\eea
which suggests the following diagram rules in the two-dimensional
transverse momentum space: For a heavy nucleus the pomerons 
carry a vanishing transverse momentum. To each collective 
gluon loop there corresponds  
$\int  d^2\bkappa \phi(\bb,x,\bkappa)$
where the propagators of $g_{R,A}$, their coupling to quarks
and the uncut 
nuclear pomeron  and coupling of the nuclear
pomeron to a nucleus are absorbed in $\phi(\bb,x,\bkappa)$.
Following Sec. VI.D, we find it convenient to introduce  
integrals over the transverse momenta $\bp_{1,2}$  of quarks 
at the $\gamma^*q\bar{q}$ vertices, and the integral over the transverse
momenta of the quark and antiquark at the unitarity cut which we
have taken in the differential form: $d^2\bp_+ d^2\bp_- = d^2\bp
d^2\bDelta$. 
\begin{figure}[!h]
\begin{center}
\includegraphics[width = 3.2cm,angle=270]{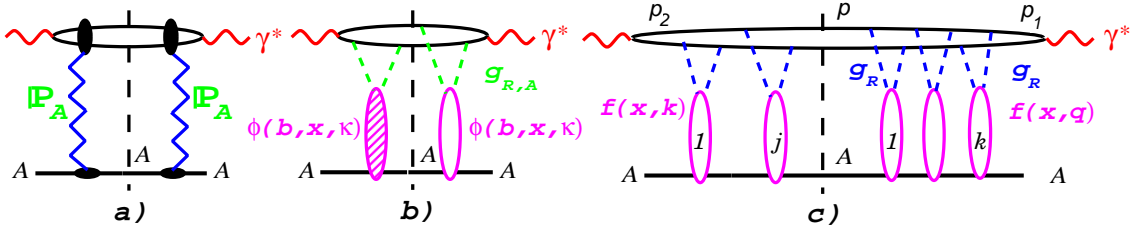}
\caption{ (a) The diffractive unitarity cut with the ground state
of the target nucleus in the intermediate state, (b) one of the 16 
diagrams for diffractive unitarity cut in terms of
couplings of coherent nuclear gluons $g_{R,A}$ to quarks and
antiquarks, (c) the total coherent diffractive cross section
in terms of the $\gamma^*(j\Pom)(k\Pom)\gamma^*$ impact factor.}
\label{fig:KMdiffractiveDIS}
\end{center}
\end{figure}

The contribution from the CD 
quark loop to the unitary cut $\gamma^*g_{R,A} g_{R,A}g_{R,A}g_{R,A}\gamma^*$ 
vertex 
$\CutD_{CD}(\Pom_A,\Pom_A)$ 
can be read from (\ref{eq:5.A.4}):
\bea
&&\CutD_{CD}(\gamma^*\to \{q\bar{q}\}_0;\Pom_A,\Pom_A;
\bp,\bp_1,\bkappa_1,\bp_2,\bkappa_2)=\delta(\bDelta)
\nonumber\\
&\times& {1\over 4}
\Psi^*(z,\bp_2)[2\delta(\bp-\bp_1) - \delta (\bp-\bkappa_1-\bp_1) -
\delta (\bp+\bkappa_1-\bp_1)]\nonumber\\
&\times&
[2\delta(\bp-\bp_2)-\delta (\bp-\bkappa_2-\bp_2) -
\delta (\bp+\bkappa_2-\bp_2)]\Psi(z,\bp_1)
\label{eq:5.A.5}
\eea
\end{widetext}
Here $\Psi(z,\bp_1)$ plays the r\^ole of the photon-quark vertex. 
On each side of the unitarity cut there are four possible couplings
of collective gluons $g_{R,A}$ to the quark and antiquark: the terms 
$\propto \delta(\bp-\bp_i)$ describe a coupling of both 
gluons from $\Pom_A$ to either quark, or antiquark, as shown by a hatched 
pomeron blob in Fig. \ref{fig:KMdiffractiveDIS}b, two other terms
describe two possible couplings of one gluon from $\Pom_A$
to
the quark, and the second gluon to the antiquark, as shown by 
on open blob in Fig. \ref{fig:KMdiffractiveDIS}b.

Making use of (\ref{eq:4.D.7}) we can expand 
(\ref{eq:5.A.4}) in contributions form the in-vacuum $j\Pom$ and $k\Pom$ 
exchanges on the two sides of the unitarity cut:
\bea
&&{d\sigma_{\nu}(\gamma^*A\to (q\bar{q})A)\over d^2\bb dz d^2\bp d^2\bDelta }
\Biggr|_{CD}=
{1\over  (2\pi)^2}\delta_{\nu 0}\nonumber\\
&\times& \sum_{j,k=1} {1\over j! k!}(-1)^{j+k}
\Big[{1\over 2}T(\bb)\Big]^{j+k} \nonumber\\
&\times& 
\int d^2\bp_1 d^2\bp_2 d^2\bk_1\dots d^2\bk_j 
d^2\bq_1\dots d^2\bq_k\nonumber\\
&\times&  
f(x,\bk_1)\dots f(x,\bk_j) f(x,\bq_1)\dots f(x,\bq_k)\nonumber\\
&\times& 
D_{CD}(j\Pom, k\Pom;\bp,\bp_2,\bp_1,\{\bk_i\},\{\bq_m\}).\nonumber\\
\label{5.B.5***}
\eea
Here the quark loop contribution to the
$\gamma^* (j\Pom)(k\Pom)\gamma^*$ vertex equals
\begin{widetext}
\bea
&&D_{CD}(\gamma^*\to
\{q\bar{q}\}_0;j\Pom,k\Pom;\bp,\bp_2,\bp_1,\{\bk_i\},\{\bq_m\})=
\delta(\bDelta)\nonumber\\
&\times & \sum\limits_{j_1,j_2,j_3,j_4=0}^{j}  
\sum\limits_{k_1,k_2,k_3,k_4=0}^{k} \delta(j-j_1-j_2-j_3-j_4)
\delta(k-k_1-k_2-k_3-k_4) \nonumber\\
&\times& {j! \over j_1! j_2! j_3! j_4!}\cdot 
{k! \over k_1! k_2! k_3! k_4!}(-1)^{j_1+j_2+k_1+k_2}\left({1\over 2}\right)^{j+k}\nonumber\\
&\times& \Psi^*(z,\bp_2) \delta (\bp-\sum\limits_1^{j_1}\bk_i+\sum\limits_{j_1+1}^{j_1+j_2}\bk_i-\bp_2)
\delta (\bp-\sum\limits_1^{k_1}\bq_m+\sum\limits_{k_1+1}^{k_1+k_2}\bq_m-\bp_1)\Psi(z,\bp_1),
\label{eq:5.A.6}
\eea
\end{widetext}
to each gluon loop there corresponds  
$\int  d^2\bkappa f(x,\bkappa)$
and the coupling of the pomeron $\Pom$ to a nucleus equals ${1\over 2}T(\bb)$.


\subsection{The universality class of dijets in higher color multiplet}
 
In the classification of Refs. \cite{Nonuniversality,QuarkGluonDijet},
production of color-octet dijets  belongs to the universality class of
excitation of dijets in higher representations --- here the octet and its
large-$N_c$ generalizations --- from initial partons in lower color
multiplets, which in DIS is the color-singlet photon. Here one
needs to solve the non-Abelian evolution for the four-parton problem,
the cross
\begin{figure}[!h]
\begin{center}
\includegraphics[width = 7.0cm]{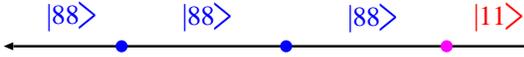}
\caption{ The sequence of color excitations for production of
color-octet dijets to the leading order in large-$N_c$ perturbation theory.}
\label{fig:Singlet-to-Octet}
\end{center}
\end{figure}
section operators in the basis of singlet-singlet, $\ket{e_1}$, and
octet-octet,  $\ket{e_2}$, four-partons states are found in Appendix A.
The large-$N_c$ expansion parameter is  the
singlet-to-octet hard transition $\Sigma_{21} \propto 1/N_c$.
To the leading order in $1/N_c$ perturbation theory, the
singlet-to-octet hard transition must be 
in the first color-excitation vertex, counting from the front face 
of the nucleus (Fig. \ref{fig:Singlet-to-Octet}).
 The further non-Abelian intranuclear evolution 
consists of the color excitation of the nucleus by color 
rotations within the octet state \cite{Nonlinear}. The 
relevant matrix element of 
${\cal T}_{\nu}(\bB',\bB)={\cal S}_{A,\nu}^{(4)}(\bB',\bB)$ would equal 
\bea
&&{\cal T}_{\nu}(\bB',\bB)= 
 \int_{0}^1  d\beta_{\nu}\dots d\beta_{1} \nonumber\\
&\times&  
\theta(1 -\beta_\nu)\theta(\beta_\nu -\beta_{\nu-1})...
\theta(\beta_2-\beta_1)\theta(\beta_1)\nonumber\\
&\times&
\textsf{S}[\bb,(1 -\beta_\nu)\hat{\Sigma}_{el}(\bC)] 
\Bigl\{ -{1\over 2}\Sigma_{22}^{ex}P_2 T(\bb)\Bigr\}\nonumber\\
&\times&
\textsf{S}[\bb,(\beta_\nu-\beta_{\nu-1})\hat{\Sigma}_{el}(\bC)]
\Bigl\{ -{1\over 2}\Sigma_{22}^{ex}P_2 T(\bb)\Bigr\}\nonumber\\
&\times& \dots \times
\Bigl\{ -{1\over 2}\Sigma_{21}^{ex}P_{ex} T(\bb)\Bigr\}
\textsf{S}[\bb,\beta_1\hat{\Sigma}_{el}(\bC)]
\nonumber\\
& =& 
 \int_{0}^1  d\beta_{\nu} \dots d\beta_{1} 
\theta(1 -\beta_\nu)\theta(\beta_\nu -\beta_{\nu-1})...\theta(\beta_1)\nonumber\\
&\times&
\textsf{S}[\bb,(1 -\beta_1)\hat{\Sigma}_{22}^{el}] \cdot
\Bigl\{ -{1\over 2}\Sigma^{ex}_{22}T(\bb)\Bigr\}^{\nu -1} \nonumber\\
&\times&{1\over 2\sqrt{N_c^2-1}}\Omega T(\bb)\cdot 
\textsf{S}[\bb,\beta_1 \Sigma_{11}^{el}] \,.
\label{eq:5.B.1}
\eea
The explicit integration over $\beta_2,..,\beta_\nu$ 
gives the factor $(1-\beta_1)^{\nu-1}/(\nu-1)!$. 
To the leading order of the 
large-$N_c$ perturbation theory $\Sigma_{22}^{el}=2\sigma_0(x)$ and
\bea
\textsf{S}[\bb,(1 -\beta_1)\hat{\Sigma}_{22}^{el}]&=& 
\textsf{S}[\bb,(1 -\beta_1)\sigma_0(x)] \nonumber\\
&\times&\cdot\textsf{S}[\bb,(1 -\beta_1)\sigma_0(x)].
\label{eq:5.B.2}
\eea
In conjunction with the expansion (\ref{eq:AppA.1.11}) 
for $\{\Sigma^{ex}_{22}\}^{\nu-1}$,
this gives rise to 
\begin{widetext}
\bea
&&{1\over (\nu-1)!} \Big[{1\over 2} (1-\beta_1)\sigma_0 T(\bb)\Big]^{\nu-1}
\textsf{S}[\bb,(1 -\beta_1)\sigma_0(x)] \cdot\textsf{S}[\bb,(1 -\beta_1)\sigma_0(x)]
\left(-\Sigma_{22}^{ex}\right)^{\nu-1}\nonumber\\
 &=& \sum_{j,k=0}^{\nu-1} \delta(\nu-1-j-k) 
\int  d^2\bkappa_1 d^2\bkappa_2 \exp[i\bkappa_1\bs +i\bkappa_2(\bs-\br +
\br')]\nonumber\\
&\times&
w_j\Big((1-\beta_1)\nu_A(\bb)\Big)
w_k\Big((1-\beta_1)\nu_A(\bb)\Big)
{d\sigma_{Qel}^{(k)}(\bkappa_1)
\over \sigma_{Qel} d^2\bkappa_1}\cdot {d\sigma_{Qel}^{(j)}(\bkappa_2)
\over \sigma_{Qel} d^2\bkappa_2}.
\label{eq:5.B.3}
\eea
\end{widetext}
The nuclear attenuation factor 
$
\textsf{S}[\bb,\beta_1 \Sigma_{11}^{el}]
= \textsf{S}[\bb,\beta_1 \sigma(x,\br)] \cdot 
\textsf{S}[\bb,\beta_1 \sigma(x,\br')]$
describes the coherent distortion of the color-singlet dipole 
in the slice $[0,\beta_1]$.
Upon the summation over color states of dijets, see Eq. (\ref{eq:AppA.1.5}),
we obtain the nonlinear $k_{\perp}$-factorization for 
topological cross sections of DIS followed by color excitation of 
$\nu$ nucleons ($\nu$ cut pomerons): 
\bea
&&{d\sigma_{\nu} (\gamma^*A\to \{q\bar{q}\}_8  X)\over d^2\bb dz d^2\bp
  d^2\bDelta} =
{1\over (2\pi)^2} T(\bb)\int_0^1 d\beta \nonumber\\
&\times&  \int  d^2\bkappa_1 d^2\bkappa_2 d^2\bkappa 
\delta(\bDelta-\bkappa_1-\bkappa_2-\bkappa)\nonumber\\
&\times& {d\sigma_{Qel}(\bkappa) \over d^2\bkappa}
\Big|\Psi(\beta;z,\bp-\bkappa_1)\nonumber\\
& -&
\Psi(\beta;z,\bp-\bkappa_1 -\bkappa)\Big|^2\nonumber\\ 
&\times& \sum_{j,k=0} \delta(\nu-1-j-k) \nonumber\\
&\times&
w_j\Big((1-\beta)\nu_A(\bb)\Big)
w_k\Big((1-\beta)\nu_A(\bb)\Big)\nonumber\\
&\times& {d\sigma_{Qel}^{(k)}(\bkappa_1) \over \sigma_{Qel} d^2\bkappa_1}
\cdot{d\sigma_{Qel}^{(j)}(\bkappa_2)
\over \sigma_{Qel} d^2\bkappa_2}.
\label{eq:5.B.4}
\eea
%

\subsection{From total to topological: the unitarity connection 
between partial and inclusive cross sections}

With the perfect hindsight, the topological cross section (\ref{eq:5.B.4})
could have been guessed from the inclusive cross section for the
color-octet dijets derived in \cite{Nonlinear}:
\bea
&&\frac{d\sigma(\gamma^*A\to \{q\bar{q}\}_8  X) }
{ d^2\bb dz d^2\bp d^2\bDelta}= \frac{1}{2(2\pi)^2} T(\bb) 
\int_0^1 d \beta\nonumber\\
&\times&
\int d^2\bkappa  d^2\bkappa_1 d^2\bkappa_2
\delta(\bDelta-\bkappa-\bkappa_1-\bkappa_2)\nonumber\\
&\times&
f(x,\bkappa)\Phi(1-\beta;\bb,x,\bkappa_1)\Phi(1-\beta;\bb,x,\bkappa_2)
\nonumber\\
&\times& \Bigl|
\Psi(\beta;z,\bp -\bkappa_1) -
\Psi(\beta; z,\bp -\bkappa_1-\bkappa)
\Bigr|^2.\nonumber\\ 
 \label{eq:5.C.1} 
\eea 
Namely, making use of
Eqs.~(\ref{eq:4.C.1}),(\ref{eq:4.C.4}),(\ref{eq:4.C.5}),
this inclusive cross section can be cast in the form
\bea
&&\frac{d\sigma(\gamma^*A\to \{q\bar{q}\}_8 X) }{ d^2\bb dz d^2\bp d^2\bDelta}
= \frac{1}{(2\pi)^2} T(\bb) 
\int_0^1 d \beta\nonumber\\
&\times& 
\int d^2\bkappa  d^2\bkappa_1 d^2\bkappa_2
\nonumber\\
&\times& \delta(\bDelta-\bkappa-\bkappa_1-\bkappa_2)
{d\sigma_{Qel}(\bkappa) \over d^2\bkappa}
\nonumber\\
&\times& \Bigl|
\Psi(\beta;z,\bp -\bkappa_1) -
\Psi(\beta; z,\bp -\bkappa_1-\bkappa)
\Bigr|^2\nonumber\\
&\times& \sum_{k=0} w_k\Big((1-\beta)\nu_A(\bb)\Big) 
{d\sigma_{Qel}^{(k)}(\bkappa_1) \over \sigma_{Qel} d^2\bkappa_1}\nonumber\\
&\times&
 \sum_{n=0}w_{n}\Big((1-\beta)\nu_A(\bb)\Big)
{d\sigma_{Qel}^{(n)}(\bkappa_1) \over \sigma_{Qel} d^2\bkappa_1}. \nonumber\\
\label{eq:5.C.2}
\eea
An obvious rearrangement of the summation
\beq
\sum_{k,n=0} = \sum_{\nu=1} \sum_{k,n=0}\delta(\nu-1-k-n)
\label{eq:5.C.3}
\eeq
gives 
\bea 
\frac{d\sigma(\gamma^*A\to  \{q\bar{q}\}_8  X) }
{ d^2\bb dz d^2\bp d^2\bDelta}
= \sum_{\nu=1}{d \sigma_\nu \bigl(\gamma^*A \to \{q\bar{q}\}_8 X\bigr)
\over d^2\bb dz d^2\bp d^2\bDelta},\nonumber\\
\label{eq:5.C.4}
\eea
where $d\sigma_\nu$ is precisely the cross section (\ref{eq:5.B.4}).
In conformity to the discussion of the unitarity in Sec. IV.C, 
the topological cross sections sum up to exactly
the inclusive cross section (\ref{eq:5.C.1}). 

What is not obvious from such a guesswork
is that the unitarity cuts are only 
allowed for pomerons entering the factor
$f(x,\bkappa)\Phi(1-\beta;\bb,x,\bkappa_1)
\Phi(1-\beta;\bb,x,\bkappa_2)$ in the inclusive cross section,
while
the initial-state coherent distortions (\ref{eq:4.D.4}) of
wave functions depend neither on $\nu$ nor $k$ and are always
described by uncut pomeron exchanges which simply do not 
affect the unitarity cuts.


\subsection{Reggeon field theory interpretation of topological
cross sections}

It is instructive to start with
the Impulse Approximation (IA)
\bea
&&{d\sigma_{IA}(\gamma^*A\to  \{q\bar{q}\}_8 X) \over d^2\bb dz  d^2\bp d^2\bDelta}=
\delta_{\nu 1} T(\bb)\nonumber\\
&\times&
{1\over (2\pi)^2}\cdot{d\sigma_{Qel}(\bDelta) \over d^2\bDelta}
\Big|\Psi(z,\bp) -
\Psi(z,\bp-\bDelta)\Big|^2\nonumber\\
&=& \delta_{\nu 1}\cdot {1\over (2\pi)^2}\cdot
{1\over 2} T(\bb)\int d^2\bkappa d^2\bp_1 d^2\bp_2 
\nonumber\\
&\times& \delta(\bDelta-\bkappa)
\Psi^*(z,\bp_2)[\delta(\bp-\bp_2)-\delta(\bp-\bp_2-\bkappa)]\nonumber\\
&\times&
[\delta(\bp-\bp_1)-\delta(\bp-\bp_1-\bkappa)]
\Psi^*(z,\bp_1)f(x,\bkappa).\nonumber\\
\label{eq:5.D.1}
\eea
Here the cut pomeron $\CutPom_{e}$ ($e=$ excitation)
corresponds to the
{\it {excitation from the color-singlet to color-octet dipole}} which is
driven by  the free-nucleon quasielastic 
scattering ${d\sigma_{Qel}(\bkappa)/d^2\bkappa}$. 
The RFT structure of  (\ref{eq:5.D.1}) is shown
in Fig. \ref{fig:DIS_Impulse}. 
\begin{widetext}
\mbox{}
\begin{figure}[!h]
\begin{center}
\includegraphics[width = 3.2cm,angle=270]{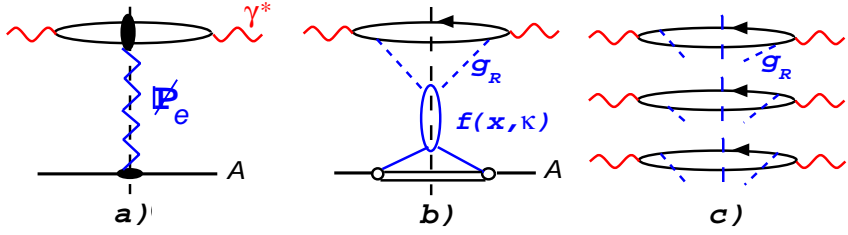}
\caption{ (a) The impulse approximation contribution
to inelastic DIS, (b)  the reggeon field theory diagrams
for the cut pomeron with one of possible coupling
of gluons to the quark loop, (c) the three remaining possible
couplings of gluons to the quark loop.}
\label{fig:DIS_Impulse}
\end{center}
\end{figure}
\mbox{}
\end{widetext}

One would identify
$\Big|\Psi(z,\bp) -
\Psi(z,\bp-\bDelta)\Big|^2$ with the unitarity-cut 
impact factor of the projectile photon, the last form
in (\ref{eq:5.D.1}) gives the impulse approximation
(free-nucleon)  $\gamma^*\CutPom_{e}\gamma^*$ 
vertex $\CutD(\{q\bar q\}_8;\CutPom_{e})$:
\bea
&&\CutD(\gamma^*\to \{q\bar q\}_8;\CutPom_{e};\bp,\bDelta,\bp_1,\bp_2,\bkappa)
= \nonumber\\
&=&\delta(\bDelta-\bkappa)\nonumber\\
&\times&
\Psi^*(z,\bp_2)[\delta(\bp-\bp_2)-\delta(\bp-\bp_2-\bkappa)]
\nonumber\\
&\times&
[\delta(\bp-\bp_1)-\delta(\bp-\bp_1-\bkappa)]
\Psi(z,\bp_1) .
\label{eq:5.D.2}
\eea

\begin{figure}[!h]
\begin{center}
\includegraphics[width = 6.0cm,angle=270]{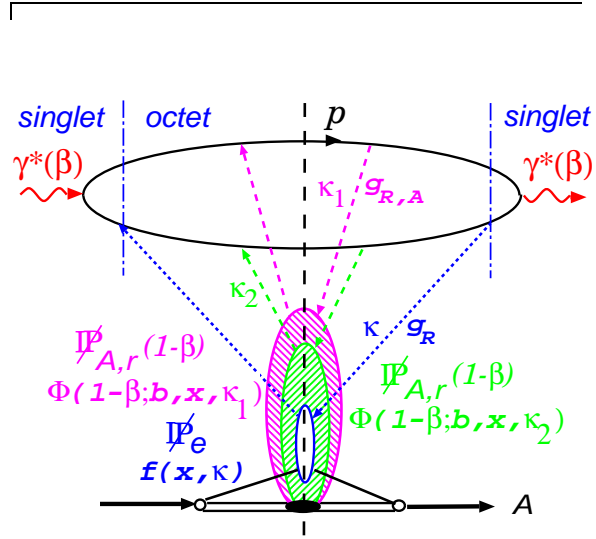}
\caption{ One of four reggeon field theory diagrams for the
unitarity cut for 
inelastic DIS in terms of
the cut in-vacuum pomeron $\CutPom_{e}$ 
(color-singlet-to-color-octet excitation) and two 
cut nuclear pomerons  $\CutPom_{A,r}$ (color rotations within
color octet dipole). Three more diagrams are obtained
recoupling in-vacuum gluons $g_R$ to the quark loop as shown
in Fig. \ref{fig:DIS_Impulse}c. The uncut-pomeron 
$\Pom_A(\beta)$ exchange
content of $\ket{\gamma^*(\beta)}$ is read from Eq. (\ref{eq:4.D.5}).} 
\label{fig:DIS_NuclearCut}
\end{center}
\end{figure}

Now notice that the color dipole properties of
$\Sigma_{22}^{ex}$ are different from those of
$\Sigma_{12}^{ex}$, see Appendix A. This invites introduction of
still another cut pomeron $\CutPom_{r}$
($r$=rotation) which
describes {\it{excitation of the nucleus
by  color rotations within
the octet dipoles}}. Specifically,
regarding its quasielastic scattering content,
see Sec. IV.C, the collective
nuclear pomeron belongs to the last category.
The RFT diagram for the
total inclusive inelastic dijet cross section
of Ref. \cite{Nonlinear} is shown in
Fig. \ref{fig:DIS_NuclearCut}. 
It is obtained
form the diagram of Fig. \ref{fig:DIS_Impulse}b by
the insertion of couplings of the quark and
antiquark to collective nuclear gluons $g_{R,A}$.
Similar insertions must be made in three
other diagrams of Fig. \ref{fig:DIS_Impulse}c. There
are two nucleus-slice-dependent cut
pomerons $\CutPom_{A,r}(1-\beta)$
accompanying the in-vacuum $\CutPom_e$ (and uncut
pomerons $\Pom_A(\beta)$ which describe the nuclear
distortions in $\Psi(\beta,z,\bp)$).
Gluons from
the free-nucleon $\CutPom_{e}$ have all four
possible couplings to both the quark and antiquarks.
In striking contrast to that, to the leading order
of large-$N_c$ perturbation theory one of the
two nuclear cut pomerons $\CutPom_{A,r}(1-\beta)$ couples
exclusively to the quark, while the second one couples
exclusively to the antiquark,
the diagrams with two gluons of the same nuclear
pomeron coupling to the quark and antiquark are
suppressed $\propto 1/N_c^2$.
Such an emergence of two
distinct cut pomerons is a deep consequence
of the non-Abelian coupled-channel
intranuclear evolution of color dipoles in QCD. In Fig.
\ref{fig:DIS_NuclearCut} we suppressed
uncut coherent $\Pom_A(\beta)$ exchanges between the
dipole and nucleus on both sides of the unitarity
cut, they are encoded in the distorted
$\Psi(\beta;z,\bp)$ the uncut multipomeron
exchange interpretation of which is given in Sec. IV,
Eq. (\ref{eq:4.D.5}).

Following (\ref{eq:5.D.1}), we define the quark loop
contribution to the 
$\gamma^* \CutPom_{A,r}\CutPom_{e}\CutPom_{A,r}\gamma^* $
vertex in terms of $\Psi(\beta;z,\bp)$,
\begin{widetext}
\bea
{d\sigma(\gamma^*A\to \{q\bar{q}\}_8 X) 
\over d^2\bb dz  d^2\bp d^2\bDelta}
&=&  {1\over (2\pi)^2}\cdot
{1\over 2} T(\bb)\int\limits_0^1 d\beta \int  d^2\bp_1 d^2\bp_2
d^2\bkappa d^2\bkappa_1 d^2\bkappa_2 
\delta(\bDelta-\bkappa_1-\bkappa_2-\bkappa)\nonumber\\
&\times& 
\Psi^*(\beta;z,\bp_2)
[\delta(\bp-\bkappa_1- \bp_2)-\delta(\bp-\bkappa_1 -\bkappa-
\bp_2)]\nonumber\\
&\times&
[\delta(\bp-\bkappa_1-\bp_1)-
\delta(\bp-\bkappa_1-\bkappa)-\bp_1)]
\Psi^*(\beta;z,\bp_1)\nonumber\\
&\times&
\Phi(1-\beta;\bb,x,\bkappa_1) 
\Phi(1-\beta;\bb,x,\bkappa_2)f(x,\bkappa), 
\label{eq:5.D.3}
\eea
which entails
\bea
&&\CutD(\gamma^*\to \{q\bar q\}_8; \CutPom_{A,r}(1-\beta),\CutPom_{e},\CutPom_{A,r}(1-\beta);\beta;
\bp,\bDelta,\bp_1,\bp_2,\bkappa,\bkappa_1,\bkappa_2)= \delta(\bDelta-\bkappa_1-\bkappa_2-\bkappa)\nonumber\\
&\times&
\Psi^*(\beta;z,\bp_2)
[\delta(\bp-\bkappa_1- \bp_2)-\delta(\bp-\bkappa_1 -\bkappa-
\bp_2)]\nonumber\\
&\times&
[\delta(\bp-\bkappa_1-\bp_1)-
\delta(\bp-\bkappa_1-\bkappa -\bp_1)]
\Psi(\beta;z,\bp_1)
\label{eq:5.D.4}
\eea

The transition to topological cross sections with $j+k$ cut 
in-vacuum pomerons $\CutPom_r$ and one cut $\CutPom_e$ is straightforward
($\nu=j+k+1$):
\bea
{d\sigma_{j+k+1}(\gamma^*A\to \{q\bar{q}\}_8 X) 
\over d^2\bb dz d^2\bp d^2\bDelta}
&=&  {1\over (2\pi)^2}\cdot
\Big[{1\over 2} T(\bb)\Big]^{j+k+1} {1\over j! k!}
\nonumber\\ &\times& 
\int\limits_0^1 d\beta (1-\beta)^{j+k} 
\textsf{S}[\bb,(1-\beta)\sigma_0(x)] 
\textsf{S}[\bb,(1-\beta)\sigma_0(x)]\nonumber\\ &\times& 
\int  d^2\bp_1 d^2\bp_2
d^2\bkappa d^2\bk_1\dots  d^2\bk_{j} 
d^2\bq_1\dots  d^2\bq_{k} 
\nonumber\\
&\times& \delta(\bDelta-\sum_{i=1}^{j}\bk_{i}-
\sum_{m=1}^{k}\bq_{m}-\bkappa)\nonumber\\
&\times&
\Psi^*(\beta;z,\bp_2)
[\delta(\bp-\sum_{i=1}^{j}\bk_{i}- \bp_2)-
\delta(\bp-\sum_{i=1}^{j}\bk_{i} -\bkappa-
\bp_2)]\nonumber\\
&\times&
[\delta(\bp-\sum_{i=1}^{j}\bk_{i}-\bp_1)-
\delta(\bp-\sum_{i=1}^{j}\bk_{i}-\bkappa-\bp_1)]
\Psi(\beta;z,\bp_1)\nonumber\\
&\times&
\prod^{j} f(x,\bk_{i}) \prod^{k} f(x,\bq_{m}).
\label{eq:5.D.5}
\eea
Here the transverse momenta $\bk_{i}$ belong to $j$
gluons $g_R$ from $j$ cut pomerons $\CutPom_{r}$ entering
$\Phi(1-\beta;\bb,x,\bkappa_1)$, and $\bkappa_1=\sum_i \bk_i$. The both
gluons from each such pomeron couple to a quark. 
Similarly, $\bq_{m}$  are the transverse momenta of $k$
gluons from $k$ cut pomerons $\CutPom_{r}$ entering
$\Phi(1-\beta;\bb,x,\bkappa_2)$, here $\bkappa_2=\sum_m \bq_m$, and
the both
gluons from each such pomeron couple to an antiquark.
Likewise, it is expedient to suppress the
familiar multiple uncut pomeron exchanges which
enter $\Psi(\beta;z,\bp_i)$. We identify
\bea 
&&\CutD(\gamma^*\to \{q\bar q\}_8; j\CutPom_{r},\CutPom_{e},k\CutPom_{r};\beta;
\bp,\bDelta,\bp_1,\bp_2,\bkappa,\{\bk_{i}\},
\{\bq_{m}\})=\nonumber\\
&=&\delta(\bDelta-\sum_{i=1}^{j}\bkappa_{1i}-
\sum_{m=1}^{k}\bkappa_{2i}-\bkappa)\textsf{S}[\bb,(1-\beta)\sigma_0(x)]
\textsf{S}[\bb,(1-\beta)\sigma_0(x)] \nonumber\\
&\times&
(1-\beta)^{j+k}\nonumber\\
&\times&
\Psi^*(\beta;z,\bp_2)
[\delta(\bp-\sum_{i=1}^{j}\bk_{i}- \bp_2)-
\delta(\bp-\sum_{i=1}^{j}\bk_{i} -\bkappa-
\bp_2)]\nonumber\\
&\times&
[\delta(\bp-\sum_{i=1}^{j}\bk_{i}-\bp_1)-
\delta(\bp-\sum_{i=1}^{j}\bk_{i}-\bkappa -\bp_1)]
\Psi(\beta;z,\bp_1)
\label{eq:5.D.6}
\eea
Here our RFT diagrams rules must be complemented by the $\beta$-integration
$\int_0^1 d\beta$.
For the sake of brevity we suppressed
the representation of the distorted wave functions and
the two absorption factors
$\textsf{S}[\bb,(1-\beta)\sigma_0(x)]$
in terms of
multiple uncut pomeron exchanges on the two
sides of the unitarity cut. The RFT diagram of Fig. \ref{fig:DIS_NuclearCut}
has a manifestly Mandelstam cut structure, the same is true
of its expansion in $\CutPom_e$ and $\CutPom_r$'s.

The special case
$j=k=0$ corresponds to the absorbed impulse
approximation. The two special cases $(j,k)=(1,0)$
and $(j,k)=(0,1)$ give rise to final states with
two cut pomerons. From the viewpoint of gluon
couplings, the function of two cut pomerons
in such a vertex
\bea
\CutD_2&=&
(1-\beta)\CutD(\gamma^*\to \{q\bar q\}_8; \CutPom_{r},\CutPom_{e},0;\beta;
\bp,\bDelta,\bp_1,\bp_2,\bkappa,\bkappa_1,0)
\nonumber\\
&+&
(1-\beta)\CutD(\gamma^*\to \{q\bar q\}_8; 0,\CutPom_{e},\CutPom_{r};\beta;
\bp,\bDelta,\bp_1,\bp_2,\bkappa,0,\bkappa_{2})
\label{eq:5.D.7}
\eea
\end{widetext}
is
very different --- singlet-to-octet excitation $\CutPom_e$
vs. octet-to-octet color rotations  $\CutPom_r$ --- and they
enter such a cut vertex in a very asymmetric
fashion. All vertices $\CutD$ vanish, and $\CutPom_e$ decouples,
at $\bkappa \to 0$ --- large wavelength gluons can not resolve the
intrinsic structure of, and can not excite to the color octet,
the initial color-single $\{q\bar{q}\}_0$.
In contrast to that, upon this excitation,
gluons of all wavelengths do contribute to
color rotations of  $\{q\bar{q}\}_8$ which
carries a net color charge and, consequently, the
$\CutD$'s are not constrained to vanish at $\bkappa_i \to 0$.
Consequently, for hard dijets $\CutPom_e$ is always in the
hard regime, while $\CutPom_r$'s are not necessarily hard.
This distinction has been noted already in Ref. \cite{Nonlinear},
it is a deep
feature of the non-Abelian intranuclear evolution
of color dipoles. Our unitarity rules with two kinds of
cut pomerons is a new result, such a two-cut-pomeron
picture
did not appear in the extensive earlier literature on
the AGK rules.


\subsection{Pseudo-diffractive color singlet dijets}

In the conventional diffractive production the target nucleus
either stays in the ground state (coherent diffraction)
or breaks up with emission of nucleons (incoherent diffraction).
Each exchanged pomeron couples to a color-singlet nucleon 
of the nucleus.
In DIS, as well as in scattering of other color-singlet projectiles, 
there exists a very interesting pseudo-diffractive channel when after
several intranuclear color-excitations the dijet ends
up in the color-singlet state \cite{Nonlinear} accompanied
by the nucleus debris also in the overall color-singlet
state. 
There is a possibility of a survival of a
large rapidity
gap between the dijet and the nucleus debris,
and this process would mimic the conventional 
incoherent diffractive dijet 
production. The color 
excitation stretches over the whole thickness of the nucleus
and a signature of 
such a pseudo-diffractive scattering would be a
multiparticle production in the nucleus fragmentation region.
\begin{figure}[!h]
\begin{center}
\includegraphics[width = 7cm]{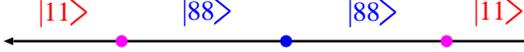}
\caption{ The sequence of large-$N_c$ perturbation theory 
color excitations  for production of the pseudo-diffractive
color-singlet dijets }
\label{fig:Singlet-to-Singlet}
\end{center}
\end{figure}

The pseudo-diffractive 
channel starts with $\nu=2$ -- the singlet-to-octet excitation on
one nucleon is compensated for by an octet-to-singlet de-excitation on
another nucleon. To higher orders, it is complemented by
color-excitation transitions between the color-octet states
as shown in Fig. \ref{fig:Singlet-to-Singlet}. We readily find
\bea
&&{\cal T}_{\nu}(\bB',\bB)= 
  \int_{0}^1  d\beta_{\nu}\dots d\beta_{1} \nonumber\\
&\times&  
\theta(1 -\beta_\nu)\theta(\beta_\nu -\beta_{\nu-1})\dots
\theta(\beta_2-\beta_1)\theta(\beta_1)\nonumber\\
&\times&
\textsf{S}[\bb,(1 -\beta_\nu)\hat{\Sigma}_{el}(\bC)] 
\Bigl\{ -{1\over 2}\Sigma_{12}^{ex}P_{ex} T(\bb)\Bigr\}\nonumber\\
&\times&
\textsf{S}[\bb,(\beta_\nu-\beta_{\nu-1})\hat{\Sigma}_{el}(\bC)]
\Bigl\{ -{1\over 2}\Sigma_{22}^{ex}P_2 T(\bb)\Bigr\}\nonumber\\
&\times& \dots\times
\Bigl\{ -{1\over 2}\Sigma_{21}^{ex}P_{ex} T(\bb)\Bigr\}
\textsf{S}[\bb,\beta_1 \hat{\Sigma}_{el}(\bC)]
\nonumber\\
& =& 
 \int_{0}^1  d\beta_{\nu}\dots d\beta_{1} 
\theta(1 -\beta_\nu)\theta(\beta_\nu -\beta_{\nu-1})...\theta(\beta_1)\nonumber\\
&\times&
\textsf{S}[\bb,(\beta_\nu -\beta_1)\hat{\Sigma}_{22}^{el}] \cdot
\Bigl\{ -{1\over 2}\Sigma^{ex}_{22}T(\bb)\Bigr\}^{\nu -2} \nonumber\\
&\times&{1\over 4(N_c^2-1)}\Omega^2 T^2(\bb)\cdot 
\textsf{S}[\bb,(\beta_1+1-\beta_\nu)
\Sigma_{11}^{el}]\nonumber\\
&=&  \int_{0}^1 d\beta \beta (1-\beta)^{\nu-2}{1 
\over (\nu-2)!} \nonumber\\
 &\times&{1\over 4(N_c^2-1)}\Omega^2 T^2(\bb)\Bigl\{ -{1\over
   2}\Sigma^{ex}_{22}T(\bb)\Bigr\}^{\nu -2} \nonumber\\
&\times& \textsf{S}[\bb,(1-\beta)\hat{\Sigma}_{22}^{el}]
\cdot \textsf{S}[\bb,\beta\Sigma_{11}^{el}]
.
\label{eq:5.E.1}
\eea
Here we lumped together the coherent distortions 
of the color-singlet dipole in the entrance,
$[0,\beta_1]$, and exit, $[1-\beta_\nu,1]$, slices of the nucleus:
$\beta = \beta_1 + (1-\beta_\nu)$.
Proceeding as in the previous section, we obtain
the pseudo-diffractive topological cross sections
\bea
&&{d\sigma_{\nu} (\gamma^*A\to  \{q\bar{q}\}_0 X)\over d^2\bb dz d^2\bp
  d^2\bDelta}
={1\over (N_c^2-1)} \nonumber\\
&\times&
{1\over (2\pi)^2} T^2(\bb)\int_0^1 d\beta \,\beta \nonumber\\
&\times&  \int  d^2\bkappa_1 d^2\bkappa_2 d^2\bkappa_3  
d^2\bkappa_4\delta(\bDelta-\bkappa_1-\bkappa_2-\bkappa)
 \nonumber\\
&\times&
{d\sigma_{Qel}(\bkappa_3) \over d^2\bkappa_3}\cdot
{d\sigma_{Qel}(\bkappa_4) \over d^2\bkappa_4}
\nonumber\\
&\times& \Big|\Psi(\beta;z,\bp-\bkappa_2)+
\Psi(\beta;z,\bp-\bkappa_2-\bkappa_3-\bkappa_4)\nonumber\\
&-&
\Psi(\beta;z,\bp-\bkappa_2-\bkappa_3)-\Psi(\beta;z,\bp-\bkappa_2-\bkappa_4) 
\Big|^2\nonumber\\
&\times& \sum_{k=0}^{\nu-2}w_k\Big((1-\beta)\nu_A(\bb)\Big)
w_{\nu-2-k}\Big((1-\beta)\nu_A(\bb)\Big)\nonumber\\
&\times& {d\sigma_{Qel}^{(k)}(\bkappa_1) \over \sigma_{Qel} d^2\bkappa_1}
\cdot{d\sigma_{Qel}^{(\nu-2-k)}(\bkappa_2)
\over \sigma_{Qel} d^2\bkappa_2}.
\label{eq:5.E.2}
\eea

For the RFT interpretation of
this result take the diagram of Fig. 
\ref{fig:DIS_NuclearCut}
and add one more exchange by the in-vacuum 
$\CutPom_{e}$ which describes de-excitation
from the color-octet to color singlet dipole. The two new gluons
$g_R$ would couple all four possible ways to the quarks loop. This
explains the structure of the generalized cut impact factor
which is proportional to
\bea
&&\Big|\Psi(\beta;z,\bp-\bkappa_2)  +
\Psi(\beta;z,\bp-\bkappa_2 -\bkappa_3-\bkappa_4)\nonumber\\
&-&\Psi(\beta;z,\bp-\bkappa_2
-\bkappa_3)
-
\Psi(\beta;z,\bp-\bkappa_2-\bkappa_4)\Big|^2\nonumber\\
\eea
and now contains 16 terms. Incidentally, a very similar 
generalized impact factor made an appearance in 
CD off nuclei \cite{3POMglue}, see
also the excitation of gluon-gluon dijets in \cite{GluonGluonDijet}.
The contribution from exchanges by cut pomerons $\CutPom_{A,r},\CutPom_r$
has exactly the same structure as in  Fig. \ref{fig:DIS_NuclearCut}.
The same is true of absorption of color octet dipole 
by multiple exchange by uncut pomerons $\Pom$. Finally, 
the coherent multiple pomeron exchanges in $\Psi(\beta;z,\bp)$ 
look the same as in Fig. \ref{fig:DIS_NuclearCut}, the extra
factor $\beta$ in the integrand is a manifestation of
lumping together the coherent distortions of color-singlet
dipoles in the entrance and exit states . 

A generalization of (\ref{eq:5.C.2}) to the
sequential singlet-to-octet-to-singlet-to-octet, and still higher order
contributions
in $1/N_c$ perturbation theory, is straightforward, we shall not dwell into that in
this communication.


\section{Unitarity rules for the single-jet spectra and Cheshire Cat grin}

Our findings on deducing the topological dijet 
cross sections in DIS from the inclusive ones 
prompt an important question: is it always
possible to go from the total cross sections to 
topological ones or to the single-parton spectra? 
The literature abounds with
such reinterpretations and it is important to assess
their credibility. Specifically, 
the often discussed pre-QCD AGK approach
does not contain salient features of our pQCD cutting rules
--- coupled-channel non-Abelian evolution of dipoles
and two kinds of cut pomerons --- which lead to 
important new properties. In this section we comment on
these distinctions. Of particular importance is the
issue of cancellations of spectator interactions when
going from the dijet to single-jet cross sections: 
spectators do not affect the single-jet spectrum but 
the CCG ---the spectator contribution to the
nucleus excitation --- persists all the way through
and is a physics observable.


\subsection{Topological vs. total cross section in DIS}

We start with two examples when going from the total to topological
and/or differential cross section works.


\subsubsection{From total to topological and differential: $qA$
scattering}

In Sec. IV.B we correctly guessed the unitarity content of 
$\phi(\bb,x,\bkappa)$, although the true distinction between
two kinds of cut pomerons $\CutPom_{r,e}$ was as yet elusive. 
We notice that by the very definition
\bea
\int d^2\bkappa \phi(\bb,x,\bkappa)&=& 1-\exp[-{1\over 2}\sigma_0 T(\bb)]
\nonumber\\
&=&
 1-\exp[-\sigma_{qN} T(\bb)]
\label{eq:6.A.1.1}
\eea
is an integrand of the quark-nucleus inelastic cross section.
Now we will try to revert the considerations of Sec. IV.B. A
series of transformations gives
\bea
 &&1-\exp[-\sigma_{qN} T(\bb)]=\nonumber\\
&=& \exp[-\sigma_{qN} T(\bb)] \{\exp[\sigma_{qN}
 T(\bb)]-1\}
\nonumber\\
&=& \exp[-\sigma_{qN} T(\bb)] \sum_{j=1} {1\over j!}[\sigma_{qN} T(\bb)]^j
\nonumber\\
&=& \exp[-\sigma_{qN} T(\bb)]{1\over j!}\Big[ {1\over 2}T(\bb)\Big]^j
\Big[\int d^2\bkappa f(x,\bkappa)\Big]^j\nonumber\\
&=&  \exp[-{1\over 2}\sigma_{0} T(\bb)]{1\over j!}\Big[ {1\over
  2}T(\bb)\Big]^j\int d^2\bkappa
\nonumber\\
&\times&  d^2\bkappa_1 \dots d^2\bkappa_j f(x,\bkappa_1)\dots
f(x,\bkappa_j) \nonumber \\
&\times&\delta(\bkappa-\sum^{j}\bkappa_i)
\label{eq:6.A.1.2}
\eea
Unfold the $d^2\bkappa$ integration and associate the $j$-fold convolution
with the $j$ cut pomeron exchange. By a miracle --- evidently, such a differential form
of the l.h.s. of (\ref{eq:6.A.1.1}) is by no means unique  --- the correct form
of $\phi(\bb,x,\bkappa)$ is recovered.


\subsubsection{From total to differential: leading quarks in DIS}

We recall the cancellation of spectator interaction effects
in the quadrature for the fully inclusive single-particle spectra of partons $b$ 
from the pQCD subprocess $a\to bc$ upon the integration over the
transverse momentum $\bp_c$ \cite{NPZcharm,SingleJet}. 
Indeed, such an integration
amounts to putting $\bb_c'=\bb_c$ in Eq. (\ref{eq:2.5}), and
in the fully inclusive case 
\bea
&&{\cal T}(\bB,\bB',\bb,\bb')=S^{(4)}_{\bar{b}\bar{c} c b}(\bb_b',\bb_c',\bb_b,\bb_c) 
\nonumber\\
&+& S^{(2)}_{\bar{a}a}(\bb',\bb) 
-
S^{(3)}_{\bar{b}\bar{c}a}(\bb,\bb_b',\bb_c')
- S^{(3)}_{\bar{a}bc}(\bb',\bb_b,\bb_c) 
\nonumber\\
&=&S_b^{\dagger}(\bb_b')S_c^{\dagger}(\bb_c) S_c(\bb_c)S_b(\bb_b)+
S_a^{\dagger}(\bb') S_a(\bb)\nonumber\\
&-& S_a^{\dagger}(\bb')S_b(\bb_b) S_c(\bb_c)-
S_b^{\dagger}(\bb_b')S_c^{\dagger}(\bb_c) S_a(\bb)\nonumber\\
&=& {\textsf S}[\bb,\sigma(x,\br-\br')]+\openone\nonumber\\
&-&{\textsf S}[\bb,\sigma(x,\br)]-
{\textsf S}[\bb,\sigma(x,\br')].
\label{eq:6.A.2.1} 
\eea 
Here we made use of  $S_a(\bb)=\openone$ for the color-singlet
projectile, $a=\gamma^*$,  and the unitarity relation $S_c^{\dagger}(\bb_c)
S_c(\bb_c)=\openone$.
The resulting leading quark spectrum is a single-channel problem.
The result is  a linear $k_\perp$-factorization quadrature 
for the single leading quark spectrum when both the coherent 
diffractive and truly inelastic DIS are lumped together,
\bea
{d\sigma (\gamma^*A\to q X)\over d^2\bb dz d^2\bp }&=&
{1\over (2\pi)^2} \int d^2\bkappa \phi(\bb,x,\bkappa)\nonumber\\
&\times&\Big|\Psi(z,\bp) -\Psi(z,\bp-\bkappa)\Big|^2,\nonumber\\
\label{eq:6.A.2.2}
\eea
which is exact for all $N_c$  
\cite{Nonlinear,NonlinearJETPLett}.

The same result can be obtained in a risky way of undoing 
the integrations in the formal Fourier representation
for the total DIS cross section:
\begin{widetext}
\bea
{d\sigma \over d^2\bb} &=& 2\int_0^1 dz  \int  d^2\br \Psi^*(z,\br)\Psi(z,\br)
\left\{ 1-\exp[-\half \sigma(x,\br)T(\bb)]\right\}\nonumber\\
&=& {1\over(2\pi)^4} \int_0^1 dz \int d^2\bkappa
\phi(\bb,x,\bkappa)[1- \exp(i\bkappa\br)][1- \exp(-i\bkappa\br)]\nonumber\\
&\times& \int d^2\bp_1 d^2\bp_2 \Psi^*(z,\bp_2)\Psi(z,\bp_1)
\exp(i\bp_2 \br -i\bp_1\br) \nonumber\\
&=& {1\over(2\pi)^2} \int_0^1 dz \int d^2\bkappa d^2\bp_1 d^2\bp_1
\Psi^*(z,\bp_2)\Psi(z,\bp_1) \phi(\bb,x,\bkappa)
\nonumber\\
&\times& [\delta(\bp_1-\bp_2)-\delta(\bp_1-\bp_2+\bkappa)
-\delta(\bp_1-\bp_2-\bkappa ) +\delta((\bp_1+\bkappa) -(\bp_2+\bkappa))].\nonumber\\
\label{eq:6.A.2.3}
\eea
\end{widetext}
At this point one needs a judicious identification of the Fourier
parameters $\bp_{1,2}$, or $\bp_{1,2}+\bkappa$, with the momentum
of the observed leading quark. Taking literally a concept of a
collective nuclear gluon $g_{R,A}$, and drawing the relevant 
four Feynman diagrams, see Fig. \ref{fig:DIS_Impulse}, 
makes such an identification unique and 
(\ref{eq:6.A.2.3}) boils down to (\ref{eq:6.A.2.2}). Such a mock
derivation does not tell, though, why one must not take for
the final state momentum $\bp$ the momentum of the quark 
somewhere in between the in-vacuum gluons of the multigluon
components of $g_{R,A}$.


\subsection{From total inclusive to topological: Cheshire Cat 
grin stays on}

Inspired by the successful,
though marred by question-marks, guesswork in Secs. VI.A.1 and
VI.A.2, one can  
substitute for $\phi(\bb,x,\bkappa)$ the expansion (\ref{eq:4.C.5})
and define
the "topological cross sections" for the single leading-quark
spectrum in DIS:
\bea
&&{d\sigma^*_\nu  (\gamma^*A\to q X)\over d^2\bb dz d^2\bp }=
{1\over (2\pi)^2} w_\nu\Big(\nu_A(\bb)\Big)\nonumber\\
&\times&
\int d^2\bkappa {d\sigma_{Qel}^{(\nu)}(\bkappa)\over d^2\bkappa}
\Big|\Psi(z,\bp) -\Psi(z,\bp-\bkappa)\Big|^2.
\nonumber \\
\label{eq:6.B.1}
\eea
This reinterpretation of the so-guessed $d\sigma^*_\nu$ as 
topological single-jet cross sections for DIS with color-excitation 
of $\nu$ nucleons ($\nu$ cut pomerons) 
would be erroneous, though. The trouble is that,
although our guesswork (\ref{eq:6.A.2.3}) proved a success, an important
distinction between the two kinds of cut pomerons $\CutPom_{r,e}$
has been lost in the final form of (\ref{eq:6.B.1}).

The correct result for the 
single-jet spectrum derives from the dijet spectrum integrating 
in (\ref{eq:5.B.4}) over the transverse momentum 
of the unobserved antiquark. This lifts 
$\delta(\bDelta-\bkappa-\bkappa_1-\bkappa_2)$
and the $\bkappa_2$ integration can be carried out
explicitly,  
\beq
\int d^2\bkappa_2{d\sigma_{Qel}^{(\nu-1-k)}(\bkappa_2)
\over \sigma_{Qel} d^2\bkappa_2}=1,
\label{eq:6.B.2}
\eeq
the result is 
\bea
&&{d\sigma_{\nu} (\gamma^*A\to q X)\over d^2\bb dz d^2\bp}=
{1\over 2(2\pi)^2} T(\bb)\int_0^1 d\beta \nonumber\\
&\times& \sum_{j,k=0} \delta(\nu-1-j-k)\nonumber\\
&\times&
w_k\Big((1-\beta)\nu_A(\bb)\Big)
w_{j}\Big((1-\beta)\nu_A(\bb)\Big)\nonumber\\
&\times&  \int  d^2\bkappa_1  d^2\bkappa  {d\sigma_{Qel}(\bkappa)\over d^2\bkappa} 
\cdot  {d\sigma_{Qel}^{(k)}(\bkappa_1)\over \sigma_{Qel}d^2\bkappa_1}\nonumber\\
&\times&\Big|\Psi(\beta;z,\bp-\bkappa_1) -\Psi(\beta;z,\bp-\bkappa_1 -\bkappa)\Big|^2
\nonumber\\
 &+&
{1\over (2\pi)^2}\delta_{\nu 0}
\left| \Psi(1;z,\bp)- \Psi(z,\bp)\right|^2.
\label{eq:6.B.3}
\eea
The RFT diagram for (\ref{eq:6.B.3}) is obtained from
Fig. \ref{fig:DIS_NuclearCut}
striking out the second nuclear cut pomeron $\CutPom_{A,r}$,
which is a graphic counterpart of cancellations of the spectator antiquark
interactions in (\ref{eq:6.A.2.1}). This makes clear the r\^ole of 
the CCG and its meaning as a physics
observable: by the unitarity the second nuclear
pomeron is gone, but it still contributes to the $\nu$ and
the target nucleus excitation: the Chesire Cat grin stays on.

Needless to say the correct topological cross sections 
$d\sigma_\nu$ are unequal to the ill-guessed $d\sigma^*_\nu$.
For instance, the $\beta$-dependence of exchanges by
uncut pomerons $\Pom$ and cut pomerons $\CutPom_r$ is entirely missed in
$d\sigma_\nu^*$.
The case of $\nu=0$ is the most striking one:  at $Q^2$ below
the saturation scale, the
coherent diffractive component of (\ref{eq:6.B.3}) makes $\approx 50$
per cent of the total DIS cross section \cite{NZZdiffr}, while the
contribution with $\nu=0$ is entirely missed in 
the ill-guessed $d\sigma^*_\nu$ of Eq.~(\ref{eq:6.B.1}).


\subsection{Isolate coherent diffraction: still more guesswork for inelastic
  $\sigma_\nu$}

Try correcting (\ref{eq:6.A.2.3}) for the
CD,
\bea 
&&\underbrace{2\left\{ 1-\exp[-\half \sigma(x,\br)T(\bb)]\right\} }_{tot}\nonumber\\
&=&\underbrace{\left\{ 1-\exp[-\half \sigma(x,\br)T(\bb)]\right\}^2
}_{coherent~~diffraction}\nonumber\\
&+&\underbrace{\left\{ 1-\exp[-\sigma(x,\br)T(\bb)]\right\}}_{inelastic},
\label{eq:6.C.1}
\eea
and focus on the inelastic cross section
\bea
{d\sigma^{in} \over d^2\bb}& =& \int_0^1 dz\int d^2\br \Psi^*(z,\br)\Psi(z,\br)
\nonumber\\
&\times&\left\{ 1-\exp[-\sigma(x,\br)T(\bb)]\right\}.
\label{eq:6.C.2}
\eea
 The substitution   
\bea
&&1-\exp[-\sigma(x,\br)T(\bb)] =\nonumber\\
&=&{1\over 2}
\int d^2\bkappa [1-\exp(i\bkappa\br)][1-\exp(-i\bkappa\br)]
\nonumber\\
&\times&
\sum_{j=1} w_{j}\Big(2\nu_A(\bb)\Big){d\sigma_{Qel}^{(\nu)}(\bkappa)
\over \sigma_{Qel} d^2 \bkappa}
\label{eq:6.C.3}
\eea
in (\ref{eq:6.C.2}) would suggest
\bea
&&{d\sigma^*_\nu \over d^2\bb dz d^2\bp }=
{1\over 2(2\pi)^2} w_\nu\Big(2\nu_A(\bb)\Big) \int_0^1 dz
\nonumber\\
&\times&
\int d^2\bkappa {d\sigma_{Qel}^{(\nu)}(\bkappa)\over d^2\bkappa}
\Big|\Psi(z,\bp) -\Psi(z,\bp-\bkappa)\Big|^2.\nonumber\\
\label{eq:6.C.4}
\eea
Once again $d\sigma_{\nu}^* \neq d\sigma_\nu$ and 
(\ref{eq:6.C.4}) must be rejected for the same reason as (\ref{eq:6.B.1}).

Still another possibility is a representation 
\bea
&&1-\exp[-\sigma(x,\br)T(\bb)] =\nonumber\\
&=& \int_0^1 d\beta
\sigma(x,\br)T(\bb)\exp[-\beta\sigma(x,\br)T(\bb)]\nonumber\\
 &=& 2T(\bb)\int_0^1 d\beta \exp[-\beta\sigma(x,\br)T(\bb)]\nonumber\\
&\times&
\int  d^2\bkappa [1-\exp(i\bkappa\br)] {d\sigma_{Qel} (\bkappa)\over d^2\bkappa}.
\label{eq:6.C.5}
\eea
Upon the re-absorption of $\exp[-\beta\sigma(x,\br)T(\bb)]= \textsf{S}[\bb,\beta\sigma(x,\br)]
\cdot \textsf{S}[\bb,\beta\sigma(x,\br)]$ into the coherent distortions of
the dipole wave functions in (\ref{eq:6.C.2}), one would find
\bea
&&{d\sigma^{in} \over d^2\bb} 
={1\over (2\pi)^2} T(\bb) \int_0^1 d\beta \int_0^1 dz
\int d^2\bp d^2\bkappa 
\nonumber\\
&\times&{d\sigma_{Qel} (\bkappa)\over d^2\bkappa}
\Big|\Psi(\beta;z,\bp) -\Psi(\beta;z,\bp-\bkappa)\Big|^2.\nonumber\\
\label{eq:6.C.6}
\eea
As far as coherent distortions of dipoles are concerned, 
it is reminiscent of 
(\ref{eq:6.B.3}), but suggests an absurd result that there only one cut 
pomeron, $d\sigma_{\nu}^* = \delta_{\nu 1}{d\sigma^{in}}$. 

The reason for all those failures is simple. Isolate CD in 
(\ref{eq:6.A.2.1}):
\begin{widetext}
\bea
{\cal T}(\bB,\bB',\bb,\bb')&=&\underbrace{1-{\textsf S}[\bb,\sigma(x,\br)]-
{\textsf S}[\bb,\sigma(x,\br')] + {\textsf S}[\bb,\sigma(x,\br)]{\textsf
  S}[\bb,\sigma(x,\br')]}_{coherent~~diffraction}\nonumber\\
&+& \underbrace{ {\textsf S}[\bb,\sigma(x,\br-\br')]-{\textsf S}[\bb,\sigma(x,\br)]{\textsf
  S}[\bb,\sigma(x,\br')]}_{inelastic}
\label{eq:6.C.7}
\eea
\end{widetext}
The single-quark spectrum from inelastic DIS would equal \cite{Nonlinear}
\bea
&&{d\sigma^{in} (\gamma^*A\to q X)\over d^2\bb dz d^2\bp }=
{1\over (2\pi)^2} \int d^2\br  d^2\br' \nonumber\\
&\times&\exp[i\bp(\br'-\br)] 
\Psi^*(z,\br')\Psi(z,\br)\nonumber\\
&\times& 
\{
{\textsf S}[\bb,\sigma(x,\br-\br')]-{\textsf S}[\bb,\sigma(x,\br)]{\textsf
  S}[\bb,\sigma(x,\br')]
\}.\nonumber\\
\label{eq:6.C.8}
\eea
The integration over the momentum of the leading quark gives $\delta(\br-\br')$
and (\ref{eq:6.C.2}) is recovered
\bea
{d\sigma^{in} \over d^2\bb} &=& \int_0^1 dz\int d^2\br \Psi^*(z,\br)\Psi(z,\br)
\nonumber\\
&\times& \left\{ 1-{\textsf S}^2[\bb,\sigma(x,\br)]]\right\}\nonumber\\
&=& \int d^2\br |\Psi(z,\br)|^2
\left\{ 1-\exp[-\sigma(x,\br)T(\bb)]\right\}.\nonumber\\
\label{eq:6.C.9}
\eea
Mimicking (\ref{eq:6.A.2.3}) did a bad service in (\ref{eq:6.C.4}) 
and (\ref{eq:6.C.6}): there is no way to guess the
crucial ${\textsf S}[\bb,\sigma(x,\br-\br')]$ from 1 in the integrand of
(\ref{eq:6.C.9}).


\subsection{The color dipole form of $\sigma_\nu$ in DIS}

In the fully integrated inelastic cross sections $\bs=0$ and $\br=\br'$,
which entails $\Sigma_{11}^{el}=\Omega =2\sigma(x,\br)$,  $
\Sigma_{22}^{el}=-\Sigma_{22}^{ex}= 2\sigma_0$ and
\bea
{d\sigma_{\nu}^{in} \over d^2\bb}&=&\int_0^1 dz \int d^2\br
|\Psi(z,\br)|^2\nonumber\\
&\times&
\int_0^1 d\beta {1\over (\nu-1)!}[(1-\beta)\sigma_0 T(\bb)]^{\nu-1} \nonumber\\
&\times&
\exp[-(1-\beta)\sigma_0 T(\bb)]
\nonumber\\
&\times&
\sigma(x,\br) T(\bb) \exp[-\beta\sigma(x,\br) T(\bb)] \nonumber\\
&=&
\int_0^1 dz \int d^2\br
|\Psi(z,\br)|^2\nonumber\\
&\times&
\int_0^1 d\beta w_{\nu-1}(2(1-\beta)\nu_A(\bb)) \nonumber\\
&\times&\sigma(x,\br) T(\bb)\exp[-\beta\sigma(x,\br) T(\bb)].
\label{eq:6.D.1}
\eea
This is a new result.
Of course, upon summing over all $\nu\geq 1$ we recover (\ref{eq:6.C.2}):
\bea
&&\sum_{\nu \geq 1} 
\int_0^1 d\beta w_{\nu-1}(2(1-\beta)\nu_A(\bb))\nonumber\\
&\times& \sigma(x,\br)
T(\bb)\exp[-\beta\sigma(x,\br) T(\bb)]
\nonumber\\
&=& \int_0^1 d\beta\sigma(x,\br)
T(\bb)\exp[-\beta\sigma(x,\br) T(\bb)]\nonumber\\
&=& 1- \exp[-\sigma(x,\br) T(\bb)].
\label{eq:6.D.2}
\eea 

Following the 1976 version of AGK rules for hadron-nucleus 
collisions \cite{Capella,TreleaniOld,Kaidalov}, one often defines 
$d\sigma_{\nu}^*$ via the expansion (for the recent review see \cite{AGKsvalka}) 
\bea
&&1-\exp[-\sigma(x,\br)T(\bb)]=\nonumber\\
&=&\exp[-\sigma(x,\br)T(\bb)]
\sum_{\nu=1} {1\over \nu!}[\sigma(x,\br)T(\bb)]^{\nu},\nonumber\\
&&{d\sigma^*_{\nu} \over d^2\bb}=\int_0^1 dz \int d^2\br
|\Psi(z,\br)|^2\nonumber\\
&\times&
{1\over \nu!}[\sigma(x,\br) T(\bb)]^\nu \exp[-\sigma(x,\br) T(\bb)].
\label{eq:6.D.3}
\eea 
Such a $d\sigma_{\nu}^*$ would coincide with the correct 
result (\ref{eq:6.D.1}) only if $\sigma_0 =\sigma(x,\br)$,
which is nonsensical in pQCD. We conclude that the
version (\ref{eq:6.D.3}) is an unwarranted one. 
This discussion casts shadow on the routinely used 
Glauber model statistics for wounded nucleon distributions
which is based on  \cite{Capella,Kaidalov}, we shall
revisit this issue elsewhere.

To summarize this discussion, it is crystal clear 
that the unitarity definition of topological cross sections
must be applied before integrating out the 
spectator-jet variables - the two procedures do
not commute because the spectator-jet integrations
(i) involve shifts of the Fourier parameter $\bp$ which
then can no longer be correctly related to the 
observed momentum of the jet, 
(ii) the distinction between two kinds of cut pomerons ---
the crucial feature of pQCD unitarity --- is missed 
upon the spectator integrations, (iii) the Chesire Cat grin
is a physics observable --- spectator 
interaction cancellations make, by virtue of
the completeness of states and unitarity relations, 
some important operators equal to unity and the
contribution of those states to topological cross
sections is missed. The whole variety of plausible
``unitarity'' reinterpretations
(\ref{eq:6.A.2.3}), (\ref{eq:6.C.2}),  (\ref{eq:6.C.6}) 
and  (\ref{eq:6.D.3})
is simply unwarranted. For a similar
failure of naive unitarity reinterpretations of
diffractive cross sections see Ref. \cite{3POMglue}.


\subsection{pQCD version of the AGK ratios}

We close this dispute by the comparison of 
different two-pomeron exchange cross sections, i.e.,
terms $\propto T^2(\bb)$ in the total cross section, coherent
diffractive cross section, the two cut pomeron exchange cross
section $\sigma_2^{in}$ and the first absorption correction to
$\sigma_1^{in}$. Such a comparison can be carried out at the level
of color dipole profile functions. The expansion of
$$\Gamma_{tot}(\bb,\br)= 2\Big\{ 1-\exp[-\half \sigma(x,\br)T(\bb)]\Big\}$$
gives
\bea
\Delta_2 \Gamma_{tot}(\Pom\Pom;\bb,\br)= -{1\over 4}[\sigma(x,\br)T(\bb)]^2,
\label{eq:6.E.1}
\eea  
while for the CD $$\Gamma_{D}(\bb,\br)= 
\Big\{ 1-\exp[-\half \sigma(x,\br)T(\bb)]\Big\}^2$$ and
\bea
\Delta_2 \Gamma_{D}(\Pom\Pom;\bb,\br)= {1\over 4}[\sigma(x,\br)T(\bb)]^2.
\label{eq:6.E.2}
\eea
The generic formula 
(\ref{eq:6.D.1}) for the profile function of $\sigma_2^{in}$ reads
\bea
&&\Gamma_{2}^{in}(\CutPom_r\CutPom_e,\Pom,\dots, \Pom;\bb,\br)= \nonumber\\
&=&[\sigma_0(x) T(\bb)]\cdot [\sigma(x,\br) T(\bb)] \nonumber\\
&\times&
\int_0^1 d\beta (1-\beta) \exp[-\beta\sigma(x,\br) T(\bb)]\nonumber\\
&\times& \exp[-(1-\beta)\sigma_0(x)  T(\bb)].
\label{eq:6.E.3}
\eea
To the second order in  $T(\bb)$ the $\beta$-dependent attenuation must
be neglected and
\bea
\Delta_2 \Gamma_{2}^{in}(\CutPom_r\CutPom_e;\bb,\br)& =& 
{1\over 2}\cdot [\sigma_0(x) T(\bb)]\nonumber\\
&\times& [\sigma(x,\br) T(\bb)],\nonumber\\
\label{eq:6.E.4}
\eea
where $\sigma(x,\br)$ and  $\sigma_0$ come from $\CutPom_e$ and $\CutPom_r$,
respectively. Finally, the generic formula (\ref{eq:6.D.1})
for the profile function 
of $\sigma_1^{in}$ reads 
\bea
&&\Gamma_{1}^{in}(\CutPom_e,\Pom,\dots, \Pom;\bb,\br)= 
\sigma(x,\br) T(\bb)\nonumber\\
&\times&\int_0^1 d\beta  \exp[-\beta\sigma(x,\br) T(\bb)]
\nonumber\\
&\times&
\exp[-(1-\beta)\sigma_0 T(\bb)].
\label{eq:6.E.5}
\eea
In order to isolate the first absorption correction, we expand the two
attenuation factors to the terms linear in $T(\bb)$ and obtain
\bea
&&\Delta_2 \Gamma_{1}^{in}(\CutPom_e\Pom;\bb,\br)= -\sigma(x,\br) T(\bb)
 \nonumber\\
&\times& \int_0^1 d\beta \Big[(1-\beta)\sigma_0(x) T(\bb) +\beta\sigma(x,\br)
T(\bb) \Big]
\nonumber\\
&=& - {1\over 2}\cdot [\sigma_0(x) T(\bb)]
\cdot [\sigma(x,\br) T(\bb)]\nonumber\\
& -&{1\over 2}[\sigma(x,\br)T(\bb)]^2.\nonumber\\
\label{eq:6.E.6}
\eea

The results for the total and CD cross sections
are model independent ones, but the two kinds of cut pomerons
inherent to pQCD make (\ref{eq:6.E.4})
and (\ref{eq:6.E.6}) distinct from the pre-QCD version of the AGK
rules. In the pre-QCD AGK rules there is only one kind of cut
pomerons and, based on (\ref{eq:6.D.3}), one would have found
(\cite{Capella,TreleaniOld,Kaidalov}, for the more
recent discussion see \cite{AGKsvalka})
\begin{widetext}
\bea
\Delta_2 \Gamma_{1}^{in}(\CutPom\Pom;\bb,\br) &=&
-2\Delta_2 \Gamma_{2}^{in}(\CutPom\CutPom;\bb,\br)
= - [\sigma(x,\br)T(\bb)]^2 \, , 
\label{eq:6.E.7}
\eea
and
\bea
\Delta_2 \Gamma_{D}(\Pom\Pom;\bb,\br)\,:\,\Delta_2
\Gamma_{1}^{in}(\CutPom\Pom;\bb,\br)
\,:\,  \Delta_2 \Gamma_{2}^{in}(\CutPom\CutPom;\bb,\br)= 1\,:\,-4\,:\,2.
\label{eq:6.E.8}
\eea
The latter relationship, as well as the cancellation
\bea
\Delta_2 \Gamma_{1}^{in}(\CutPom\Pom;\bb,\br) + 2 \Delta_2
\Gamma_{2}^{in}(\CutPom\CutPom;\bb,\br)=0,
\label{eq:6.E.9}
\eea
break down in the pQCD world, although the sum rule
\bea
\Delta_2 \Gamma_{tot}(\Pom\Pom;\bb,\br)= \Delta_2 \Gamma_{D}(\Pom\Pom;\bb,\br) +\Delta_2
\Gamma_{1}^{in}(\CutPom\Pom;\bb,\br)+ \Delta_2
\Gamma_{2}^{in}(\CutPom\CutPom;\bb,\br)
\label{eq:6.E.10}
\eea
is still retained. In the pre-QCD version of eq. (\ref{eq:6.E.10}) there is
only one kind of cut pomerons, while the two kinds of cut
pomerons in our pQCD cutting rules can be read from eqs.
(\ref{eq:6.E.4}) and (\ref{eq:6.E.6}).
\end{widetext}


\section{Phenomenological applications of topological
cross sections}

We only briefly sketch possible phenomenological 
applications of the derived topological cross sections
for inelastic DIS. 
Much guidance comes from the earlier literature
on hadron-nucleus interactions 
\cite{Capella,Kaidalov}. 
We list several observables of increasing complexity.


\subsection{Multiproduction in the backward (nucleus)
hemisphere}

Consider first the minimal bias events.
We refer to rapidities $\eta < \eta_A$ as the nucleus
or backward (B) hemisphere. 
Arguably, the multiplicity of hadrons in the backward hemisphere,
$n_B$, will be proportional to $\nu$:
\beq
n_{B} \approx \nu \langle n_{B}\rangle_{pN},
\label{eq:7.A.1}
\eeq
which allows to evaluate $\nu$ on the event-by-event basis.
(Strictly speaking, the backward hemisphere is infested by 
intranuclear cascading of secondary hadrons 
formed inside the nucleus \cite{NikDav,KolyaUFN}
and Eq. (\ref{eq:7.A.1}) needs further qualification
within the nonperturbative hadronization models, we do
not dwell into that.)

Then our result (\ref{eq:6.D.1}) for the topological cross 
sections can readily be applied for a modeling of $n_{B}$
distributions for inelastic DIS off nuclei. If $P_n(\langle n_{B}\rangle_{pN})$
is the multiplicity distribution for a free nucleon target,
then the corresponding $n_B$ distribution for $\nu$ cut pomerons
will be a $\nu$-fold convolution of  $P_n(\langle n_{B}\rangle_{pN})$.
For instance, the moments $
p_k=\langle (\nu-1)\dots (\nu-k)\rangle$ can readily be evaluated:
\bea
&&\sum_{\nu> k} {d\sigma_{\nu} \over d^2\bb}(\nu-1)\dots (\nu-k)
=\nonumber\\
&=&
\int_0^1 dz \int d^2\br
|\Psi(z,\br)|^2
\int_0^1 d\beta [2(1-\beta)\nu_A(\bb)]^k \nonumber\\
&\times&\sigma(x,\br)T(\bb)
\exp[-\beta\sigma(x,\br)T(\bb)].
\nonumber\\
\label{eq:7.A.2}
\eea
A manifest dependence of the moments $p_k$ on the infrared-sensitive
$\sigma_0$ is not disturbing as it comes along with the 
related infrared-sensitive conversion of the multiplicity of partons to 
the observed multiplicity of hadrons.


\subsection{Long-range rapidity correlations between leading and backward particles:
breaking of limiting fragmentation
}


\subsubsection{Radiationless breaking of limiting fragmentation}

Consider now the $\bp$ integrated leading jet spectra.
For leading jets $z$ coincides with the Feynman variable $x_F$.
The familiar concept of limiting fragmentation amounts to a 
target independence of the $x_F$ spectra. There are two sources 
of the breaking of limiting fragmentation in DIS off nuclei.
As we have seen above, the partition of the energy-momentum
between $k$ cut pomerons $\CutPom_r$ which couple to the 
observed quark contributes to its quenching. We shall discuss
it in the next section, here we comment on the radiationless
quenching of leading jets  --- a mechanism inherent to the pQCD
color dipole approach to DIS. 

The $x_F$ spectrum of leading quark jets is obtained from (\ref{eq:6.D.1})
undoing the $z$ integration:
\bea
{d\sigma_{\nu} \over d^2\bb dx_F}&=& \int d^2\br
|\Psi(x_F,\br)|^2\nonumber\\
&\times&
\int_0^1 d\beta w_{\nu-1}(2(1-\beta)\nu_A(\bb)) \nonumber\\
&\times&\sigma(x,\br) T(\bb)\exp[-\beta\sigma(x,\br) T(\bb)].\nonumber\\
\label{eq:7.B.1.1}
\eea
Here we recall that for transverse and longitudinal photons and the flavor $f$
\cite{NZ91} 
\bea
&&|\Psi_T(x_F,\br)|^2 ={6\alpha_{em} \over (2\pi)^2}e_f^2 \nonumber\\
&\times&\{
[(1-x_F)^2+x_F^2]\varepsilon^2 K_1^2(\varepsilon r)+
m_f^2K_0^2(\varepsilon r)\},
\nonumber\\
&&|\Psi_L(x_F,\br)|^2 ={6\alpha_{em} \over (2\pi)^2}e_f^2\nonumber\\
&\times&
4Q^2 x_F^2(1-x_F)^2 K_0^2(\varepsilon r)
\label{eq:7.B.1.2}
\eea
where $Q^2$ is the virtuality of the photon, $\alpha_{em}$ is the QED fine
structure
constant, $m_f$ and $e_f$ are the quark mass and electric charge, $K_{0,1}(x)$ are the Bessel
functions and 
\beq
\varepsilon^2 =  x_F(1-x_F) Q^2+m_f^2.
\label{eq:7.B.1.3}
\eeq
By virtue of (\ref{eq:7.B.1.3}) the resulting $x_F$ distribution depends
on the $\br$ dependence of the nuclear factor. The latter is controlled by
the nuclear impact parameter distribution which depends on $\nu$.
It can easily be shown, that the larger is $\nu$ the stronger is the
suppression of the contribution from large dipoles and from $x_F$ close
to 1. Consequently, even without any nonperturbative energy loss, 
 the $x_F$ spectrum of leading quarks would depend on $\nu$,
which we dub the radiationless breaking of limiting fragmentation.


\subsubsection{Nonperturbative quenching of forward jets 
in DIS off nuclei}

As discussed in the Introduction 
each excited nucleon adds more secondary particles and 
more energy in the nucleus fragmentation region. 
This flow of energy into the nucleus 
hemisphere is
a nonperturbative contribution to the quenching of forward jets.
It is complementary to radiationless quenching in the Born
approximation, Sec. VII B.1, and quenching by the 
QCD Landau-Pomeranchuk-Migdal (LPM) effect to higher orders
discussed in \cite{SlavaLPM,SlavaReview,VirtualReal}.  

The universality class of CD DIS is unique
for the lack of the energy loss. Fragmentation of coherent
diffractive dijets produced off nuclei will be exactly the 
same as of dijets from free-nucleon interactions. 

In inelastic DIS the color excitation centers will be
scattered along the whole thickness of the nucleus at
a given impact parameter $\bb$. Besides these color centers
and projectile partons, color strings will
be stretched between the color centers themselves. In a dilute
nucleonic gas, the energy stored in the latter string 
will be of the order of the nuclear radius times the string 
tension and will be substantially, $\sim A^{1/3}$ times,
larger than the elastic recoil
energy of nuclear quarks discussed in \cite{Recoil}. 
The two extreme scenarios for the energy flow from the
projectile system to a color-excited nucleus have already been
described in the Introduction. In the scenario of 
Fig. \ref{fig:CutPomerons}b, color strings stretched 
between color-excited nucleons and the projectile system
hadronize independently into small-$p_\perp$ hadrons, i.e., 
the underlying minimal bias event. Different versions of
this scenario \cite{Capella,Kaidalov} are widely used
in the Monte-Carlo hadronization codes for nuclear
interactions \cite{VENUS}. We have shown that, at large $N_c$,
in DIS quasielastic 
rescatterings of the quark and antiquark are 
independent of each other.
If $\langle x_{F,1} \rangle$ is the average fraction of
the initial energy carried by the quasielastically scattered
parton, then after $k$ sequential scatterings of the
observed quark
\beq
\langle x_{F,k} \rangle \sim \langle x_{F,1} \rangle^k,
\label{eq:7.B.2.1}
\eeq
for instance, see \cite{KNP}.
The equipartition of energy between
$\nu$ strings, 
\beq
\langle x_{F,\nu} \rangle \sim {1\over \nu} \langle x_F \rangle,
\label{eq:7.B.2.2}
\eeq
is less likely. 

The formation length arguments suggest a different pattern of
quenching. In the fragmentation of the color string between 
the propagating quark  which changed its color and 
the color-excited nucleon, the secondary hadron 
of momentum $x_F E_q$ is formed from  the quark-antiquark pair
produced in the chromo-electric field of the string. 
In the comoving frame the hadron formation time $\tau \sim 1/m_\perp$, 
where $m_\perp$ is the transverse 
mass of the prompt hadron, in the laboratory frame 
it is stretched by
the $\gamma$-factor of the parton (\cite{Kancheli,NZfusion}
for the review see \cite{KolyaUFN}),
\beq
l_f \approx {x_F E_q \over m_\perp^2}.
\label{eq:7.B.2.3}
\eeq
The second quasielastic scattering would take place at a distance
\beq
l_{abs} \approx {1\over n_A \sigma_0}.
\label{eq:7.B.2.4}
\eeq
The energy flow $E_{ex}$ into the hadronization between
two consecutive interactions can be estimated as
\beq
 E_{ex} \approx l_{abs}  m_\perp^2.
\label{eq:7.B.2.5}
\eeq
The energy flow $\Delta_k E$ 
after $k$ quasielastic rescatterings can be estimated as 
\beq
\Delta_k E \sim k E_{ex}.
\label{eq:7.B.2.6}
\eeq
It does not increase with the energy of the beam. 
In this scenario
\beq
\langle x_{F,k}\rangle \approx \langle x_{F,1}
\rangle \left(1-k {E_{ex} \over E_q}\right)
\label{eq:7.B.2.7}
\eeq
and the impact of nonperturbative nuclear
quenching would diminish with the beam energy.
Such a scenario  is a part of
the so-called  yo-yo, or the inside-outside cascade, models 
\cite{Gyulassy,HIJING,Lund}.

In still another version of the formation length model, 
the hadronization in the nucleus region does not affect the 
leading jet production at all. The hadronization of the 
color string degrees of freedom with with the formation 
length $l_f \gsim R_A$, i.e., with the momenta larger that 
$k_A \approx R_A m_\perp^2$, proceeds way beyond the target
nucleus. Then the yield of such hadrons will be independent
of the target nucleus fragmentation. The energy 
which flows into the excitation of the nucleus will cause a depletion of 
the density of secondary hadrons with the momenta $p \sim R_A m_\perp^2$,
while the spectra of faster secondary hadrons will not be affected
by excitation of the nucleus. Such a scenario with 
$\langle x_{F,k}\rangle = \langle x_F\rangle$ 
was suggested long time ago within the pre-QCD parton and multi peripheral
models \cite{NZfusion,NikDav,KolyaUFN}.

The separation between different scenarios can only be done 
experimentally. First, one must evaluate a quenching 
of the leading parton
spectrum for the perturbative LPM effect \cite{VirtualReal}. 
The nonperturbative quenching will be an extra to the LPM effect,
it is characterized by $z_k = \langle x_{F,k}\rangle/\langle x_{F,1}\rangle$.
If $D(x_F)$ is the fragmentation function 
for jets produced off free nucleons, then after $k$ quasielastic 
rescatterings of the leading parton  
\beq
D_k(x_F)\approx {1\over z_k}D({x_F\over z_k}).
\label{eq:7.B.2.8}
\eeq
The manifest dependence of the fragmentation function $D_k(x_F)$
on $k$ breaks the limiting fragmentation.
At a fixed $\nu$, the recoiling jet will be described by the
fragmentation function $D_{\nu-1-k}(x_F)$.


\subsubsection{Going more differential: backward production
in semiinclusive DIS}

If the nonperturbative quenching is under quantitative control
one would be able to reconstruct the parton level jet cross sections.
Then, by virtue of (\ref{eq:7.A.1}), and its extension 
to the rapidity and 
transverse momentum spectra in backward multiproduction,
\beq
{dn_B\over d\eta} \approx  \nu { dn\over d\eta}\Bigr|_{pN},
\label{eq:7.B.3.1}
\eeq
our topological cross sections
(\ref{eq:6.B.3}), (\ref{eq:7.B.1.1}) describe a correlation between 
the leading (antiquark) quark jet and multiproduction in the
backward hemisphere. Generalizing the arguments of Sec. VII.A, 
we can predict from (\ref{eq:7.B.1.1}) the backward multiplicity 
distributions as a function of $x_F$ of the leading jet. Similarly, 
(\ref{eq:6.B.3}) can be used to predict the $(x_F,\bp)$ dependence
of $n_B$ distributions.

The other way around, (\ref{eq:6.B.3}) predicts the $\nu$, i.e., $n_B$,
 dependence
of the leading particle spectrum. Here the crucial ingredient
is ${d\sigma_{Qel}^{(k)}(\bkappa)/ d^2 \bkappa }$.
The detailed studies of the collective glue $f^{(k)}(x,\bkappa)$
are found in Refs. \cite{NSSdijet,Nonlinear}. The onset of the 
unitarity constraints in fully inclusive DIS is controlled
by the so-called saturation scale
\beq
Q_A^2(\bb,x)\approx {4\pi^2 \over N_c}\alpha_S(Q_A^2)G(x,Q_A^2)T(\bb). 
\label{eq:7.B.3.2}
\eeq
In the semiinclusive case the relevant $k$ dependent 
scale is \cite{Nonlinear}
\beq
Q_{A,k}^2(\bb,x)\approx {k\over \nu_A(\bb)} Q_A^2(\bb,x).
\label{eq:7.B.3.3}
\eeq
For small momenta, $\bkappa^2 \lsim Q_{A,k}^2(\bb,x)$, the nuclear
attenuation effects take over and
\bea
{d\sigma_{Qel}^{(k)}(\bkappa) \over d^2\bkappa } \approx
{\sigma_{Qel}\over \pi}\cdot { Q_{A,k}^2(\bb,x)\over
  [\bkappa^2+Q_{A,k}^2(\bb,x)]^2} \nonumber \\
\approx {\sigma_{Qel}\over \pi Q_{A,k}^2(\bb,x) }
\propto {1\over k}.
\label{eq:7.B.3.4}
\eea
Consequently, for semi-hard $\bp^2 \lsim  Q_A^2(\bb,x)$ the contribution
from large $k$ is suppressed.

For large momenta, $\bkappa^2 \gsim Q_{A,k}^2(\bb,x)$,  the
$k$-dependence of the multiple quasielastic scattering cross
section is given by \cite{NSSdijet}
\bea
&&{d\sigma_{Qel}^{(k)}(\bkappa) \over d^2\bkappa } =
k{d\sigma_{Qel}(\bkappa) \over d^2\bkappa }
\Biggl[1+ \nonumber\\
&+&{k-1 \over \nu_A(\bb)}\cdot {\gamma^2 \over 2}\cdot 
{\alpha_S(\bkappa^2) G(x_A, \bkappa^2) \over \alpha_S(Q_A^2)G(x_A,Q_A^2)}
\cdot {Q_A^2(\bb,x_A) \over \bkappa^2}\Biggr], \nonumber\\
\label{eq:7.B.3.5}
\eea
where $\gamma\approx 2 $ is the exponent of the large-$\bkappa$ behavior
${d\sigma_{Qel}(\bkappa)/ d^2 \bkappa } \sim (\bkappa^2)^{-\gamma}$.
To the leading twist, the $k$-fold hard quasielastic scattering 
is dominated by the  single-hard scattering 
on any of the $k$ nucleons, on the background of soft scatterings
on remaining $(k-1)$ nucleons. This hierarchy goes on: one of those 
$(k-1)$ rescatterings is again hard and gives rise to the 
anti shadowing nuclear higher twist correction. The latter is behind the
nuclear Cronin effect (see \cite{SingleJet} and references therein)
and our Eq. (\ref{eq:6.B.3}) enables us to predict the $\nu$ dependent
Cronin effect for leading jets, the numerical studies will be 
reported elsewhere.


\subsubsection{Still more differential: decorrelation of quark-antiquark 
dijets in DIS}

Finally, in the fully differential form, Eq. (\ref{eq:5.B.4}) describes
the $\nu$, i.e., centrality, dependence of
the out-of-plane (azimuthal, acoplanarity) and in-plane  decorrelation
of quark-antiquark dijets produced in the current fragmentation 
region of DIS off nuclei. The relevant discussion for the inclusive
case is found in \cite{Nonlinear}.


\subsection{Eliminating the  Cheshire Cat grin: isolation of
cut pomerons in single-jet spectrum by
multiplicity re-summations}

The re-summation over the spectator antiquark interactions,
i.e., over $j$, in Eq. (\ref{eq:6.B.3}) gives a particularly
simple result for the single quark spectrum (now $k$
is a total number of the non-spectator $\CutPom_e$ and $\CutPom_r$'s): 
\bea
&&{d\sigma^{(k)} (\gamma^*A\to q X)\over d^2\bb dz d^2\bp}= 
\sum_{\nu> k} {d\sigma_{\nu} (\gamma^*A\to q X)\over d^2\bb dz d^2\bp}
\nonumber\\&=&
{1\over 2(2\pi)^2} T(\bb)\int_0^1 d\beta \nonumber\\
&\times&w_{k-1}\Big((1-\beta)\nu_A(\bb)\Big)\int  
d^2\bkappa_1  d^2\bkappa  \nonumber\\
&\times&{d\sigma_{Qel}(\bkappa)\over d^2\bkappa} 
\cdot  {d\sigma_{Qel}^{(k-1)}(\bkappa_1)\over \sigma_{Qel}d^2\bkappa_1}\nonumber\\
&\times&\Big|\Psi(\beta;z,\bp-\bkappa_1) -\Psi(\beta;z,\bp-\bkappa_1 -\bkappa)\Big|^2.
\label{eq:7.C.1}
\eea
Including the effect of $j$ color excitations by the spectator quark,
the backward multiplicity would equal   
\beq
n_B(j,k)\approx \nu\langle n_B\rangle_{pN}=(j+k)\langle n_B\rangle_{pN}
\label{eq:7.C.2}
\eeq
and this re-summation over $j$ amounts to the re-summation over the backward
multiplicities 
\beq
n_B > k\langle n_B\rangle_{pN}.
\label{eq:7.C.3}
\eeq
{\it {In a paradoxical way, such a multiplicity re-summation,
crude though it is, eliminates the CCG and isolates
the topological cross section with $k$ cut pomerons,
$\CutPom_e + (k-1)\CutPom_r$, complemented by absorption 
for the exchange by 
uncut pomerons $\Pom$ contained in $w_{k-1}$. }}
This way one can test experimentally all the 
properties of $d\sigma_{Qel}^{(k-1)}$ discussed in Sec. VII.B.3.
We strongly urge this multiplicity re-summation technique
as an efficient tool to extract the CCG-unbiased single parton
production properties. 
Evidently, one should not stretch this re-summation to
multiplicities at the tail of the multiplicity distributions.


\section{Nonlinear $k_{\perp}$-factorization for 
topological cross sections in quark-gluon dijet production 
}

We follow closely the technique of Ref. 
\cite{QuarkGluonDijet}. The elastic and color-excitation 
scattering operators for
production of quark-gluon dijets in quark-nucleon and
quark-nucleus collisions are reported in Appendix B.
Here-below $\Psi_{qg}(z,\br)$ will be the wave function of the 
$qg$ Fock state of the quark.
One can think of the Additive Quark Model scenario 
for the dilute and small projectile interacting with the dense 
large target. For instance, a heavy quarkonium 
is well approximated by the $q\bar{q}$ Fock state. The
intrinsic transverse momentum of quarks in the quarkonium
is small and comoving spectator antiquark can not give
rise to final states with a high-$p_{\perp}$ quark and 
high-$p_{\perp}$ gluon \cite{PionDijet}. We re-sum 
over all powers of the large parameter in the problem - the 
nuclear thickness.
The comoving antiquark of the 
projectile will contribute to the excitation of the 
nucleus, but the effects of its interaction on
the dijet spectrum cancel out upon the integration 
over the antiquark transverse momentum \cite{NPZcharm,SingleJet}. 
Neither will it
affect the energy flow from the $qg$ dijet to a nucleus,
which can be treated at the level of the quark-nucleus
collisions as the $q\to qg$ excitation. The $qg$ dijet
can be in either color-triplet or higher dimension 
--- sextet and 15-plet --- representations. We keep
these $SU(3)$ labeling of multiplets at arbitrary $N_c$.


\subsection{Dijets in the color-triplet representation}


\subsubsection{Color excitation distributions}

We start with the 
universality class of dijets in the same representation
as the
incident parton, i.e., with the color-triplet channel of the 4-parton interaction.
To the leading order in $1/N_c$ perturbation, only the diagonal 
part of the color-excitation operator (\ref{eq:AppB.2.4}) contributes
to the color-triplet dijet production and the 
calculation of
$\bra{3\bar{3}}{\textsf S}_{A,\nu}^{(4)}(\bC)\ket{3\bar{3}}$
simplifies to a single-channel problem. 
The  summation over all orderings of the elastic and color-excitation
scatterings of the $qg$ dipole  i.e., the $\beta$ integrations
in (\ref{eq:3.F.3}) can be performed analytically with the result
\bea
\bra{3\bar{3}}{\cal S}^{(4)}_{A,\nu}(\bC,\bb)\ket{3\bar{3}}& =& 
{1\over \nu!} \Big[-\half \Sigma^{ex}_{11} T(\bb)\Big]^{\nu}\nonumber\\ 
&\times&\textsf{S}[\bb,\Sigma^{el}_{11}(\bC)].
\label{eq:8.A.1.1}
\eea
The cross section operators from eqs. (\ref{eq:AppB.2.2},\ref{eq:AppB.2.5}) give
\bea
&&\bra{3\bar{3}}{\cal S}^{(4)}_{A,\nu}(\bC,\bb)\ket{3\bar{3}} =\nonumber\\
&=&
\int d^2\bkappa  {d\sigma_{Qel}^{(\nu)}(\bkappa)\over \sigma_{Qel}d^2\bkappa} 
\exp[i\bkappa(\bs+\br-\br')]\nonumber\\
&\times& w_{\nu}\Big(\nu_A(\bb)\Big)
\textsf{S}[\bb,\sigma(x,\br)]\textsf{S}[\bb,\sigma(x,\br')]\Bigr\}.\nonumber\\
\label{eq:8.A.1.2}
\eea
The factor $\textsf{S}[\bb,\sigma_0(x)]$ in 
$
\textsf{S}[\bb,\Sigma^{el}_{11}(\br,\br')]=
\textsf{S}[\bb,\sigma_0(x)]\textsf{S}[\bb,\sigma(x,\br)]\textsf{S}[\bb,\sigma(x,\br')]
$
was crucial for this identification
of $ w_{\nu}\Big(\nu_A(\bb)\Big)$.

The related contributions from the 2-parton and 3-parton states are given
by  elastic and color-excitation operators from
eqs. (\ref{eq:AppB.1.1},\ref{eq:AppB.1.2},
\ref{eq:AppB.1.3}) ($n=2,3$),
\bea
\bra{3\bar{3}}{\cal S}^{(n)}_{A,\nu}(\bC)\ket{3\bar{3}}& =& 
{1\over \nu!} \Big[-\half \Sigma_{ex}^{(n)} T(\bb)\Big]^{\nu} 
\nonumber\\ 
&\times&\textsf{S}[\bb,\Sigma_{el}^{(n)}(\bC)],
\label{eq:8.A.1.3}
\eea
with the result
\begin{widetext}
\bea
&&\bra{3\bar{3}}{\cal S}^{(2)}_{A,\nu}(\bb,\bb')\ket{3\bar{3}} =
\int d^2\bkappa  {d\sigma_{Qel}^{(\nu)}(\bkappa)\over \sigma_{Qel}d^2\bkappa} 
\exp[i\bkappa(\bs+(1-z)\br-(1-z)\br')] w_{\nu}\Big(\nu_A(\bb)\Big),\nonumber\\
&&\bra{3\bar{3}}{\cal S}^{(2)}_{A,\nu}(\bB,\bb')\ket{3\bar{3}} =
\int d^2\bkappa  {d\sigma_{Qel}^{(\nu)}(\bkappa)\over \sigma_{Qel}d^2\bkappa} 
\exp[i\bkappa(\bs+\br-(1-z)\br')]
w_{\nu}\Big(\nu_A(\bb)\Big)\textsf{S}[\bb,\sigma(x,\br)],
\nonumber\\
&&\bra{3\bar{3}}{\cal S}^{(2)}_{A,\nu}(\bB',\bb)\ket{3\bar{3}} =
\int d^2\bkappa  {d\sigma_{Qel}^{(\nu)}(\bkappa)\over \sigma_{Qel}d^2\bkappa} 
\exp[i\bkappa(\bs+(1-z)\br-\br')]
w_{\nu}\Big(\nu_A(\bb)\Big)\textsf{S}[\bb,\sigma(x,-\br')].
\label{eq:8.A.1.4}
\eea
\end{widetext}
We combine Eqs. (\ref{eq:8.A.1.2}), (\ref{eq:8.A.1.4}), and 
in a now familiar pattern, re-absorb $\textsf{S}[\bb,\sigma(x,\br)]$
and $\textsf{S}[\bb,\sigma(x,\br')]$ into intranuclear distortion
of the $qg$ dipoles. The resulting nonlinear $k_\perp$-factorization 
for the color-triplet dijet cross section wit $\nu$-color-excited nucleons
reads 
\bea
&&{d\sigma_\nu   \bigl(q \to\{ qg\}_3\bigr) 
\over d^2\bb dz d^2\bDelta d^2\bp } =\nonumber\\
&=& {1 \over (2 \pi)^2}  
 w_{\nu}\Big(\nu_A(\bb)\Big){d\sigma_{Qel}^{(\nu)}(\bDelta)\over \sigma_{Qel}
d^2\bDelta}\nonumber\\
&\times&
\left| \Psi_{qg}(1;z,\bp)- \Psi_{qg}(z,\bp-z\bDelta)\right|^2.
\label{eq:8.A.1.5} 
\eea

Following the considerations of Sec. V.C,
this result could have been obtained
from the inclusive triplet-dijet cross section of Ref. \cite{QuarkGluonDijet},
\bea
&&{d \sigma \bigl(q \to \{ qg\}_3\bigr) \over d^2\bb dz d^2\bDelta d^2\bp
  }=
{1 \over (2 \pi)^2}\phi(\bb,x,\bDelta)\nonumber\\
&\times&
\left| \Psi_{qg}(1;z,\bp)- \Psi_{qg}(z,\bp-z\bDelta)\right|^2,
\label{eq:8.A.1.6}
\eea 
by taking for $\phi(\bb,x,\bDelta)$ the familiar multiple scattering expansion.
Such a guesswork leaves open a question about multipomeron exchanges
in $\Psi_{qg}(1;z,\bp)$, though.

\subsubsection{Reggeon field theory interpretation}

The 
reggeon field theory diagrams for the inclusive triplet
cross section (\ref{eq:8.A.1.6}) are shown in Fig.
\ref{fig:QGtripletnew}. Its large-$N_c$ 
structure is best seen
in the composite
quark-antiquark $Q\overline{Q}$ representation for the gluon,
$g^{\alpha}_{\beta}=\overline{Q}^{\alpha}Q_{\beta}$, where $Q$ and $\overline{Q}$ 
propagate at the same impact parameter. At large-$N_c$, the triplet
state  $\{qg\}_3$ has the structure $Q\{\overline{Q}q\}_0$ with 
the color-singlet $\{\overline{Q}q\}_0$
\cite{QuarkGluonDijet}.

The correspondence to the four terms in $\left| \Psi_{qg}(1;z,\bp)- 
\Psi_{qg}(z,\bp-z\bDelta)\right|^2$ is as follows:
The diagram of Fig. \ref{fig:QGtripletnew}a corresponds 
to $|\Psi_{qg}(z,\bp-z \bDelta)|^2$. It originates from 
$\textsf{S}^{(2)}$, i.e., the scattering of the quark is 
followed by the in-vacuum 
fragmentation into the dijet behind the nucleus.
As such, it is free of intranuclear
absorption and distortion
of the color dipole. There is an evident shift of the argument of the
wave function for the incident quark with the transverse
momentum $\bDelta$.
The diagram of Fig.\ref{fig:QGtripletnew}b corresponds 
to $|\Psi_{qg}(1;z,\bp)|^2$ and describes interactions
of the $qg$ pair formed before the target
and originates from $\textsf{S}^{(4)}$.
The open blobs on the two sides of the unitarity cut indicate the
operator, Eq. (\ref{eq:4.D.5}), of the 
intranuclear absorption and distortions of $\Psi_{qg}(1;z,\bp)$ --- 
it is the color-singlet $\{\overline{Q}q\}_0$ dipole which is distorted coherently. 
The diagrams of Figs. \ref{fig:QGtripletnew}c,d
describe the interference of the two mechanisms
and originate from $\textsf{S}^{(3)}$ for the two possible sets of the
three-body states.

The contribution from the intermediate state $\{qg\}_3$ to the 
$ q\CutPom_{A,r}q$ vertex  $\CutD(q\to \{qg\}_3;\CutPom_{A,r})$
equals
\begin{widetext}
\mbox{}
\begin{figure}[!h]
\begin{center}
\includegraphics[width = 3.2cm,angle=270]{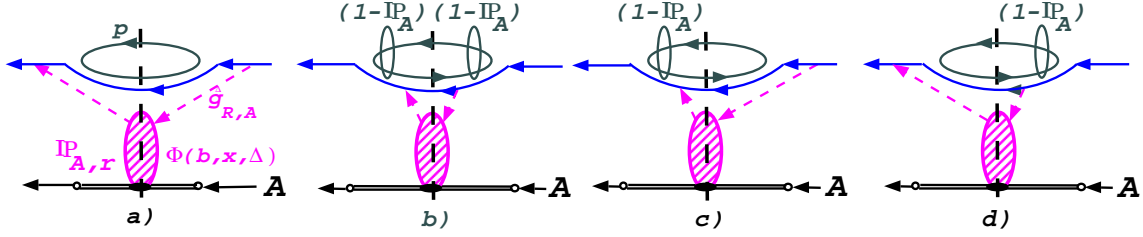}
\caption{ The large-$N_c$ reggeon field theory diagrams for 
the production of $qg$ dijets on the color-triplet sate.
The gluon is shown in the composite quark-antiquark representation.
The correspondence to the excitation vertex $\propto 
 \left| \Psi_{qg}(1;z,\bp)- \Psi_{qg}(z,\bp-z\bDelta)\right|^2$
is described in the text: (a) $\Longrightarrow$ $|\Psi_{qg}(z,\bp-z\bDelta|^2$,
(b) $\Longrightarrow$ $|\Psi_{qg}(z,\bp)|^2$, 
(c) $\Longrightarrow$ $\Psi^*_{qg}(1;z,\bp)\Psi_{qg}(z,\bp-z\bDelta)$,
 (d)$\Longrightarrow$ $\Psi^*_{qg}(z,\bp-z\bDelta) \Psi_{qg}(1;z,\bp)$.}
\label{fig:QGtripletnew}
\end{center}
\end{figure}
\mbox{}
\bea
&&\CutD(q\to \{qg\}_3; \CutPom_{A,r};
\bp,\bDelta,\bp_1,\bp_2,\bkappa)= 
\delta(\bDelta-\bkappa)\nonumber\\
&\times&
[\Psi^*_{qg}(1;z,\bp_2)\delta(\bp -\bp_2)-\Psi^*_{qg}(z,\bp_2)\delta(\bp-z\bkappa-
\bp_2)]\nonumber\\
&\times&
[\Psi_{qg}(1;z,\bp_1)\delta(\bp-\bp_1)-
\Psi_{qg}(z,\bp_1)\delta(\bp-z\bkappa-\bp_1)].
\label{eq:8.A.2.1}
\eea
Its expansion in terms of the in-vacuum cut pomerons $\CutPom_r$ reads:
\bea
&&\CutD(q\to \{qg\}_3; \nu\CutPom_{r};\beta;
\bp,\bDelta,\bp_1,\bp_2,\{\bkappa_i\})= 
\delta(\bDelta-\sum^{\nu}\bkappa_i)\textsf{S}[\bb,\sigma_0]\nonumber\\
&\times&
[\Psi^*_{qg}(1;z,\bp_2)\delta(\bp-\bp_2)-\Psi^*_{qg}(z,\bp_2)\delta(\bp-z\sum^{\nu}\bkappa_i-
\bp_2)]\nonumber\\
&\times&
[\Psi_{qg}(1;z,\bp_1)\delta(\bp-\bp_1)-
\Psi_{qg}(z,\bp_1)\delta(\bp-z\sum^{\nu}\bkappa_i-\bp_1)].
\label{eq:8.A.2.2}
\eea
\end{widetext}
One can expand further in terms of the uncut in-vacuum pomerons $\Pom$
in $\textsf{S}[\bb,\sigma_0]$ and $\Psi_{qg}(1;z,\bp)$, we leave this as 
an exercise.


\subsubsection{Nonperturbative quenching of forward jets in the
triplet channel}

The quenching of color-triplet quark-gluon jets is entirely 
controlled by the quark energy loss before a hard fragmentation
$q\to qg$, which is parameterized in terms of $z_{\nu-1}$
defined in Sec. VII.B2 -- compared to the free-nucleon target, here we have
$\nu-1$ extra quasielastic rescatterings. 
This suggests a  hadronization of the quark and gluon jets with the
modified quark and gluon fragmentation functions 
\beq
D_{q(g),\nu} = {1\over z_{\nu-1}} D_{q(g)}({x_F\over z_{\nu-1}}).
\label{eq:8.A.3.1} 
\eeq


\subsubsection{Integrated topological cross sections in the
triplet channel}

It is convenient to derive the integrated topological triplet
cross sections starting from Eqs. (\ref{eq:8.A.1.2})-(\ref{eq:8.A.1.4}).
The integrations over the jet momenta give $\br'=\br$ and $\bs=0$.
Taking the relevant elastic and excitation scattering
operators from Appendix B,
one readily finds
\begin{widetext}
\bea
&&{d\sigma_{\nu} (q\to \{qg\}_3) \over d^2\bb}=
 \int_0^1 dz \int d^2\br \Psi^*_{qg}(z,\br)\Psi_{qg}(z,\br)
w_{\nu}\Big(\nu_A(\bb)\Big)
\int d^2\bkappa  {d\sigma_{Qel}^{(\nu)}(\bkappa)\over
  \sigma_{Qel}d^2\bkappa}\nonumber\\
&\times& \Bigr\{
\Big(1- \exp\Big[-{1\over 2} \sigma(x,\br)T(\bb)\Big]\Big)^2
+ 2 \exp\Big[-{1\over 2} \sigma(x,\br)T(\bb)\Big]
(1-\exp[i(1-z)\bkappa \br])\Big\} \nonumber\\
&=& \int_0^1 dz \int d^2\br \Psi^*_{qg}(z,\br)\Psi_{qg}(z,\br) w_{\nu}\Big(\nu_A(\bb)\Big)\nonumber\\
&\times&
\Bigr\{
\Big(1- \exp\Big[-{1\over 2} \sigma(x,\br)T(\bb)\Big]\Big)^2
+2 \exp\Big[-{1\over 2} \sigma(x,\br)T(\bb)\Big]
\Big[1-\Big(1-{\sigma(x,(1-z)\br)\over \sigma_0}\Big)^\nu \Big]\Big\} \nonumber\\
&=&\int_0^1 dz \int d^2\br \Psi^*_{qg}(z,\br)\Psi_{qg}(z,\br)
\Biggl\{w_{\nu}\Big(\nu_A(\bb)\Big)\Big(1- \exp\Big[-{1\over 2}
\sigma(x,\br)T(\bb)\Big]\Big)^2
\nonumber\\
&+& 2\exp\Big[-{1\over 2} \sigma(x,\br)T(\bb)\Big]\sum_{k=1}^{\nu}{1\over
  k!}(-1)^{k-1}
w_{\nu-k}\Big(\nu_A(\bb)\Big) \Big[{1\over 2}
\sigma(x,(1-z)\br)T(\bb)\Big]^k\Biggr\}.
\nonumber\\
\label{eq:8.A.4.1}
\eea
\end{widetext}
For $\nu=1$ one has an absorption screened impulse approximation.
It is proportional to $\sigma(x,(1-z)\br)$, in agreement to the color
multiplet decomposition of the free-nucleon cross section in
Eq. (85) of Ref. \cite{QuarkGluonDijet}.


\subsection{The universality class of coherent diffraction}

The component of (\ref{eq:8.A.1.5}) with $\nu=0$ gives the
CD excitation of color-triplet dijets:
\bea
&&\left.{d  \sigma \bigl(q \to \{qg\}_3\bigr) 
\over d^2\bb dz d^2\bDelta d^2\bp }\right|_{coher} =
{1 \over (2 \pi)^2} \delta(\bDelta)\nonumber\\
&\times&
\left| \Psi_{qg}(1;z,\bp)- \Psi_{qg}(z,\bp)\right|^2 
\textsf{S}[\bb,\sigma_0].
\label{eq:8.B.1} 
\eea 
This is a known result \cite{QuarkGluonDijet}, 
the above derivation sheds more light on the connection
between the diffractive production and the generic color-triplet 
excitation: Eq.
(\ref{eq:8.B.1}) can be obtained putting $\nu=0$ in  
(\ref{eq:8.A.1.5}). Incidentally, the     $\nu=0$ 
result (\ref{eq:8.A.4.1})
for the integrated cross section gives the
integrated diffraction cross section.


\subsection{The universality class of dijets in higher
color multiplets}


\subsubsection{Color excitation distributions} 

Excitation of quark-gluon dijets in higher representations, 
the sextet and 15-plet, is a
property of the 4-parton problem, and belongs to the same
universality class as an excitation of color-octet dijets in DIS,
the only difference is that now the incident quark carries a net color.
The leading contribution to the cross section comes from 
the off-diagonal $\hat{\Sigma}_{ex}$, is of the first
order in $\Omega$ and ${\cal O}(N_c^{-1})$, but this 
suppression is compensated for by the large number of
final states, see Eq. (\ref{eq:AppB.2.7}). 

Let there be
$m$ color-excitation interactions in the initial color-triplet $qg$, 
followed by the triplet-to-$\{6+15\}$ transition at
the depth $\beta\equiv \beta_{m+1}$, followed by
$k$ color-excitation interactions in the final state. 
The calculation of the matrix
element $\bra{6\bar{6} + 15\, \overline{15}}{\text S}^{(4)} (\bB,\bB') \ket{3\bar{3}}$
requires the evaluation of 
\bea 
&&  \int_{0}^1  d\beta_{\nu} \dots d\beta \dots d\beta_{1} \nonumber\\
&\times& 
\theta(1 -\beta_\nu)\theta(\beta_\nu
-\beta_{\nu-1}) \dots \theta(\beta_1)\nonumber\\
&\times& \sum_{m,k=0} \delta(\nu-1-m-k)\nonumber\\
&\times& \exp\Bigl[-{1\over 2}(1 -\beta)\Sigma_{22}^{el}T(\bb)\Bigr]
\cdot\left[-{1\over 2} \Sigma_{22}^{ex} T(\bb)\right]^{k}\nonumber\\
&\times& 
{1\over 2N_c} \Omega  T(\bb) \nonumber\\
&\times& 
 \exp\left[-{1\over 2}\beta\Sigma_{11}^{el}T(\bb)\right]
\cdot\left[-{1\over 2} \Sigma_{11}^{ex} T(\bb)\right]^{m}.
\label{eq:8.C.1.1}
\eea

The $\beta_i$ integrations excepting $\beta$ give the factors
$(1-\beta)^k/k! $ for the final state, 
and $\beta^{m}/m!$ for the initial state.  The initial-state
attenuation $\textsf{S}[\bb,\beta\Sigma_{11}]$ can be decomposed 
following (\ref{eq:8.A.1.4}), the familiar identification of 
the coherent distortion of dipoles in the slice $[0,\beta]$ and
of the probability of $m$-fold quasielastic scattering
of the incident quark follow.  

The gluon Casimir $C_A$ enters manifestly the Fourier representation
for  $\Sigma_{22}^{ex}$ of Eq. (\ref{eq:AppB.2.6}) and 
\bea
&&\Bigl(-\Sigma_{22}^{ex}\Bigr)^k =\Bigl\{\int
d^2\bkappa f(x, \bkappa) \exp(i\bkappa\bs)+\nonumber\\
&+&{C_A\over C_F} \int
d^2\bkappa f(x,\bkappa) \exp[i\bkappa(\bs +\br -\br')] \Bigr\}^k  \nonumber\\
&=& 
\sigma_0^k\sum_{j,n=0}\delta(k-j-n)
{k! \over j! n!} \nonumber\\
&\times& \int  d^2\bkappa_1 d^2\bkappa_2 
\exp[i\bkappa_1\bs +i\bkappa_2(\bs-\br + \br')]\nonumber\\
&\times&
 \left({C_A\over C_F}\right)^n
{d\sigma_{Qel}^{(n)}(\bkappa_2)
\over \sigma_{Qel}d^2\bkappa_2} \cdot {d\sigma_{Qel}^{(j)}(\bkappa_1)
\over \sigma_{Qel}d^2\bkappa_1}
\,.
\label{eq:8.C.1.2}
\eea
The decomposition of the final state attenuation,
\bea
\textsf{S}[\bb,(1 -\beta)\Sigma_{22}^{el}]&=&
 \textsf{S}[\bb,(1-\beta)\sigma_0] \nonumber\\
&\times& 
\textsf{S}[\bb,{C_A\over C_F}(1-\beta)\sigma_0],
\label{eq:8.C.1.3}
\eea
allows an identification of the product 
$ w_n \Big({C_A\over C_F}(1 -\beta)\nu_A(\bb)\Big)
w_{j} \Big((1 -\beta)\nu_A(\bb)\Big)$.

Upon the projection onto final states, Eq. (\ref{eq:AppB.2.7}), 
the final nonlinear $k_\perp$-factorization formula for
the spectrum of the sextet and 15-plet dijets in final states with
color excitation of $\nu$ nucleons in the target nucleus reads
\bea
&&{d \sigma_\nu \bigl(q \to\{ qg\}|_{6+15}\bigr) 
\over d^2\bb dz d^2\bDelta d^2\bp }
 = {1 \over (2 \pi)^2} T(\bb)\int_0^1 d\beta  \nonumber\\
&\times& \int  d^2\bkappa d^2\bkappa_1 d^2\bkappa_2 d^2\bkappa_3
\nonumber\\
&\times&\delta(\bkappa +\bkappa_1+\bkappa_2+\bkappa_3-\bDelta)\nonumber\\
&\times& 
 \left|\Psi_{qg}(\beta;z,\bp-\bkappa_1)-
\Psi_{qg}(\beta;z,\bp-\bkappa_1-\bkappa)\right|^2\nonumber\\
&\times& 
\sum_{j,k,m,n=0}\delta(\nu-1-m-k) \delta(k-j-n)
\nonumber\\
&\times&
w_{m}\Big(\beta\nu_A(\bb)\Big) w_n \Big({C_A\over C_F}(1 -\beta)\nu_A(\bb)\Big)\nonumber\\
&\times&
w_{j} \Big((1 -\beta)\nu_A(\bb)\Big)\nonumber\\
&\times& 
{d\sigma_{Qel}(\bkappa)
\over d^2\bkappa}\cdot
{d\sigma_{Qel}^{(m)}(\bkappa_3)
\over \sigma_{Qel}d^2\bkappa_3} \nonumber\\
&\times& 
 {d\sigma_{Qel}^{(n)}(\bkappa_2)
\over \sigma_{Qel}d^2\bkappa_2}
 \cdot {d\sigma_{Qel}^{(j)}(\bkappa_1)
\over \sigma_{Qel}d^2\bkappa_1}.
\label{eq:8.C.1.4} 
\eea
 This result could have been obtained, following the considerations of Sec. V.C,
from the inclusive cross section of Ref. \cite{QuarkGluonDijet}:
\bea
&&{d \sigma \bigl(q \to \{ qg\}|_{6+15}\bigr) \over d^2\bb dz 
d^2\bDelta d^2\bp }
 = {1 \over (2 \pi)^2} T(\bb)\int_0^1 d\beta  \nonumber\\
&\times& \int  d^2\bkappa d^2\bkappa_1 d^2\bkappa_2 d^2\bkappa_3
\delta(\bDelta-\bkappa -\bkappa_1-\bkappa_2-\bkappa_3)\nonumber\\
&\times& \left|\Psi_{qg}(\beta;z,\bp-\bkappa_1)-
\Psi_{qg}(\beta;z,\bp-\bkappa_1-\bkappa)\right|^2\nonumber\\
&\times& f(x,\bkappa)
\Phi(1-\beta;\bb,x,\bkappa_1)\nonumber\\
&\times&\Phi({C_A\over C_F}(1-\beta);\bb,x,\bkappa_2)
\Phi(\beta;\bb,x,\bkappa_3).
\label{eq:8.C.1.5} 
\eea 


\subsubsection{Reggeon field theory interpretation}

\begin{figure}[!h]
\begin{center}
\includegraphics[width = 7.5cm]{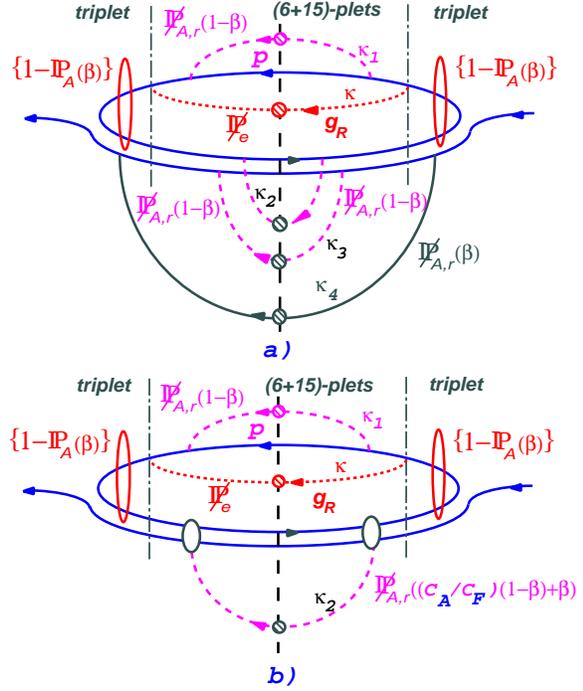}
\caption{The large-$N_c$ structure of cut nuclear pomeron RFT diagram for 
production of quark-gluon
dijets in higher multiplets (sextet and 15-plet in $SU(3)_c$
and their large-$N_c$ generalizations). The bottom diagram (b) is the RFT 
representation for the diagram (a) upon the convolution
(\ref{eq:8.C.2.4}). Gluons are shown in the 
composite quark-antiquark representation.
For a more detailed description see the
text. }
\label{fig:QGsextet2}
\end{center}
\end{figure}

With hindsight, we realize that 
the  nonlinear $k_\perp$ factorization
has a remarkable built-in separation of the uncut, 
and two kinds of cut, pomerons.
First we 
recall the large-$N_c$ property $C_A=2C_F$ and the
convolution property 
\bea
\Biggl(\Phi(\beta_1;\bb)\otimes \Phi(\beta_2;\bb)\Biggr)(\bkappa)=
\Phi(\beta_1+\beta_2;\bb,x,\bkappa).\nonumber\\
 \label{eq:8.C.2.1} 
\eea
Anticipating the composite quark-antiquark representation
for the large-$N_c$ gluon, we make use of (\ref{eq:8.C.2.1}) to represent 
\bea
&&\Phi({C_A\over  C_F}(1-\beta);\bb,x,\bkappa)=\Phi(2(1-\beta);\bb,x,\bkappa)\nonumber\\
&=&\int  d^2\bkappa_1 d^2\bkappa_2 \delta(\bkappa -\bkappa_1-\bkappa_2)\nonumber\\
&\times& 
\Phi(1-\beta;\bb,x,\bkappa_1)\Phi(1-\beta;\bb,x,\bkappa_2),
\label{eq:8.C.2.2} 
\eea 
where we made use of the large-$N_c$ property $C_A=2C_F$.
Then we  cast (\ref{eq:8.C.1.5}) in the form
\bea
&&{d \sigma \bigl(q \to \{ qg\}|_{6+15}\bigr) \over d^2\bb dz 
d^2\bDelta d^2\bp }
 = {1 \over (2 \pi)^2} T(\bb)\int_0^1 d\beta  \nonumber\\
&\times& \int  d^2\bkappa d^2\bkappa_1 d^2\bkappa_2 d^2\bkappa_3 d^2\bkappa_4\nonumber\\
&\times& 
\delta(\bDelta-\bkappa -\bkappa_1-\bkappa_2-\bkappa_3-\bkappa_4)\nonumber\\
&\times& \left|\Psi_{qg}(\beta;z,\bp-\bkappa_1)-
\Psi_{qg}(\beta;z,\bp-\bkappa_1-\bkappa)\right|^2\nonumber\\
&\times& f(x,\bkappa)
\Phi(1-\beta;\bb,x,\bkappa_1)\Phi(1-\beta;\bb,x,\bkappa_2)\nonumber\\
&\times& \Phi(1-\beta;\bb,x,\bkappa_3)
\Phi(\beta;\bb,x,\bkappa_4),\nonumber\\
\label{eq:8.C.2.3} 
\eea
which is custom-tailored for identification of interactions of 
the composite $Q\overline{Q}$ 
gluon with different pomerons.

The corresponding unitarity-cut RFT diagram is shown in
Fig. \ref{fig:QGsextet2}a. In this very busy plot only the gluon --- either
$g_{R,A}$ or $g_R$ --- loops are shown, the dashed cut circles
stand for the corresponding gluon densities. We show only one of the four possible
coupling of the in-vacuum gluon $g_R$ from the cut pomeron $\CutPom_e$
to the quark loop. Only the pomeron operator structure of
distortions of $\Psi_{qg}(\beta;z,\bp)$ 
in the slice $[0,\beta]$ is indicated, see Eq. (\ref{eq:4.D.5}),
these distortions do not depend on the multiplicity of cut pomerons.
The transitions at the depth $\beta$ 
from the initial state color-triplet $qg$ to the 
higher representations --- sextet and 15-plet --- are indicated by 
short-dashed vertical line on the rhs and lhs of the unitarity cut.

In the slice $[0,\beta]$ color rotations  only
within the triplet state are allowed. Correspondingly, as 
we did learn in Sec. VIII.A.2, only the quark $Q$ of the composite
gluon would couple to the nuclear glue $g_{R,A}$: the triplet
property of this quark is retained after any multiplicity of
rescatterings, while the color-singlet $\{\overline{Q}q\}_0$ state is
distorted coherently. The higher --- 
sextet and 15-plet --- representations are obtained
after excitation of  the color-singlet $\{\overline{Q}q\}_0$ to the
color-octet $\{\overline{Q}q\}_8$, upon which all three partons 
interact independently.

In the practical calculations it is advisable to  make use of
the convolution property (\ref{eq:8.C.2.1}) the other way around,
\bea
&&\Bigl(\Phi(1-\beta;\bb)\otimes \Phi(1-\beta;\bb)
\otimes\Phi(\beta;\bb)\Bigr)(\bkappa)
\nonumber\\
&=&\Phi(\beta^*;\bb,x,\bkappa),
 \label{eq:8.C.2.4} 
\eea
where 
\beq
\beta^*={C_A\over C_F}(1-\beta)+\beta,
 \label{eq:8.C.2.5} 
\eeq
which defines the collective nuclear  cut pomeron
$\CutPom_{A,r}(\beta^*)$.
The corresponding RFT diagram is shown in
Fig. \ref{fig:QGsextet2}b. The above clearcut origin of different nuclear
cut pomerons is obscured by such a short-hand form of the spectrum,
but it brings the spectrum to a close similarity with the spectrum of dijets
in inelastic DIS and is convenient for the RFT representation of topological
cross sections in terms of multiple exchanges by in-vacuum cut pomerons
$\CutPom_r$. For instance, the contribution from
intermediate state $\{ qg \}_{6+15}$
 to the 
$q\CutPom_{e}\CutPom_{A,r}(\beta^*)q$ vertex
would equal
\begin{widetext}
\bea  
&&\CutD (q \to \{ qg \}_{6+15}; 
\CutPom_{A,r}(1-\beta),\CutPom_{e},
\CutPom_{A,r}(\beta^*);
\beta;
\bp,\bDelta,\bp_1,\bp_2,\bkappa,
\bkappa_1 ,
\bkappa_2 )=\delta(\bDelta-\bkappa_1-\bkappa_2-\bkappa)\nonumber\\
&\times&
\Psi^*_{qg}(\beta;z,\bp_2)
[\delta(\bp-\bkappa_1 - \bp_2)-
\delta(\bp-\bkappa_1  -\bkappa-\bp_2)]\nonumber\\
&\times&
[\delta(\bp-\bkappa_1-\bp_1)-
\delta(\bp-\bkappa_1-\bkappa)-\bp_1)]
\Psi_{qg}(\beta;z,\bp_1),
\label{eq:8.C.2.6}
\eea
and its expansion in terms of the in-vacuum cut pomerons $\CutPom_r$ reads
\bea  
&&\CutD(q\to \{ qg\}_{6+15}; j\CutPom_{r},\CutPom_{e},
k\CutPom_{r};\beta;
\bp,\bDelta,\bp_1,\bp_2,\bkappa,\{\bk_{i}\},
\{\bq_{m}\})=\nonumber\\
&=&\delta(\bDelta-\sum_{i=1}^{j}\bk_{i}-
\sum_{m=1}^{k}\bq_{m}-\bkappa)\textsf{S}[\bb,(1-\beta)\sigma_0(x)] 
\textsf{S}\Big[\bb, \beta^*\sigma_0(x)\Big]\nonumber\\
&\times&
(\beta^*)^k(1-\beta)^j
\nonumber\\
&\times&
\Psi^*_{qg}(\beta;z,\bp_2)
[\delta(\bp-\sum_{i=1}^{j}\bk_{i}- \bp_2)-
\delta(\bp-\sum_{i=1}^{j}\bk_{i} -\bkappa-
\bp_2)]\nonumber\\
&\times&
[\delta(\bp-\sum_{i=1}^{j}\bk_{i}-\bp_1)-
\delta(\bp-\sum_{i=1}^{j}\bk_{i}-\bkappa -\bp_1)]
\Psi_{qg}(\beta;z,\bp_1)
\label{eq:8.C.2.7}
\eea
\end{widetext}


\subsubsection{Nonperturbative quenching of leading jets}

In comparison to DIS, a new feature is that both the
initial and final state interactions of the $qg$ system 
 contribute to the 
nonperturbative quenching of jets. 
One would associate
color excitations in the slice $[0,\beta]$ of the nucleus 
with the $m$-fold quasielastic scattering of the
incident quark. It is followed by hard excitation $q\to qg$ 
at the depth $\beta$. The quark and gluon jet
hadronization (fragmentation) functions are different 
(for the review see \cite{DreminGary}) and a correct
partition of $n$-fold
color excitations in the slice $[\beta,1]$ between the 
quark and gluon quasielastic rescatterings is 
important. Evidently,  
$w_k \Big({C_A\over C_F}(1 -\beta)\nu_A(\bb)\Big)$
must be regarded as a probability of the $k$-fold 
rescattering of the gluon, while 
$w_{j} \Big((1 -\beta)\nu_A(\bb)\Big)$ describes 
the $j$-fold rescattering of the quark.
The nucleon excited into the color-octet state would
arguably hadronize independently of the projectile parton.
Along the tree of quark interactions we have $(m+j)$
quasielastic scatterings and  
the average energy of the final state quark will be scaled down by
the factor $z_q=z_{m+j}$ which must be used in the
modified fragmentation function (\ref{eq:7.B.2.8}), while
the energy of gluons scales down by the factor 
$z_g = z_{m+k}$.


\subsubsection{Integrated topological cross sections}

Taking the relevant elastic and excitation scattering operators
for $\bs=0$ and $\br=\br'$, we obtain
\bea
&&{d\sigma_{\nu} (q\to \{qg\}_{6+15}) \over d^2\bb}=
\nonumber\\
&=&\int_0^1 dz \int d^2\br \Psi^*_{qg}(z,\br)\Psi_{qg}(z,\br)\nonumber\\
&\times&\int_0^1 d\beta \sum_{j,k=0} \delta(\nu-1-j-k)\nonumber\\
&\times&
w_{j}\Big(\beta \nu_A(\bb)\Big)w_{k}\Big({C_{15}\over
  C_F}(1-\beta)\nu_A(\bb)\Big)\nonumber\\
&\times&
\sigma(x,\br)\exp\Big[-\beta
\sigma(x,\br)T(\bb)\Big]
\label{eq:8.C.4.1}
\eea
Here the Casimir $C_{15}=C_{6}$ makes an explicit appearance.

There is a strong distinction between 
the topological cross section for the triplet and sextet plus
15-plet sectors. It is a clearcut manifestation of the
non-Abelian coupled-channel intranuclear evolution of
color dipoles in pQCD.

Regarding the practical applications of  (\ref{eq:8.A.4.1}) and 
(\ref{eq:8.C.4.1}) to the proton-nucleus
collisions, one can start with the oversimplified 
Additive Quark Model scenario for
multiproduction \cite{KolyaUFN,Anisovich}. Here (\ref{eq:4.C.5})
is the Born cross section for the  
constituent quark-nucleus interaction. This Born cross section 
must be corrected for pQCD virtual emission as described
in Ref. \cite{VirtualReal}. Then our Eqs. (\ref{eq:8.A.4.1})
and (\ref{eq:8.C.4.1}) describe the real emission radiative
corrections to the constituent quark-nucleus interaction.


\section{Topological cross sections for gluon-gluon dijet production}


\subsection{Color-diagonal and color-excitation interactions}

The underlying pQCD subprocess is $gg_t \to gg$, where $g$ 
is the gluon from the projectile hadron. A full derivation of
irreducible color-representations for digluon states,
their nomenclature, and the 
cross-section operators $\hat \Sigma^{(n)}$ for color-singlet 
two-gluon, three-gluon and four-gluon states at
arbitrary $N_c$ is found in Ref. \cite{GluonGluonDijet}.

Make notice of the important change of the notations: hereafter the basic 
quantity is a cross section 
for the color-singlet octet-octet dipole
\bea
\sigma_{gg}(x,\br) &=& {C_A \over C_F} \int d^2\bkappa f (x,\bkappa)
[1-\exp(i\bkappa\br)]\nonumber\\
& =& {C_A \over C_F}\sigma_{q\bar{q}}(x,\br)\,.
\label{eq:9.A.1}
\eea
Similarly, we define the nuclear thickness 
\beq
\nu_{A,gg}(\bb)=\half \sigma_{gg,0}(x)T(\bb) = \half \cdot  {C_A \over
  C_F}\sigma_{\bar{q}q,0}(x)T(\bb).
\label{eq:9.A.2}
\eeq
The differential cross section of quasielastic gluon-nucleon scattering
equals
\beq
{d\sigma_{g,Qel}\over d^2\bkappa} = {C_A \over
  C_F}\cdot \half f(x,\bkappa)
\label{eq:9.A.3}
\eeq
and 
\beq
\sigma_{g,Qel}=\int  d^2\bkappa {d\sigma_{g,Qel}\over d^2\bkappa} = {C_A \over
  C_F}\cdot \sigma_{q,Qel}
\label{eq:9.A.4}
\eeq
The octet-octet dipole nuclear $\textsf{S}$-matrix 
defines the octet-octet collective 
nuclear glue
\bea
&&\Phi_{gg}(\bb,x,\bkappa_1) =\nonumber\\&=&{1\over (2\pi)^2}\int d^2\br
  \textsf{S}[\bb,\sigma_{gg}(x,\br)]\exp[-i\bkappa\cdot \br]\nonumber\\
&=&\sum_{j=0} w_j\Big(\nu_A(\bb)\Big)
{d\sigma_{g,Qel}^{(j)}\over \sigma_{g,Qel} d^2\bkappa}.
\label{eq:9.A.5}
\eea
There is an obvious identity 
\bea
{d\sigma_{g,Qel}^{(j)}\over \sigma_{g,Qel} d^2\bkappa}
={d\sigma_{Qel}^{(j)}\over \sigma_{Qel} d^2\bkappa}.
\label{eq:9.A.6}
\eea
Finally, now $\Psi_{gg}(z,\bp)$ would be the wave function of the $gg$ Fock
state of the gluon, its relation to the splitting function $P_{gg}(z)$ is found
in Ref. \cite{SingleJet}.


\subsection{Topological cross sections for gluon-gluon dijets}

The subsequent derivation of the dijet cross sections would repeat
that for the quark-gluon dijets and we sketch it only briefly.
The incident parton is the color-octet gluon, and the excitation of 
color-octet digluons 
is driven by color rotations within the octet sector 
and is ${\cal O}(N_c^0)$. Excitation of
dijets in higher multiplets --- two decuplets, 27 and $R_7$ --- 
is driven by the off-diagonal $\hat{\omega}$
and is also ${\cal O}(N_c^0)$: the large number of states in higher 
multiplets overcomes the suppression of their excitation. 
Excitation of color-singlet dijets is ${\cal O}(N_c^{-2})$.
Invoking the considerations of Sec. V.E,
we can derive the topological dijet cross sections directly 
from our results for the inclusive cross sections \cite{GluonGluonDijet},
we confine ourselves to the two most important examples.
We report the leading terms of the large-$N_c$ perturbation theory.


\subsubsection{The universality class of color multiplets of
the projectile parton: color-octet sector}

For dijets in the color-octet final state the inclusive cross section equals 
\bea
&&{d \sigma (g \to \{g_1g_2 \}_{8_A + 8_S} ) 
\over d^2\bb dz d^2\bp d^2\bDelta }=\nonumber\\
& =& 
 {1 \over 2(2 \pi)^2} \int d^2 \bkappa_1 
\int d^2 \bkappa_2 \delta^{(2)}(\bDelta- \bkappa_1 - \bkappa_2)
\nonumber\\ 
&\times&
\Big\{ |\Psi_{gg}(1;z,\bp-\bkappa_1) - \Psi_{gg}(z, \bp - z (\bkappa_1+\bkappa_2))|^2
\nonumber\\
& +
&|\Psi_{gg}(1;z,\bp-\bkappa_2) - \Psi_{gg}(z, \bp - z (\bkappa_1+\bkappa_2))|^2
\Big\} 
\nonumber\\ 
&\times&
\Phi_g(\bb,x,\bkappa_2) \Phi_g(\bb,x,\bkappa_1). 
\label{eq:9.B.1.1}
\eea
(Here the symmetrization over $\bkappa_{1,2}$ is optional.)
Its major subtlety is that the nonlinear $k_\perp$-factorization
quadrature is formulated in terms of the collective glue
$\Phi_g(\bb,x,\bkappa_1)$
for dipoles interacting with one half of the octet-octet dipole
cross section, 
\bea
\Phi_g(\bb,x,\bkappa_1) &=&{1\over (2\pi)^2}\int d^2\br 
\textsf{S}[\bb,\half\sigma_{gg}(x,\br)]\nonumber\\
&=&\sum_{j=0} w_j\Big(\half\nu_{A,gg}(\bb)\Big)
{d\sigma_{g,Qel}^{(j)}\over \sigma_{g,Qel} d^2\bkappa}.\nonumber\\
\label{eq:9.B.1.2}
\eea
At large $N_c$ we have an equality
$\Phi_g(\bb,x,\bkappa_1)=\Phi(\bb,x,\bkappa_1)$.
The topological cross sections derive from the expansion
\bea 
&&\Phi_g(\bb,x,\bkappa_2) \Phi_g(\bb,x,\bkappa_1)=\nonumber\\
& =&
\sum_{\nu=0} \sum_{k=0} \sum_{n=0}  
\delta(\nu-k-n)\nonumber\\
&\times&
w_k\Big(\half\nu_{A,gg}(\bb)\Big)w_n\Big(\half\nu_{A,gg}(\bb)\Big)\nonumber\\
&\times&
{d\sigma_{g,Qel}^{(k)}(\bkappa_1)\over \sigma_{g,Qel} d^2\bkappa_1}\cdot
{d\sigma_{g,Qel}^{(n)}(\bkappa_2)\over \sigma_{g,Qel} d^2\bkappa_2}.
 \label{eq:9.B.1.3}
\eea 
The term with $\nu=0$ gives the CD $g\to gg$.

\begin{figure}[!h]
\begin{center}
\includegraphics[width = 6cm,angle=270]{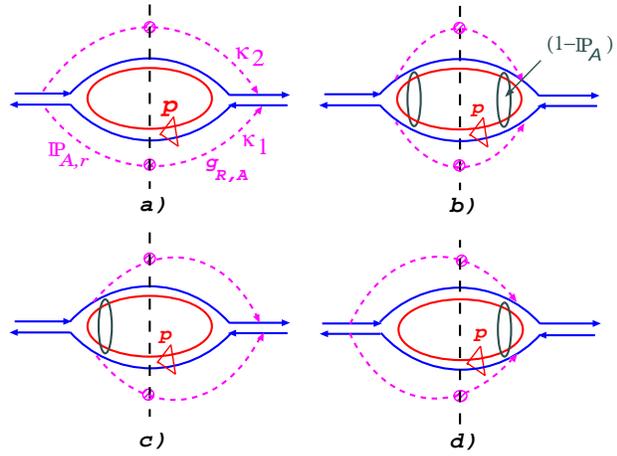}
\caption{The large-$N_c$ structure of cut nuclear pomeron diagram for 
production of color-octet gluon-gluon
dijets. Gluons are shown in the composite quark-antiquark representation,
the shaded circles stand for the collective nuclear gluon densities.
The correspondence to the excitation vertex $\propto 
 \left| \Psi_{gg}(1;z,\bp-\bkappa_1)- \Psi_{gg}(z,\bp-z(\bkappa_1+\bkappa_2))\right|^2$
is as follows: (a) $\Longrightarrow$ $|\Psi_{gg}(z,\bp-z(\bkappa_1+\bkappa_2))|^2$,
(b) $\Longrightarrow$  $|\Psi_{gg}(1;z,\bp-\bkappa_1)|^2$,
 (c) $\Longrightarrow$  $\Psi^*_{gg}(1;z,\bp-\bkappa_1)
\Psi_{gg}(z,\bp-z(\bkappa_1+\bkappa_2))$,
 (d) $\Longrightarrow$  
$\Psi^*_{gg}(z,\bp-z(\bkappa_1+\bkappa_2)) \Psi_{gg}(1;z,\bp-\bkappa_1)$.}
\label{fig:GGoctet}
\end{center}
\end{figure}

Recall the large-$N_c$ equality
$\Phi_g(\bb,x,\bkappa_1)=\Phi(\bb,x,\bkappa_1)$.
Then the emergence of $\Phi_g(\bb,x,\bkappa_1)$ entails a simple RFT
representation for (\ref{eq:9.B.1.1}) in terms of the
composite quark-antiquark representation in Fig. \ref{fig:GGoctet}.
Here the central quark loop describes
a propagation of the color-singlet quark-antiquark. It is the
wave function of this color-singlet dipole which is
coherently distorted by intranuclear propagation, as shown by
vertical open blobs in diagrams
\ref{fig:GGoctet}b,c,d containing interactions of the 
preformed gluon-gluon state.

\begin{figure}[!h]
\begin{center}
\includegraphics[width = 5.5cm,angle=270]{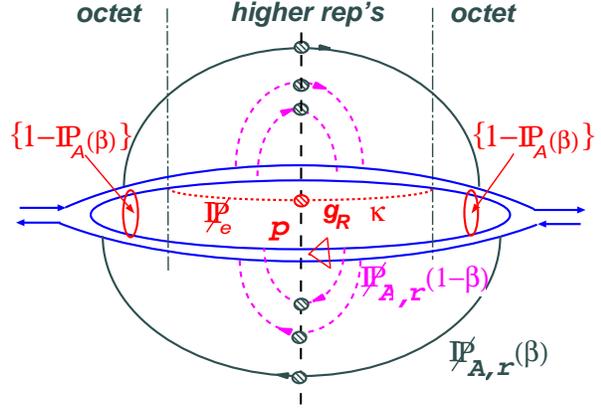}
\caption{The large-$N_c$ structure of cut nuclear pomeron diagram for 
production of  gluon-gluon
dijets in higher representations. Gluons are shown in the composite quark-antiquark representation.
There are three more diagrams with recouplings of the gluon $g_R$ form
$\CutPom_e$ within
the internal quark loop.}
\label{fig:GGhigherRep}
\end{center}
\end{figure}

\begin{widetext}

The contribution from the gluon loop 
with the color-octet intermediate state 
$g\to \{gg\}_8$ to the $g\CutPom_{A,r},\CutPom_{A,r}g$ vertex  
is a straightforward
generalization of Eq. (\ref{eq:8.A.2.1}) for color-triplet 
excitation $q\to \{qg\}_3$:
\bea
&&\CutD(g\to \{gg\}_8; \CutPom_{A,r},\CutPom_{A,r};
\bp,\bDelta,\bp_1,\bp_2,\bkappa_1,\bkappa_2)= 
\delta(\bDelta-\bkappa_1-\bkappa_1)\textsf{S}[\bb,\sigma_0]\nonumber\\
&\times&
[\Psi^*_{gg}(1;z,\bp_2)\delta(\bp -\bkappa_1-\bp_2)-\Psi^*_{gg}(z,\bp_2)\delta(\bp-z(\bkappa_1+\bkappa_2)
- \bp_2)]\nonumber\\
&\times&
[\Psi_{gg}(1;z,\bp_1)\delta(\bp -\bkappa_1-\bp_1)-
\Psi_{gg}(z,\bp_1)\delta(\bp-z(\bkappa_1+\bkappa_2)-\bp_1)].
\label{eq:9.B.1.4}
\eea
The pattern of transformation from (\ref{eq:9.B.1.4}) to the
expansion in terms of the in-vacuum cut pomerons $\CutPom_r$ 
is the same as in the transformation from (\ref{eq:8.A.2.1})
to (\ref{eq:8.A.2.2}), we leave this as an exercise.

The fully integrated topological cross section can readily be
derived:
\bea
&&{d\sigma_{\nu} (g \to \{gg\}_8) \over d^2\bb}
 = \int_0^1 dz \int d^2\br| \Psi_{gg}(z,\br)|^2 w_{\nu}\Big(\nu_A(\bb)\Big)\nonumber\\
&\times&
\Bigr\{
\Big(1- \exp\Big[-{1\over 2} \sigma_{gg}(x,\br)T(\bb)\Big]\Big)^2\nonumber\\
&+& 
2 \exp\Big[-{1\over 2} \sigma_{gg}(x,\br)T(\bb)\Big]
\Big[1-\Big(1-{\sigma_{gg}(x,(1-z)\br)+\sigma_{gg}(x,z\br) \over 2\sigma_{gg,0}(x)}\Big)^\nu ]\Big\} \nonumber\\
&=&\int_0^1 dz \int d^2\br |\Psi_{gg}(z,\br)|^2
\Biggl\{w_{\nu}\Big(\nu_A(\bb)\Big)\Big(1- \exp\Big[-{1\over 2}
\sigma_{gg}(x,\br)T(\bb)\Big]\Big)^2
+ 2\exp\Big[-{1\over 2} \sigma_{gg}(x,\br)T(\bb)\Big]
\nonumber \\
&\times&
\sum_{k=1}^{\nu}{1\over
  k!}(-1)^{k-1}
w_{\nu-k}\Big(\nu_A(\bb)\Big) 
\Big[{1\over 4} [\sigma_{gg}(x,(1-z)\br)+
\sigma_{gg}(x,z\br)]T(\bb)\Big]^k\Biggr\}.
\label{eq:9.B.1.5}
\eea
In conformity to the concept of universality classes, 
there is a striking similarity to (\ref{eq:8.A.4.1}).


\subsubsection{The universality class of higher representations: $\dim(\textsf{R}) =
{\cal O}(N_c^4)$.}

We cite the result \cite{GluonGluonDijet}
\bea
&&{d \sigma (g \to \{g_1g_2 \}_{10+ \overline{10} + 27 + R_7 } ) 
\over d^2\bb dz d^2\bp d^2\bDelta } = 
{1 \over 4 (2\pi)^2}\
 \int_0^1 d\beta 
\int d^2\bkappa_4 d^2\bkappa_3 d^2\bkappa_2 d^2 \bkappa_1 d^2\bkappa   \nonumber\\
&\times&
\delta^{(2)}(\bDelta-\bkappa- \bkappa_1 - \bkappa_2 -\bkappa_3 -\bkappa_4) 
\nonumber\\
& \times& 
\Big\{|\Psi_{gg}(\beta;z, \bp-\bkappa_2-\bkappa_4) - 
\Psi_{gg}(\beta;z, \bp-\bkappa_2 -\bkappa_4-\bkappa)|^2 \nonumber\\
&+&|\Psi_{gg}(\beta;z, \bp-\bkappa_1-\bkappa_3) - 
\Psi_{gg}(\beta;z, \bp-\bkappa_1 -\bkappa_3-\bkappa)|^2\Big\}
\nonumber \\
& \times&
\Phi_g(\beta;\bb,x,\bkappa_1)
 \Phi_g(\beta;\bb,x,\bkappa_2)
\Phi_g\Big({C_{27}\over C_A}(1-\beta);\bb,x,\bkappa_3\Big)
\Phi_g\Big({C_{27}\over
  C_A}(1-\beta);\bb,x,\bkappa_4\Big){d\sigma_{g,Qel}(\bkappa)\over
  d^2\bkappa}.
\nonumber \\
\label{eq:9.B.2.1}
\eea
\end{widetext}
Here the distorted wave function
is evaluated with the nuclear glue $\Phi_g$, i.e., in the
short-hand notation, 
$\Psi_{gg}(\beta;z, \bp) = (1-\Pom_{A,g}(\beta)\otimes)\Psi_{gg}(z,\bp)$.
At large $N_c$ one could have used ${C_{27}= 2C_A}$, 
we keep the explicit ratio of quadratic Casimirs on purpose to
elucidate the color-representation dependence of the initial
and final state interaction effects. Bearing in mind the
quark-antiquark description of large-$N_c$ gluons, one can
deconvolute
\bea
&&\Phi({C_{27}\over C_A}(1-\beta);\bb,x,\bkappa)=\Phi(2(1-\beta);\bb,x,\bkappa)\nonumber\\
&=&\int  d^2\bkappa_1 d^2\bkappa_2 \delta(\bkappa -\bkappa_1-\bkappa_2)\nonumber\\
&&\times
\Phi(1-\beta;\bb,x,\bkappa_1)\Phi(1-\beta;\bb,x,\bkappa_2),
\label{eq:9.B.2.2} 
\eea 
In the integrand of the 
nonlinear $k_\perp$ factorization quadrature 
this gives rise to the product
$$
\Phi_g(\beta;\bb,x,\bkappa_1)
 \Phi_g(\beta;\bb,x,\bkappa_2)\prod_{i=3}^6 \Phi_g\Big(1-\beta;\bb,x,\bkappa_i\Big)
.
$$
In the now familiar reinterpretation, the product $\Phi_g(\beta;\bb,x,\bkappa_1)
 \Phi_g(\beta;\bb,x,\bkappa_2)$ describes color rotations of the
color-octet $\{gg\}_8$ dipole. These color rotations are accompanied by a
coherent distortion of the wave function of the color 
singlet quark-antiquark pair. The product 
$$\prod_{i=3}^6 \Phi_g\Big(1-\beta;\bb,x,\bkappa_i\Big)$$ 
describes rotations within the space of higher dimensional 
representations after the $gg$ dipole has been excited at the depth $\beta$ 
from the octet to those higher representations. 
The corresponding RFT diagram is shown
in Fig. \ref{fig:GGhigherRep}.

Upon this identification, one can convolute back to the two nuclear cut pomerons
$\CutPom_{A,r}(\beta^*)$, where
\beq
\beta^*={C_{27}\over C_A}(1-\beta)+\beta.
\label{eq:9.B.2.3}
\eeq
The 
$g\CutPom_{e}\CutPom_{A,r}(\beta^*)
\CutPom_{A,r}(\beta^*)g$ vertex
would equal
\begin{widetext}
\bea  
&&\CutD (g \to \{ gg \}_{10+\overline{10}+27+R_7}; 
\CutPom_{A,r}(\beta^*),\CutPom_{e},
\CutPom_{A,r}(\beta^*)
;\beta;
\bp,\bDelta,\bp_1,\bp_2,\bkappa,
\bkappa_1,
\bkappa_2 )=\nonumber\\
&=&\delta(\bDelta-\bkappa_1-\bkappa_2-\bkappa)\nonumber\\
&\times&
\Psi^*_{gg}(\beta;z,\bp_2)
[\delta(\bp-\bkappa_1 - \bp_2)-
\delta(\bp-\bkappa_1  -\bkappa-\bp_2)]\nonumber\\
&\times&
[\delta(\bp-\bkappa_1-\bp_1)-
\delta(\bp-\bkappa_1-\bkappa -\bp_1)]
\Psi_{gg}(\beta;z,\bp_1),
\label{eq:9.B.2.4}
\eea
and its expansion in terms of the in-vacuum cut pomerons $\CutPom_r$ 
would read
\bea  
&&\CutD(g \to \{ gg \}_{10+\overline{10}+27+R_7}); j\CutPom_{r},\CutPom_{e},
k\CutPom_{r};\beta;
\bp,\bDelta,\bp_1,\bp_2,\bkappa,\{\bk_{i}\},
\{\bq_{m}\})=\nonumber\\
&=&\delta(\bDelta-\sum_{i=1}^{j}\bk_{i}-
\sum_{m=1}^{k}\bq_{m}-\bkappa)\textsf{S}[\bb,\beta^*\sigma_0(x)] 
\textsf{S}[\bb,\beta^*\sigma_0(x)]\nonumber\\
&\times&
(\beta^*)^{k+j}
\nonumber\\
&\times&
\Psi^*_{gg}(\beta;z,\bp_2)
[\delta(\bp-\sum_{i=1}^{j}\bk_{i}- \bp_2)-
\delta(\bp-\sum_{i=1}^{j}\bk_{i} -\bkappa-
\bp_2)]\nonumber\\
&\times&
[\delta(\bp-\sum_{i=1}^{j}\bk_{i}-\bp_1)-
\delta(\bp-\sum_{i=1}^{j}\bk_{i}-\bkappa -\bp_1)]
\Psi_{gg}(\beta;z,\bp_1).
\label{eq:9.B.2.5}
\eea
\end{widetext} 

Now we cite the integrated topological 
cross sections for the higher representation sector:
\bea
&&{d\sigma_{\nu} (g \to \{ gg \}_{10+\overline{10}+27+R_7}) \over d^2\bb}= T(\bb)
 \int_0^1 dz \nonumber\\
&\times&\int d^2\br |\Psi_{gg}(z,\br)|^2
\int_0^1 d\beta \sum_{j,k=0} \delta(\nu-1-k-j)\nonumber\\
&\times&
w_{k}\Big({C_{27}\over C_A}(1-\beta)\nu_{A,gg}(\bb)\Big)
w_{j}\Big(\beta\nu_{A,gg}(\bb)\Big)\nonumber\\
&\times&\sigma_{gg}(x,\br)
\Big(1-{\sigma_{gg}(x,\br)\over \sigma_{gg,0}(x)}\Big)^j
\nonumber \\
&\times& 
\exp\Big[- {1 \over 2}\sigma_{gg}(x,\br)T(\bb)\Big].
\label{eq:9.B.2.6}
\eea
Taking $\nu=1$ and neglecting absorption, one would recover
the impulse approximation result, cf. eq. (87) 
of Ref. \cite{GluonGluonDijet}.


\section{Topological cross sections for open charm production
in hadron-nucleus collisions}

For the sake of completeness we report the results for
this last remaining hard pQCD subprocess. It is also of
great practical significance since 
Bremsstrahlung by heavy quarks is weaker than by
light quarks \cite{Djordjevic} and one can utilize 
nuclear quenching of heavy flavors for the determination
of the nonperturbative energy flow to a nucleus. 
 The underlying pQCD
process is $gg_t \to c\bar{c}$. Interactions of the
$4$-parton system, $c\bar{c}c'\bar{c}'$, are described
by the same two-channel operator as in DIS
\cite{Nonlinear}. Besides the open heavy flavor (for the sake of
definiteness we speak of the open charm), the results
are equally applicable to mid-rapidity hard light quark-antiquark
dijets. 

The averaging over colors of the initial
gluon and summation over the color
states of dijets in the final state is of the form 
\bea
&&{1\over \sqrt{\textsf{dim[8]}}}\sum_f\bra{f}=\sum_{R}\sqrt{{\textsf{dim}[R]}
\over \textsf{dim[8]}}\bra{R\bar{R}}=\nonumber\\
&=&{1\over \sqrt{N_c^2-1}} \bra{e_1}+\bra{e_2}.
\label{eq:10.1}
\eea
In the large $N_c$ perturbation theory,
the reaction is dominated by color rotations within the space
of color-octet $\{c\bar{c}\}_8$. It differs from the 
excitation of color-triplet quark-gluon 
and color-octet gluon-gluon dijets: open charm is excited 
from the incident parton already in the highest possible
representation. In this respect, the open charm is a
universality class of its own.

In close similarity
to (\ref{eq:8.A.1.2}) we obtain 
\bea
\bra{88}{\cal S}^{(4)}_{A,\nu}(\bC,\bb)\ket{88}& = &
{1\over \nu!} \Big[-\half \Sigma^{ex}_{22} T(\bb)\Big]^{\nu} \nonumber\\
\textsf{S}[\bb,\Sigma^{el}_{22}(\bC)].
\label{eq:10.2}
\eea
In the evaluation of two- and three-body states we
notice that  
\bea
\Sigma^{el}_{22}(\bC)&=&\Sigma_{el}^{(3)}(\bB,\bb')=\Sigma_{el}^{(3)}(\bb,\bB')
=\Sigma_{el}^{(2)}(\bb,\bb')\nonumber\\
& =& {C_A\over C_F}\sigma_0(x)=2\sigma_0(x).
\label{eq:10.3}
\eea
Consequently, in the expansion over color excitations we
encounter
\bea
&&[-\Sigma^{ex}_{22}(\bC)]^\nu+ [-\Sigma_{ex}^{(2)}(\bb,\bb')]^\nu
\nonumber\\
&-&[-\Sigma_{ex}^{(3)}(\bB,\bb')]^\nu-[-\Sigma_{ex}^{(3)}(\bb,\bB')]^\nu=\nonumber\\
&=&\sigma_0^\nu \sum_{k=0}^{\nu}
{\nu! \over k! (\nu-k)!} \int  d^2\bkappa_1 d^2\bkappa_2 \nonumber\\
&\times& {d\sigma_{Qel}^{(k)}(\bkappa_1)
\over \sigma_{Qel}d^2\bkappa_1} \cdot {d\sigma_{Qel}^{(\nu-k)}(\bkappa_2)
\over \sigma_{Qel}d^2\bkappa_2}\exp(i(\bkappa_1+\bkappa_2)\bs)
\nonumber\\
&\times&\Bigl\{ \exp[-i\bkappa_2\br]- 
\exp[-i(\bkappa_1+ \bkappa_2)(1-z)\br] \Big\} \nonumber\\
&\times& \Big\{ \exp[i\bkappa_2\br']- 
\exp[+i(\bkappa_1+ \bkappa_2)(1-z)\br'] \Big\},\nonumber\\
\label{eq:10.4}
\eea
which gives 
\bea
&&{d \sigma_\nu \bigl(g \to\{c\bar{c}\}_{8}\bigr) 
\over d^2\bb dz d^2\bDelta d^2\bp }
 = 
 {1 \over 2(2 \pi)^2} \int d^2 \bkappa_1 
 d^2 \bkappa_2  \nonumber \\
&\times& \delta^{(2)}  (\bDelta- \bkappa_1 - \bkappa_2)
\nonumber\\ 
&\times&
\Big\{ |\Psi_{c\bar{c}}(z,\bp-\bkappa_1) - \Psi_{c\bar{c}}(z, \bp - z(\bkappa_1 + \bkappa_2))|^2
\nonumber\\
& +
&|\Psi_{c\bar{c}}(z,\bp-\bkappa_2) - \Psi_{c\bar{c}}(z, \bp - z (\bkappa_1 + \bkappa_2))|^2
\Big\} 
\nonumber\\ 
&\times&\sum_{k=0}^{\nu} w_{\nu-k}\Big(\nu_A(\bb)\Big) 
w_k \Big(\nu_A(\bb)\Big)\nonumber\\ 
&\times&
 {d\sigma_{Qel}^{(\nu-k)}(\bkappa_2)
\over \sigma_{Qel}d^2\bkappa_2}
 \cdot {d\sigma_{Qel}^{(k)}(\bkappa_1)
\over \sigma_{Qel}d^2\bkappa_1}
\label{eq:10.5}
\eea
Of course, this result could have been derived from the inclusive spectrum
\cite{SingleJet,Nonuniversality,Paradigm,GluonGluonDijet}
\bea
&&{d \sigma_\nu \bigl(g \to\{c\bar{c}\}_{8}\bigr) 
\over d^2\bb dz d^2\bDelta d^2\bp }
 = 
 {1 \over 2(2 \pi)^2} \int d^2 \bkappa_1 
 d^2 \bkappa_2 \nonumber\\ 
&\times&\delta^{(2)}(\bDelta- \bkappa_1 - \bkappa_2)
\nonumber\\ 
&\times&
\Big\{ |\Psi_{c\bar{c}}(z,\bp-\bkappa_1) - \Psi_{c\bar{c}}(z, \bp - z(\bkappa_1 + \bkappa_2))|^2
\nonumber\\
& +&
|\Psi_{c\bar{c}}(z,\bp-\bkappa_2) - \Psi_{c\bar{c}}(z, \bp - z (\bkappa_1 + \bkappa_2))|^2
\Big\} 
\nonumber\\ 
&\times&\Phi(\bb, x,\bkappa_1) \Phi(\bb, x,\bkappa_2).
\label{eq:10.6}
\eea

\begin{figure}[!h]
\begin{center}
\includegraphics[width = 5.5cm,angle=270]{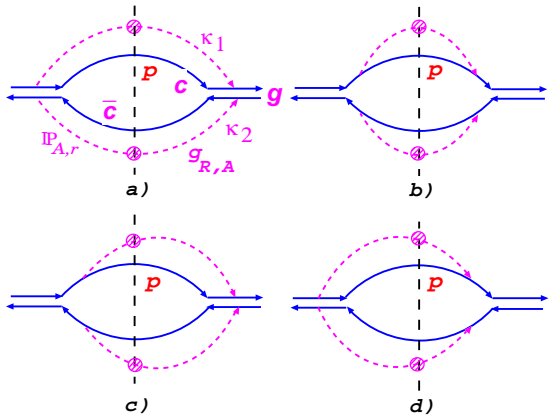}
\caption{The large-$N_c$ structure of cut nuclear pomeron diagram for 
production of  open charm. The incident gluon is 
shown in the composite quark-antiquark representation.
}
\label{fig:CCopen}
\end{center}
\end{figure}

The RFT diagrams for (\ref{eq:10.6}) are shown in Fig. \ref{fig:CCopen}.
There is a very close similarity to the case of color-octet digluons,
Fig. \ref{fig:GGoctet},
the absence of the color singlet quark-antiquark dipole explains the
absence of coherent distortions of the dipole wave function. The 
charm loop contribution to the $g\CutPom_{A,r}\CutPom_{A,r}g$
vertex is readily obtained from Eq. (\ref{eq:9.B.1.4}) by
the substitution $\Psi_{gg}(1;z,\bp)\to \Psi_{c\bar{c}}(z,\bp)$, we skip citing it here.

The ${\cal O}(N_c^{-2})$ excitation of open charm in the 
color-singlet representation is entirely similar to the
excitation of color-octet dijets in DIS. We leave the
derivation of these topological cross sections as an exercise.


\section{Topological cross sections for mid-rapidity gluons and
the $t$-channel multipomeron transitions}

We briefly comment here on 
topological cross sections with mid-rapidity gluons
in quark-nucleus and gluon-nucleus collisions which are 
of special interest from the RFT
perspective, more detailed discussion will be 
reported elsewhere. We consider $z\ll 1$, but $\eta \gsim \eta_A$.
The basic RFT diagrams for the relevant unitarity cut are the same
as shown before. Such a gluon is separated from the projectile 
parton by a large rapidity gap, and it is 
tempting to interpret the emerging spectra in terms 
of the unitarity cuts of diagrams with the $t$-channel
transition of the
in-vacuum cut pomeron $\CutPom_r$ on the beam side to
multipomeron states on the nucleus side, as indicated in
Fig. \ref{fig:MidRapidityGlue}c. Here we sketch briefly 
the nonlinear $k_\perp$ factorization results for the
emerging multipomeron vertices.


\subsection{Single mid-rapidity gluon spectra: linear $k_\perp$ factorization}

Here we start with the seemingly 
simple RFT properties of single mid-rapidity gluon 
spectra. Then, in the the spirit of Sec. VI, we recall the pitfalls
of the unitarity cut interpretation of the single-gluon cross
sections. Hereafter $\bp \equiv \bp_g$ and $z\equiv z_g$, we use the
results from Ref. \cite{SingleJet} and follow the discussion
in ref. \cite{VirtualReal}.

We need the small-$z$ limit  
\bea
&&|\Psi_{ga}(z,\bp)-\Psi_{ga}(z,\bp-\bkappa)|^2 =
\nonumber \\
&=&
2\alpha_S P_{ga}(z)\left|{\bp \over \bp^2+\mu_g^2}-
{\bp -\bkappa \over (\bp-\bkappa)^2+\mu_g^2}\right|^2
\nonumber\\
 &=&
 2\alpha_S P_{ga}(z)K(\bp,\bp-\bkappa)\nonumber\\
 &=&
{4\alpha_S C_a \over z}K(\bp,\bp-\bkappa),
\label{eq:11.A.1}
\eea
where $\mu_g$ is the (optional) infrared
regularization and we used the explicit form of the splitting
function $ P_{ga}(z)$ for soft gluons. 
For the transverse momenta above the infrared parameter 
$\mu_g$ we can use
\beq
K(\bp,\bp-\bkappa)={\bkappa^2 \over \bp^2(\bp-\bkappa)^2}.
\label{eq:11.A.2}
\eeq
We also recall that 
\beq
{dG_a (z,\bp-\bkappa) \over d^2\bp} = 
2\alpha_S zP_{ga}(z)\cdot{1\over (\bp-\bkappa)^2}
\label{eq:11.A.3}
\eeq
is the unintegrated glue in the 
incident parton $a$, at $z \ll 1$ it does not depend on
the virtuality of the incident parton $a$.
We define the nuclear counterpart of Eq. (\ref{eq:3.A.6}):
\bea
\bkappa^2\phi_{gg}(\bb,x_A,\bkappa)= {4\pi \alpha_s\over N_c}\cdot  
{dG_{A,gg}(\bb,x_A,\bkappa)\over d^2\bb d^2\bkappa},\nonumber\\
\label{eq:11.A.4}
\eea 

The spectator interaction cancellations are known to entail
the linear $k_\perp$-factorization form of the
quadrature for the single
mid-rapidity gluon spectrum \cite{SingleJet,KovchegovMueller}.
Making use of (\ref{eq:11.A.2})-(\ref{eq:11.A.4}), one can cast
it in the beam-target symmetric form:
\bea
&&{ (2\pi)^2 d\sigma_A\over d\eta_g d^2\bp
d^2\bb}\Biggr|_{a\to ag} =\nonumber\\
 &=& 
z\int d^2\bkappa \phi_{gg}(\bb,x_A,\bkappa)\nonumber\\
&\times&|\Psi_{ga}(z,\bp) -
\Psi_{ga}(z,\bp-\bkappa)|^2 \nonumber\\
&=& {4\pi \alpha_s\over N_c}\int d^2\bkappa d^2\bp_a V_{BFKL}(\bp;\bkappa,\bp_a) \nonumber\\
&\times&
{dG_a (z,\bp_a) \over  d^2\bp_a}\cdot {dG_{A,gg}(\bb,x_A,\bkappa)\over d^2\bb d^2\bkappa}.
\label{eq:11.A.5}
\eea
The remaining factor 
\bea
V_{BFKL}(\bp;\bkappa,\bp_a)={1\over \bp^2}\delta(\bp -\bp_a-\bkappa)
\label{eq:11.A.6}
\eea 
is familiar from the square of 
the BFKL vertex of radiation of the gluon by a
reggeized $t$-channel gluon, or a fusion $g_R(\bp_a)+g_R(\bkappa)\to g(\bp)$
\cite{BFKL}. The absence of any
nuclear renormalization of this vertex is noteworthy. 
Notice a correspondence
\bea
&&V_{BFKL}(\bp;\bkappa,\bp_a)= {1 \over  2\alpha_S P_{ga}(z)} \cdot
{(\bp-\bkappa)^2\over \bkappa^2}\nonumber\\
&\times&  |\Psi_{ga}(z,\bp)-\Psi_{ga}(z,\bp-\bkappa)|^2\delta(\bp -\bp_a-\bkappa),\nonumber\\
\label{eq:11.A.7}
\eea 
which can be used in all other cases for the identifications
(\ref{eq:11.A.3}) and (\ref{eq:11.A.4}).

\begin{figure}[!h]
\begin{center}
\includegraphics[width = 3.5cm,height=7cm,angle=270]{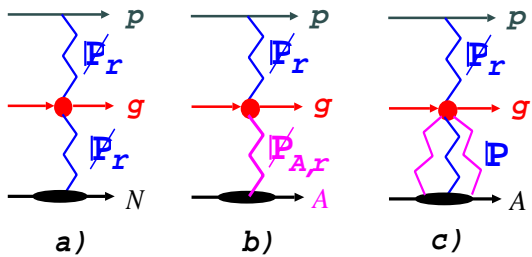}
\caption{The  Kancheli-Mueller optical theorem for the mid-rapidity
gluon spectrum: (a) for the free-nucleon target
in terms of cut in-vacuum pomerons, (b) for the nuclear target
with a cut nuclear pomeron on the nucleus side, (c) the
anticipated multiple in-vacuum pomeron exchange on the nucleus side.
}
\label{fig:MidRapidityGlue}
\end{center}
\end{figure}

When viewed from the Kancheli-Mueller optical theorem perspective, 
Eq. (\ref{eq:11.A.5}) invites the cut pomeron expansion for
 $\phi_{gg}(\bb,x_A,\bkappa)$ and the cut pomeron structure 
shown in Fig. \ref{fig:MidRapidityGlue}. One should not be
mislead, though: Sec. VI was a good warning of pitfalls of the unitarity cut
reinterpretation of $\phi_{gg}(\bb,x_A,\bkappa)$ in Eq. (\ref{eq:11.A.5}).
Such an operational expansion in the multipomeron exchanges
is possible\footnote{Such an expansion 
for the recoil multiplicity distribution
appeared, for instance, in Ref. \cite{Gelis}, which
was posted after this manuscript was finalized.},  
but it would lack a direct connection to color excitations 
of the nucleus.
Now we move to the cut-pomeron structure 
which follows from the nonlinear
$k_\perp$ factorization results reported in previous sections.

\subsection{Single mid-rapidity gluon spectra: cut pomeron structure}


\subsubsection{Spectator interactions and Cheshire Cat grin}

We consider first the universality class of dijets in the
same color representation as the incident quark.
For the incident quarks Eq. (\ref{eq:8.A.1.6}) gives
\bea
&&{d\sigma_\nu   \bigl(q \to\{ qg\}_3\bigr) 
\over d^2\bb d\eta_g  d^2\bp } = {1 \over (2 \pi)^2}  
 w_{\nu}\Big(\nu_A(\bb)\Big)
\nonumber\\
&\times& \int d^2\bkappa
{d\sigma_{Qel}^{(\nu)}(\bkappa)\over \sigma_{Qel}
d^2\bkappa}
z\left| \Psi_{qg}(1;z,\bp-\bkappa)- \Psi_{qg}(z,\bp)\right|^2.\nonumber\\
\label{eq:11.B.1.1} 
\eea 
Here we changed from the quark to gluon momentum and took
a limit $z\equiv z_g\to 0$, which simplifies $\Psi_{qg}(z,\bp-z\bkappa)
\to \Psi_{qg}(z,\bp)$.  The similar result for incident gluons is
(for the sake of a comparison with Eq. (\ref{eq:11.B.1.1}) we
make use of the large $N_c$--property $ \nu_{A,gg}(\bb)= 2\nu_{A}(\bb)$)
\bea
&&{d \sigma (g \to \{gg \}_{8_A + 8_S} ) 
\over d^2\bb d\eta_g d^2\bp  } = \nonumber\\
&=& {1 \over (2 \pi)^2}
\sum_{\nu=0} \sum_{j,k=0}   
\delta(\nu-j-k)\nonumber\\
&\times&
w_k\Big(\nu_A(\bb)\Big)w_j\Big(\nu_A(\bb)\Big)\nonumber\\
&\times&
\int d^2\bkappa_1 \int d^2 \bkappa_2 
{d\sigma_{g,Qel}^{(j)}(\bkappa_1)\over \sigma_{g,Qel} d^2\bkappa_1}\cdot
{d\sigma_{g,Qel}^{(k)}(\bkappa_2)\over \sigma_{g,Qel} d^2\bkappa_2}\nonumber\\
&\times&
z\left| \Psi_{gg}(1;z,\bp-\bkappa_1)- \Psi_{gg}(z,\bp)\right|^2.
\label{eq:11.B.1.2} 
\eea 

Upon the integration over $\bkappa_2$ the momentum structure of
(\ref{eq:11.B.1.2}) will be exactly the same as in (\ref{eq:11.B.1.1}),
the difference is in the CCG --- the contribution $k$
to the total multiplicity of cut pomerons. The difference between the
two cases is evident from a close comparison of Fig. \ref{fig:QGtripletnew}
and \ref{fig:GGoctet}. In Fig. \ref{fig:QGtripletnew}
the nuclear cut pomeron $\CutPom_{A,r}$ couples only to the 
quark of the gluon in the quark-antiquark representation, the spectator
quark interacts only with uncut pomerons and does not contribute
to the nucleus excitation. In contrast to that,
in Fig. \ref{fig:QGtripletnew} the spectator gluon too couples to 
$\CutPom_{A,r}$ and contributes exactly the above $k$ to the nucleus excitation.
Neither of the two topological cross sections  to the mid-rapidity gluon
spectrum can be cast in the linear $k_\perp$ factorizable form
suggested by Eq. (\ref{eq:11.A.5}). Even for the incident quark, 
when spectator interactions do not contribute to the nucleus excitation,
the uncut pomeron exchanges in  $\Psi_{qg}(1;z,\bp-\bkappa)$ make
Eq. (\ref{eq:11.B.1.1}) dissimilar to (\ref{eq:11.A.5}). As we learned 
in Sec. VI, in the general
case it is impossible to guess out
the correct unitarity structure brought into the problem by the
CCG of the integrated out spectator interactions. 

Now comes the universality class of dijets in higher color representations.
Here taking $z\to 0$ does not bring in any principal simplifications,
but working with the maximally convoluted form of the nuclear glue
does help. For the incident quarks we find
\bea
&&{d \sigma_\nu \bigl(q \to\{ qg\}|_{6+15}\bigr) 
\over d^2\bb d\eta_g d^2\bp }
 = {1 \over (2 \pi)^2} T(\bb)\int_0^1 d\beta  \nonumber\\
&\times& \sum_{j,k \geq 0}\delta(\nu-1-j-k) \nonumber\\
&\times& 
w_{k}\Big((1-\beta)\nu_A(\bb)\Big)\nonumber\\
&\times&
 w_j \Big(\Big[{C_A\over C_F}(1 -\beta)+\beta\Big]\nu_A(\bb)\Big)\nonumber\\
&\times&
\int  d^2\bkappa d^2\bkappa_1 d^2\bkappa_2 
{d\sigma_{Qel}(\bkappa)
\over d^2\bkappa} \nonumber\\
&\times&
 {d\sigma_{Qel}^{(k)}(\bkappa_2)
\over \sigma_{Qel}d^2\bkappa_2}\cdot
  {d\sigma_{Qel}^{(j)}(\bkappa_1)
\over \sigma_{Qel}d^2\bkappa_1}.
\nonumber\\
&\times&
 \left|\Psi_{qg}(\beta;z,\bp-\bkappa_1)-
\Psi_{qg}(\beta;z,\bp-\bkappa_1-\bkappa)\right|^2.\nonumber\\
\label{eq:11.B.1.3} 
\eea
The dependence on the momentum $\bkappa_2$ of the spectator quark 
integrates out entirely, see
Eq. (\ref{eq:6.B.2}), but CCG --- the
contribution $k$ to the multiplicity $\nu$ of excited nucleons ---
stays on, which is made obvious by Fig. \ref{fig:QGsextet2}.

In conformance to the universality class concept, the story repeats 
itself for gluons. Working again with the maximally convoluted form
of the collective nuclear glue, we have 
\bea
&&{d \sigma_\nu \bigl(g \to\{ gg\}|_{10+\overline{10}+27+R_7}\bigr) 
\over d^2\bb d\eta_g d^2\bp } =\nonumber\\
&=& {1 \over (2 \pi)^2} T(\bb)\int_0^1 d\beta  \nonumber\\
&\times& \sum_{j,k\geq 0}\delta(\nu-1-j-k) 
\nonumber\\
&\times&
w_{k}\Big(\Big[{C_{27}\over C_A}(1 -\beta)+\beta\Big]\nu_A(\bb)\Big) \nonumber\\
&\times&
w_j \Big(\Big[{C_{27}\over C_A}(1 -\beta)+\beta\Big]\nu_A(\bb)\Big)
\nonumber\\
&\times&\int  d^2\bkappa d^2\bkappa_1 d^2\bkappa_2 
{d\sigma_{Qel}(\bkappa)
\over d^2\bkappa}\nonumber\\
&\times& {d\sigma_{Qel}^{(k)}(\bkappa_2)
\over \sigma_{Qel}d^2\bkappa_2}\cdot
  {d\sigma_{Qel}^{(j)}(\bkappa_1)
\over \sigma_{Qel}d^2\bkappa_1}.
\nonumber\\
&\times&
\left|\Psi_{gg}(\beta;z,\bp-\bkappa_1)-
\Psi_{gg}(\beta;z,\bp-\bkappa_1-\bkappa)\right|^2.\nonumber\\
\label{eq:11.B.1.4} 
\eea
Again the dependence on $\bkappa_2$ integrates out.
The change of the projectile parton from
the quark to a gluon changes substantially 
the $\beta$ dependence of the CCG --- $w_{k}(\beta^*\nu_A(\bb))$
for incident gluons vs. $w_{k}((1-\beta)\nu_A(\bb))$ for incident 
quarks. We
reiterate the same comment that neither (\ref{eq:11.B.1.3})
nor (\ref{eq:11.B.1.4}) can be made reminiscent of
Eq. (\ref{eq:11.A.5}).


\subsection{What are the multipomeron vertices?}

The point that spectator partons contribute to the
excitation of the nucleus can not be contested --- 
within QCD as a gauge theory one
can not delimitate interactions of colored spectator
partons. Then it is not surprising that this CCG depends
on the process. It is not surprising that the
pattern of coherent distortions by uncut
pomeron exchanges depends on the color structure of 
final states. In conjunction with the discussion
in Sec. VI, this casts a shadow on the possibility
of the formulation of a Kancheli-Mueller optical 
theorem, Fig. \ref{fig:MidRapidityGlue}c, 
in terms of universal multipomeron 
couplings.

Still sort of a universality is recovered at the expense of
losing a direct contact with event-by-event observables
and going to the multiplicity re-summed cross sections.
Following the discussion in Sec. VII.C, we define
the multiplicity re-summed topological cross section of
processes with at least $j$ cut pomerons,
\beq
d\sigma^{(j)} = \sum_{\nu > j} d\sigma_\nu.
\label{eq:11.C.1}
\eeq
Here $j$ is the mid-rapidity gluon contribution to
the nucleus excitation, unlike $\nu$ it can not be determined 
experimentally on an event-by-event basis. Following
Sec. VII.C one would argue, that such a re-summation 
eliminates CCG not only from the spectator quark of
the hard dijet, but also from the 
comoving partons of the parent hadron the incident 
parton, $q$ or $g$, belongs to. 
As we shall see in a minute, such a multiplicity
re-summed
$d\sigma^{(j)}$ has very interesting properties and we
urge the multiplicity re-summation analysis of the experimental data.

First, we notice that at $z\ll 1$
\bea
&&\left| \Psi_{gg}(1;z,\bp-\bkappa_1)- \Psi_{gg}(z,\bp)\right|^2=
\nonumber\\
&=&
{C_A\over  C_F}\left| \Psi_{qg}(1;z,\bp-\bkappa_1)- \Psi_{qg}(z,\bp)\right|^2.\nonumber\\
\label{eq:11.C.2}
\eea
Then, apart from this ratio of Casimirs, upon such a 
multiplicity re-summation the 
integrand of Eq. (\ref{eq:11.B.1.2}) 
would become identical to that of (\ref{eq:11.B.1.1}) and
\beq
d\sigma^{(j)}(g\to \{gg\}_8)= {C_A\over  C_F}d\sigma^{(j)}(q\to \{qg\}_3)
\label{eq:11.C.3}
\eeq
Here $C_A/C_F$ is a ratio of couplings of the in-vacuum pomeron to
the beam gluon and quark --- a part of the generic Regge
factorization. Then, according to Fig. \ref{fig:QGtripletnew},
and (\ref{eq:11.B.1.1}), the Kancheli-Mueller diagram for the 
universality class of dijets in the color representation of the incident
parton will contain the following set of pomeron transitions
in the $t$-channel:
\begin{widetext}
\bea
&&\Psi^*_{ga}(z,\bp)\Big\{ \CutPom_r \to
\CutPom_{A,r}\Big\}\Psi_{ga}(z,\bp)+\nonumber\\
&+&
\Psi^*_{ga}(z,\bp-\bkappa_1)\Big(1-\otimes \Pom_A\Big)\Big\{ \CutPom_r \to
\CutPom_{A,r}\Big\}\Big(1-\Pom_A\otimes \Big)\Psi_{ga}(z,\bp-\bkappa_1)\nonumber\\
&-&\Psi^*_{ga}(z,\bp-\bkappa_1)\Big(1-\otimes \Pom_A\Big)\Big\{ \CutPom_r \to
\CutPom_{A,r}\Big\}\Psi_{ga}(z,\bp)\nonumber\\
&-&
\Psi^*_{ga}(z,\bp)\Big\{ \CutPom_r \to
\CutPom_{A,r}\Big\}\Big(1-\Pom_A\otimes \Big)\Psi_{ga}(z,\bp-\bkappa_1)
\label{eq:11.C.4}
\eea 
\end{widetext}
Here the position of $\otimes$ indicates the action of the distortion
operator on either $\Psi_{ga}(z,\bp)$ or $\Psi^*_{ga}(z,\bp)$.
The representation of these vertices in terms of the in-vacuum and
coherent nuclear gluons can easily be read from the above cited 
$\CutD$'s, we leave this as an exercise. We only notice, that in
the expansion of $\CutPom_{A,r}$ in  $j\CutPom_r$ the term with
$j=1$ is 
the absorbed impulse approximation, see Eq. (\ref{eq:8.A.2.2}).
Next comes the triple-pomeron 
transition $\CutPom_r \to \CutPom_r\CutPom_r$. All such transitions 
$\CutPom_r \to j\CutPom_r$ 
are furnished by uncut pomeron exchanges in the absorption factor
and in the wave function distortions as indicated in (\ref{eq:11.C.4}).

The expansion of (\ref{eq:11.C.4}) in $\CutPom_r$'s  contains also the term with
$j=0$, i.e., the diffractive cut. This diffractive cut defines 
the $t$-cannel transition $\CutPom_r \to \Pom_A \Pom_A$, and
the corresponding transitions $\CutPom_r \to (k\Pom) (m\Pom)$
with $k$ and $m$ uncut in-vacuum pomerons on the two sides
of the unitarity cut, respectively.

The similar re-summation of topological cross sections
in the universality class of higher color representations
leads to the counterpart of the Regge factorization (\ref{eq:11.C.3}):
\bea
&&d\sigma^{(j)}(g\to \{gg\}_{(10+\overline{10}+27+R_7)})
\nonumber\\
&=& {C_A\over  C_F}d\sigma^{(j)}(q\to \{qg\}_{6+15}).
\label{eq:11.C.5}
\eea
Here the crucial point is the large-$N_c$ equality
\beq
{C_A\over C_F}={C_{27}\over C_A},
\label{eq:11.C.6}
\eeq                                                               
by which $\beta^*$ as a function of $\beta$ 
is the same for both channels. All the wave functions 
are coherently distorted in the same slice of the nucleus and, 
suppressing the
wave functions, we have the following structure of
$t$-channel transitions:
\bea
&& \CutPom_r \to \nonumber\\
&\to& \Big(1-\otimes \Pom_A(\beta)\Big)
\CutPom_{A,r}(\beta^*) \CutPom_e (1-\Pom_A(\beta)\otimes ).
\nonumber\\
\label{eq:11.C.7}
\eea 
The distinction between the two major universality classes
persists even after the multiplicity re-summations.
 
The absorbed impulse approximation starts with the
transition $\CutPom_r \to \CutPom_e$. Next comes the
triple pomeron transition $\CutPom_r \to\CutPom_r \CutPom_e$
which is evidently different from 
$\CutPom_r \to\CutPom_r \CutPom_r$ in the previous 
universality class, and from the diffractive 
transition cut $\CutPom_r \to\Pom \Pom$. One can readily
verify that from the afore cited $\CutD$'s. 
Such a variety
of triple-pomeron vertices is a new observation not
found in the previous literature. 
One can readily read from the same results for  $\CutD$'s
the
triple-pomeron vertices $\CutPom_r \to \CutPom_r \Pom$
and $\CutPom_r \to \CutPom_e \Pom$.
The experimentally observed $d\sigma^{(j)}$ will be a sum 
of topological cross sections for the two universality classes.

The pseudo-diffractive channels, when dijets in the
color representation of the beam parton are produced 
via the intermediate state in higher multiplets,
defines the triple-pomeron vertex $\CutPom_r \to\CutPom_e \CutPom_e$.
Such a vertex is ${\cal O} (N_c^{-2})$, though.

Our results for $\CutD$ in the open-charm
production can readily be reformulated in terms of the
quark loop contribution to the triple pomeron vertex
$\CutPom_r \to \CutPom_{A,r} \CutPom_{A,r}$ and its
derivatives in terms of multiple in-vacuum $\CutPom_r$'s.
Finally, we notice that the multipomeron vertices 
$\CutPom_r \to j\CutPom_{r}+ k\CutPom_{e}+ m\Pom$ 
{\it{must be regarded as local ones in the rapidity, 
rather than a branching tree (fan) of triple-pomeron vertices.}}
A detailed discussion and comparison of all those 
vertices, including the one which appeared in the
nonlinear evolution for the collective nuclear glue
\cite{VirtualReal}, will be reported elsewhere.


\section{Conclusions}

Starting from the first principles of pQCD, 
we derived nonlinear $k_\perp$-factorization 
quadratures for hard 
scattering 
off nuclei with a fixed multiplicity of color-excited nucleons.
Within pQCD each color-excited nucleon is  
associated with a cut pomeron, and our results
must be regarded as a pQCD version of the AGK unitarity rules. 
On the technical side, our
$\textsf{S}$-matrix formalism allows a consistent separation of
color-diagonal and color-excitation  
rescatterings in the course of the non-Abelian intranuclear evolution
of color dipoles. An indispensable feature of the coupled-channel
non-Abelian evolution of dipoles are the two distinct classes of cut
pomerons  --- excitation $\CutPom_e$ and
color-rotation $\CutPom_r$.  This distinction of $\CutPom_r$ and
$\CutPom_e$ permeates the topological cross sections for all 
projectiles and is a new finding not contained in the early
literature on the AGK rules.

We reported the reggeon field theory interpretation
of topological cross sections for different reaction
universality 
classes. Here the concept of the collective nuclear glue as a coherent state
of the in-vacuum (reggeized) gluons proves to be extremely useful.
 After the r\^ole of the uncut, $\Pom$, and
the two cut, $\CutPom_r$ and $\CutPom_e$, pomerons 
has been identified, one would readily deduce
the topological cross sections from the known nonlinear 
$k_\perp$ factorization results for the inclusive dijet spectra
\cite{Nonlinear,PionDijet,SingleJet,Nonuniversality,Paradigm,QuarkGluonDijet,GluonGluonDijet}. 
This is a unique example when going from the total to 
topological cross sections does work. We demonstrate that,
in contrast to the common wisdom,
in the general case it is not possible to guess the 
topological cross sections from the Glauber formulas for
the total cross section. This is the case, for instance,
with the topological cross sections in DIS off nuclei
for which our results differ from the earlier discussed
Glauber model guesses. 

In DIS 
the topological cross sections are tagged by the multiplicity
of hadrons in the backward (nucleus) hemisphere and all our
results can be subjected to the direct experimental test.
On the phenomenological side, there is a long 
shopping list of
possible numerical predictions for 
long-range rapidity correlations between leading 
quark-antiquark dijets 
in the photon hemisphere and
multiproduction in the backward hemisphere. One notable
example is a nonperturbative contribution to the 
quenching of leading jets.
In the single jet problem, both in  DIS and  hadron induced
processes, the effects of spectator, and comover, interactions can be
eliminated by the multiplicity re-summation technique
of Sec. VII.C, which might become an important tool
in the analysis of the experimental data. 

In this communication we confined
ourselves to tree diagrams of the RFT, an extension to pomeron loops 
requires  a further scrutiny. Even at the tree level
we found very interesting properties of the topological
cross sections.  We focused on topological cross sections for 
lowest order pQCD hard final states. Based on our experience
with small-$x$ radiative corrections \cite{VirtualReal},
we maintain that principal features of the derived 
topological cross
sections must  be stable against small-$x$ evolution. 
In particular, our finding that the multipomeron vertices are local in
rapidity, rather than a branching tree (fan) of triple--Pomeron
vertices must persist to higher orders of the LL$(1/x)$--evolution. 
The numerical applications of the reported formalism 
will be reported elsewhere.

\section*{Acknowledgments}

We are grateful to B.G. Zakharov for helpful discussions.


\section{Appendix A: Color-diagonal and color-excitation interactions
in DIS and open charm production}
\setcounter{equation}{0}
\renewcommand{\theequation}{A.\arabic{equation}} 


\subsection{Color-singlet projectile: DIS}

The assignment of partons is $a=\gamma^*$, $b=q$ and $c=\bar{q}$.
The results for the multiparton cross section operators 
for all relevant partonic pQCD subprocesses are found in 
\cite{NZ94,Nonlinear,QuarkGluonDijet,GluonGluonDijet,VirtualReal},
here we only need to describe $\hat{\Sigma}_{el}(\bC)$.

First we notice, that Eq. (\ref{eq:3.D.2})
entails a property
\beq
\hat{\Sigma}_{el}(\bC)= \hat{\Sigma}_{R,el}(\bB)+\hat{\Sigma}_{R',el}(\bB').
\label{eq:AppA.1.1}
\eeq
The simplest case is of a single parton $b$,
\beq
\sigma_{b,el}={C_b \over 2C_F}\sigma_0(x),
\label{eq:AppA.1.2}
\eeq
while for the two-parton system $bc$ in the color 
representation $R$ one has
\bea
\sigma_{R,el}(x) &=& {C_b + C_c - C_R \over 2C_F}\sigma(x,\br)+ 
{C_R \over 2C_F}\sigma_0(x).\nonumber\\
\label{eq:AppA.1.3}
\eea
In DIS the incoming parton is an Abelian photon which does not couple 
to a gluon.  For the 2-body, $\gamma^{*'}\gamma^{*}$, and 3-body, 
$\gamma^{*'}q\bar{q}$ and  $\gamma^{*}q'\bar{q}'$, states this entails
\bea
{\cal S}_{A,\nu}^{(2)}(\bb',\bb)&=&  \delta_{\nu 0} \openone
,\nonumber\\
{\cal S}_{A,\nu}^{(3)}(\bB',\bb)&=&  \delta_{\nu 0} 
 \textsf{S}[\bb, \sigma(x,\br')],\nonumber\\
{\cal S}_{A,\nu}^{(3)}(\bb',\bB)&=&  \delta_{\nu 0}
 \textsf{S}[\bb, \sigma(x,\br)].
\label{eq:AppA.1.4}
\eea

The non-Abelian  evolution of color-singlet 4-parton systems  
$q\bar{q}q'\bar{q}'$ is a coupled two-channel problem in the basis 
of states $\ket{e_1}=\ket{1\bar{1}}$ and $\ket{e_2}=\ket{8\bar{8}}$. 
The summation over the final state dijets is of the form 
\bea
&&\sum_f\bra{f}=\sum_{R}\sqrt{\textsf{dim[R]}}\bra{R\bar{R}}=\nonumber\\
&=&\bra{e_1}+ \sqrt{N_c^2-1}\bra{e_2}.
\label{eq:AppA.1.5}
\eea
The corresponding $\hat{\Sigma}^{(4)}(\bC)$ for DIS was derived in
\cite{Nonlinear}. Making use of Eqs. (\ref{eq:AppA.1.1})-(\ref{eq:AppA.1.3}),
we decompose it into the color-diagonal and color-excitation
 components: 
\bea
\hat{\Sigma}^{(4)}_{el}(\bC)&=& \Sigma_{11}^{el}\ket{e_1}\bra{e_1}+\Sigma_{22}^{el}
\ket{e_2}\bra{e_2},\nonumber\\
\Sigma_{11}^{el} & =& \sigma(x,\br)+
\sigma(x,\br'),\nonumber\\
\Sigma^{el}_{22} & =& {C_A \over C_F} \sigma_0(x)\nonumber\\
&-& {1 \over N_c^2-1}[\sigma(x,\br)+
\sigma(x,\br')],
\label{eq:AppA.1.6}
\eea
and
\bea
\hat{\Sigma}^{(4)}_{ex}(\bC) &=& \Sigma_{12}^{ex}
[\ket{e_1}\bra{e_2} +\ket{e_2}\bra{e_1}]+\Sigma_{22}^{ex}
\ket{e_2}\bra{e_2} \, ,
\nonumber\\
&=& \Sigma_{12}^{ex} \, P_{ex} + \Sigma_{22}^{ex} \, P_2 
\nonumber \\
 \Sigma_{22}^{ex}     &=& {C_A \over 2C_F} [ \sigma(x,\bs)+\sigma(x,\bs-\br+\br') -2\sigma_0]
\nonumber\\
&+& {2 \over N_c^2-1}\Omega,
\nonumber\\
\Sigma_{21}^{ex} & = & 
-{1 \over \sqrt{N_c^2-1}}\Omega.
\label{eq:AppA.1.7}
\eea
The operator
\beq
\Omega=
\sigma(x,\bs-\br)+\sigma(x,\bs+\br')-\sigma(x,\bs)-\sigma(x,\bs-\br+\br')
\label{eq:AppA.1.8}
\eeq
will be encountered in all the transitions of color
dipoles between color representations of different dimension
\cite{QuarkGluonDijet,GluonGluonDijet}. There is a clear cut $1/N_c$ hierarchy: 
singlet-to-octet transitions are $N_c$ suppressed, while the octet-to-octet
rotations are not.  In the $k_\perp$ factorization form 
\bea
\Omega& =& \int d^2\bkappa f(x,\bkappa)\nonumber\\
&\times&
[1-\exp(-i\bkappa\br)][1-\exp(i\bkappa\br')]\exp(i\bkappa\bs)\nonumber\\
&=&\sigma_0(x) \int d^2\bkappa {1\over  \sigma_{Qel}}
{d\sigma_{Qel}(\bkappa) \over d^2\bkappa}\nonumber\\
&\times&
[1-\exp(-i\bkappa\br)][1-\exp(i\bkappa\br')]\exp(i\bkappa\bs).\nonumber\\
\label{eq:AppA.1.9} 
\eea  
At large $N_c$, the color rotations within the color-octet states are described by 
\bea
&-&\Sigma_{22}^{ex}= 2\sigma_0(x) - \sigma(x,\bs)-\sigma(x,\bs-\br + \br')
\nonumber \\
& =&
\int d^2\bkappa f(x,\bkappa) \Big\{\exp[i\bkappa\bs] + 
\exp[i\bkappa(\bs-\br + \br')]\Big\}.\nonumber \\
\label{eq:AppA.1.10}
\eea
The identification (\ref{eq:4.C.3}) leads to
\bea
&&\left(-\Sigma_{22}^{ex}\right)^\nu = \sigma_0^\nu(x)\sum_{k=0}^{\nu}
{\nu! \over k! (\nu-k)!} \nonumber\\
&\times&
\int  d^2\bkappa_1 d^2\bkappa_2 \exp[i\bkappa_1\bs +i\bkappa_2(\bs-\br +
\br')]\nonumber\\
&\times&
{d\sigma_{Qel}^{(k)}(\bkappa_1)
\over \sigma_{Qel} d^2\bkappa_1}\cdot {d\sigma_{Qel}^{(\nu-k)}(\bkappa_2)
\over \sigma_{Qel} d^2\bkappa_2}.\nonumber\\
\label{eq:AppA.1.11}
\eea


\subsection{Color-octet projectile: open charm production}

The assignment of partons is $a=g$, $b=q$ and $c=\bar{q}$. 
The results hold for any flavor, for the
sake of definiteness we speak of the open charm excitation $g\to c\bar{c}$,
the early discussion is found in \cite{SingleJet,Nonuniversality,Paradigm}.
Here $\hat{\Sigma}^{(4)}(\bC)$ is exactly the same as in DIS.
Here the incident parton has a net color, which 
changes the  2-body, $gg$, and 3-body, $gc'\bar{c}'$ and
 $g'c\bar{c}$, interactions which are  all single-channel problems. 
We only list the results which did not
appear before:
\begin{widetext}
\bea
\Sigma^{(2)}(\bb,\bb')&=&{C_A\over C_F}\sigma(x,\bs-(1-z)\br+(1-z)\br')\nonumber\\
\Sigma_{el}^{(2)}(\bb,\bb') &=& {C_A\over C_F}\sigma_0(x)\nonumber\\
\Sigma_{ex}^{(2)}(\bb,\bb')&=&{C_A\over
 C_F}[\sigma(x,\bs-(1-z)\br+(1-z)\br')-\sigma_0(x)]
\nonumber\\
\Sigma^{(3)}(\bB,\bb') &=&{C_A\over 2 C_F}\bigl[\sigma(x,\bs+(1-z)\br')+
\sigma(x,\bs-\br+(1-z)\br')
-\sigma(x,\br)\bigr]+\sigma(x,\br), \nonumber\\
\Sigma_{el}^{(3)}(\bB,\bb') &=&  {C_A\over C_F}\sigma_0(x)+{2C_F-C_A\over
 2C_F}\sigma(\br)\nonumber\\
\Sigma_{ex}^{(3)}(\bB,\bb') &=&{C_A\over 2 C_F}\bigl[\sigma(x,\bs+(1-z)\br')+
\sigma(x,\bs-\br+(1-z)\br')-2\sigma_0(x)]\nonumber\\
\Sigma^{(3)}(\bb,\bB') &=&{C_A\over 2 C_F}\bigl[\sigma(x,\bs-(1-z)\br)+
\sigma(x,\bs+\br'-(1-z)\br)
-\sigma(x,\br')\bigr]+\sigma(x,\br'), \nonumber\\
\Sigma_{el}^{(3)}(\bb,\bB') &=&  {C_A\over C_F}\sigma_0(x)+{2C_F-C_A\over
 2C_F}\sigma(\br')\nonumber\\
\Sigma_{ex}^{(3)}(\bb,\bB')& =& {C_A\over 2 C_F}\bigl[\sigma(x,\bs-(1-z)\br)+
\sigma(x,\bs+\br'-(1-z)\br)-2\sigma_0(x)].
\label{eq:AppA.2.1} 
\eea 
\end{widetext}
The corresponding ${\cal S}_{A,\nu}^{(4)}$ are given by 
Eqs.~(\ref{eq:AppA.1.6})-(\ref{eq:AppA.1.8}).


\section{Appendix B: Color-diagonal and color-excitation interactions
for excitation of the quark-gluon dijets }
\setcounter{equation}{0}
\renewcommand{\theequation}{B.\arabic{equation}}


\subsection{2-parton $(q\bar{q})$ and 3-parton $(qg\bar{q})$ states}

The assignment of partons is  $a=q$, $b=q$ and $c=g$.
The overall color-singlet quark-antiquark and quark-gluon-antiquark 
sates have a unique color structure and their intranuclear 
propagation is a single-channel problem. In contrast
to the photon in DIS, the incident quark has a net color charge and
both pure elastic and color-excitation interactions are possible in
all three channels.

In the color-singlet system $\bar{q}qg$ the $qg$ dipole is in
the color-triplet state.
We simply borrow the results for the 2-parton and 3-parton cross
sections from Ref. \cite{QuarkGluonDijet}
\begin{widetext}
\bea
\Sigma^{(2)}(\bb,\bb') &=& \sigma(x,\bs +(1-z)\br-(1-z)\br'), \nonumber\\
\Sigma^{(3)}(\bB,\bb') &=&{C_A\over 2 C_F}\bigl[\sigma(x,\br)+
\sigma(x,\bs+\br-(1-z)\br')
-\sigma(x,\bs-(1-z)\br')\bigr]+\sigma(x,\bs-(1-z)\br'), \nonumber\\
\Sigma^{(3)}(\bb,\bB')&=&{C_A\over 2 C_F}\bigl[\sigma(x,-\br')+
\sigma(x,\bs-\br'+(1-z)\br)
-\sigma(x,\bs+(1-z)\br)\bigr]+\sigma(x,\bs+(1-z)\br).
\label{eq:AppB.1.1} 
\nonumber \\
\eea 
The elastic cross sections are given by Eqs. (\ref{eq:AppA.1.1})-(\ref{eq:AppA.1.3}),
\bea
\Sigma_{el}^{(2)}(\bb,\bb')&=& \sigma_0(x),\nonumber\\
\Sigma_{el}^{(3)}(\bB,\bb')&=& \sigma_0(x) + {C_A\over 2C_F} \sigma(x,\br),\nonumber\\
\Sigma_{el}^{(3)}(\bb,\bB')&=& \sigma_0(x) + {C_A\over 2C_F} \sigma(x,-\br'),
\label{eq:AppB.1.2} 
\eea 
and the excitation cross sections equal 
\bea
\Sigma_{ex}^{(2)}(\bb,\bb')&=& \sigma(x,\bs + (1-z) \br-(1-z)\br')-\sigma_0(x),
\nonumber\\
\Sigma_{ex}^{(3)}(\bB,\bb')&=& \sigma(x,\bs+\br-(1-z)\br')- \sigma_0(x)
+\Big({C_A\over 2
  C_F}-1\Big)[\sigma(x,\bs+\br-(1-z)\br')-\sigma(x,\bs-(1-z)\br')]\nonumber\\
\Sigma_{ex}^{(3)}(\bB',\bb)&=& \sigma(x,\bs-\br'+(1-z)\br)- \sigma_0(x)
+ \Big({C_A\over 2C_F}-1\Big)[\sigma(x,\bs-\br'+(1-z)\br)-\sigma(x,\bs+(1-z)\br')].\nonumber\\
\label{eq:AppB.1.3} 
\eea
The results for the leading order in $1/N_c$ expansion 
are obtained putting $C_A=2C_F$.
\end{widetext}

\subsubsection{4-parton $(q\bar{q}gg')$ states}

We use the triplet, sextet and 15-plet $qg$ states 
and four-body states $\ket{E_1}=\ket{3\overline{3}},~
\ket{E_2}=\ket{6\overline{6}},~
\ket{E_3}=\ket{15\,\overline{15}}$.
The diagonal matrix $\hat{\Sigma}_{R,el}$ is defined by Eq.  (\ref{eq:AppA.1.1}),
for $\Sigma_{3,el}(\br)$ see Eq.~(\ref{eq:AppB.1.2}),
for higher representations 
\bea
\Sigma_{6,el}(\br)&=&{1\over 2} \left[{C_6\over C_F}\sigma_0(x) + {N_c\over
    N_c^2-1}\sigma(x,\br)\right], \nonumber\\
C_6&=& {3N_c+1\over N_c+1}C_F,  \nonumber\\
\Sigma_{15,el}(\br)&=&{1\over 2} \left[{C_{15}\over C_F}\sigma_0(x) - {N_c\over
    N_c^2-1} \sigma(x,-\br)\right],
\nonumber\\C_{15}&=& {3N_c-1\over N_c-1}C_F.
\label{eq:AppB.2.1}
\eea
In the large-$N_c$ approximation $C_6=C_{15} = C_A+C_F$ and we find
\bea
\Sigma_{11}^{el}(\br,\br')&=&\sigma_0(x) + \sigma(x,\br)+ \sigma(x,-\br')\,,\nonumber\\
\Sigma_{22}^{el}(\br,\br')&=&{C_6\over C_F}\sigma_0(x) = 
\left[1+ {C_A\over C_F}\right]\sigma_0(x) 
\nonumber \\ 
&=&3\sigma_0(x)\,,\nonumber\\
\Sigma_{33}^{el}(\br,\br')&=&{C_{15}\over C_F}\sigma_0(x) = 
\left[1+ {C_A\over C_F}\right]\sigma_0(x) \nonumber\\
&=& 3\sigma_0(x)=\Sigma_{22}^{el}(\br,\br')\,.
\label{eq:AppB.2.2}
\eea
A transformation to the basis 
\bea
\ket{e_1}&=&\ket{E_1},\nonumber\\
\ket{e_2}&=& 
{1\over \sqrt{2}} (\ket{E_2}+\ket{E_3}),\nonumber\\
\ket{e_3}&=& 
{1\over \sqrt{2}} (\ket{E_2}-\ket{E_3}),
\label{eq:AppB.2.3}
\eea
leads to the following 4-body operator $\hat{\Sigma}^{(4)}$
\cite{QuarkGluonDijet}:
\bea
\hat{\Sigma}_{el}^{(4)}&=&
{\textsf{diag}}(\Sigma^{el}_{11},\Sigma^{el}_{22},\Sigma^{el}_{22})
\nonumber\\
\hat{\Sigma}_{ex}^{(4)}& =& {\textsf{diag}}\{\Sigma_{11}^{ex},\Sigma_{22}^{ex},\Sigma_{33}^{ex}\}
\nonumber\\ &-& {1\over N_c}\Omega [\ket{e_1}\bra{e_2}+\ket{e_2}\bra{e_1}] .
\label{eq:AppB.2.4}
\eea
Here $\Omega$ is the familiar transition operator of Eq.~(\ref{eq:AppA.1.8}),
and the diagonal elements equal
\bea
\Sigma_{11}^{ex}  & = & \sigma(x,\bs +\br -\br') -\sigma_0(x)\nonumber\\
& =& -\sigma_0(x) \int
d^2\bkappa  f(x,\bkappa)\exp[i\bkappa(\bs +\br -\br')]\, ,\nonumber\\
\Sigma_{22}^{ex} & = &  {C_A\over C_F}\sigma(x,\bs + \br
-\br')+\sigma(x,\bs)\nonumber\\
&-&
 \left[ {C_A\over C_F}+1\right]
 \sigma_0(x) =\nonumber\\
& = & -\sigma_0(x)\Big\{\int
d^2\bkappa f(x,\bkappa) \exp(i\bkappa\bs)\nonumber\\
&+&{C_A\over C_F}\int
d^2\bkappa f(x,\bkappa) \exp[i\bkappa(\bs +\br -\br')]\Big\}
 ,\nonumber\\
\Sigma_{33}^{ex} &=& \lambda_{2}+\Omega.
\label{eq:AppB.2.5}
\eea
The large $N_c$ hierarchy in $\hat{\Sigma}_{ex}^{(4)}$ is the same as in DIS: 
the diagonal elements $\Sigma_{ii}^{ex}$ which describe
color rotations of dipoles within the same color
representation, are ${\cal O}(1)$, whereas color excitation transitions 
between representations of different size are  ${\cal O}(N_c^{-1})$.
To the considered order in $1/N_c$, the state $\ket{e_3}$ 
decouples from the intranuclear evolution \cite{QuarkGluonDijet}.

The projection onto the final states, 
\bea
&&\sum_f\bra{f}=\sum_{R}\sqrt{\dim(R)}\bra{R\bar{R}}=\nonumber\\
&=&\sqrt{N_c}\bra{3\bar{3}}+ \sqrt{{1\over 2} N_c(N_c+1)(N_c-2)}\bra{6\bar{6}}\nonumber\\
&+&\sqrt{{1\over 2} N_c(N_c-1)(N_c+2)}\bra{15\,\overline{15}},
\label{eq:AppB.2.6}
\eea
 at large $N_c$ reads as
\bea
\sum_f\bra{f}&=&\sum_{R}\sqrt{\dim(R)}\bra{R\bar{R}}\nonumber\\
&=&\sqrt{N_c}\bra{e_1}+ (\sqrt{N_c})^3\bra{e_2}.
\label{eq:AppB.2.7}
\eea


\section{Appendix C: Color-diagonal and color-excitation interactions
for excitation of the gluon-gluon dijets }
\setcounter{equation}{0}
\renewcommand{\theequation}{C.\arabic{equation}}

Interactions of the two- and three-gluon
systems are single-channel problems:
\bea
\Sigma_{el}^{(2)}(\bb,\bb')&=& \sigma_{gg,0}(x),\nonumber\\
\Sigma_{ex}^{(2)}(\bb,\bb') &=& \sigma_{gg}(x,\bs +z\br-z\br')-\sigma_{gg,0}(x).\nonumber\\
\Sigma_{el}^{(3)}(\bB,\bb')&=& \sigma_{gg,0}(x) + \half\sigma_{gg}(x,\br),\nonumber\\
\Sigma_{ex}^{(3)}(\bB,\bb')&=&\half[\sigma_{gg}(x,\bs+\br-z\br')-\sigma_{gg,0}(x)]\nonumber\\
&+& 
\half[\sigma_{gg}(x,\bs-z\br')-\sigma_{gg,0}(x)], \nonumber\\
\Sigma_{el}^{(3)}(\bb,\bB')&=& \sigma_{gg,0}(x) + \half\sigma_{gg}(x,-\br'),\nonumber\\
\Sigma_{ex}^{(3)}(\bb,\bB')&=&\half[\sigma_{gg}(x,\bs-\br'+z\br)-\sigma_{gg,0}(x)]
\nonumber\\
&+& 
\half[\sigma_{gg}(x,\bs+z\br)-\sigma_{gg,0}(x)], \nonumber\\
\label{eq:AppC.1}
\eea
For the 4-gluon states we use the results of Ref. \cite{GluonGluonDijet}. 
The Clebsch-Gordan series for
the product of two gluon (adjoint) states reads
\bea
&&(N_c^2-1) \times (N_c^2-1)= \\
&=&1 + (N_c^2-1)_A + (N_c^2-1)_S \nonumber  \\
&+& {(N_c^2-4)(N_c^2-1) \over 4} +
\Big[{(N_c^2-4)(N_c^2-1) \over 4}\Big]^* \nonumber \\
&+&  {N_c^2 (N_c+3)(N_c-1) \over 4} + {N_c^2 (N_c-3)(N_c+1) \over 4} \nonumber \\
&=& 1 + 8_A + 8_S + 10 + \overline{10} + 27 + R_7,
\label{eq:AppC.2}
\eea
where we named the representations by their $SU(3)$ dimensions,
except for one of the symmetric representations that vanishes for $N_c=3$,
and will be referred to as  $R_7$. These representations do
naturally group according to their dimension: the singlet state 
$\ket{1}$ with $\textsf{dim[1]}=1$, two octets with 
$\textsf{dim}[8_{A,S}]={\cal O}(N_c^2)$ and four higher 
representations $\ket{R}=\ket{10},\ket{\overline{10}},\ket{27},
\ket{R_7}$ with the dimension $\textsf{dim[R]}={\cal O}(N_c^4)$.
The non-Abelian evolution of tetra-gluons is a forbidding multichannel 
problem: a full menagerie of possible color singlet tetra-gluons 
includes 9 states, of which 3 states, $\ket{(8_S 8_A)^{(\pm)}}$
and $(\ket{10\,\overline{10}}- \ket{\overline{10}\,10})/\sqrt{2}$,
decouple from our problem exactly. Excitation of one more state,
$(\ket{27\,27}- \ket{R_7 R_7})/\sqrt{2}$ is ${\cal O}(N_c^{-2})$
and it decouples in the considered leading order of large-$N_c$
perturbation theory. 

In the basis of remaining 5 states,
\bea
\ket{e_1}&=&\ket{11}, \nonumber\\
\ket{e_2} & =& {1 \over \sqrt{2}} ( \ket{8_A 8_A} + \ket{8_S
8_S} ), \nonumber \\
\ket{e_3} &=&{1 \over \sqrt{2}} ( \ket{8_A 8_A} - \ket{8_S 8_S} ) \nonumber \\
\ket{e_4} &=& {1\over 2} \Big( \ket{10\overline{10}} +
\ket{\overline{10}10} + \ket{2727} + \ket{R_7 R_7} \Big )
\nonumber \\
\ket{e_5} &=&  {1\over 2} \Big(\ket{10\overline{10}} +
\ket{\overline{10}10} - \ket{2727} - \ket{R_7 R_7} \Big),
\nonumber \\
\label{eq:AppC.3}
\eea
the color-diagonal and color-excitation operators take the form
\bea
\hat{\Sigma}_{el}^{(4)}&=&
\textsf{diag}\left(\Sigma_{11}^{el},\Sigma_{22}^{el},\Sigma_{33}^{el},
\Sigma_{44}^{el},\Sigma_{55}^{el}\right),\nonumber\\
\Sigma_{11}^{el}& =& \sigma_{gg}(x,\br)+\sigma_{gg}(x,-\br'),\nonumber\\
\Sigma_{22}^{el}& =& \sigma_{gg,0}(x)+\half [\sigma_{gg}(x,\br)+\sigma_{gg}(x,-\br')],\nonumber\\
\Sigma_{33}^{el}& =& \sigma_{gg,0}(x)+\half [\sigma_{gg}(x,\br)+\sigma_{gg}(x,-\br')],\nonumber\\
\Sigma_{44}^{el}&=& {C_{27}\over C_A} \sigma_{gg,0}(x),\nonumber\\
\Sigma_{55}^{el}&=& {C_{27}\over C_A} \sigma_{gg,0}(x).
 \label{eq:AppC.4}
\eea
In the considered large-$N_c$ approximation, quadratic Casimirs
for all higher multiplets are equal to each other:
\bea  
C_{27}=C_{R_7}=C_{10}=2C_A.
 \label{eq:AppC.5}
\eea
We expand the color-excitation operator into the diagonal and
off diagonal components:
\begin{widetext}
\bea
\hat{\Sigma}_{ex}^{(4)}&=&\textsf{diag}\{
\Sigma_{11}^{ex},\Sigma_{22}^{ex},\Sigma_{33}^{ex},\Sigma_{44}^{ex},\Sigma_{55}^{ex}\}+\hat{\omega},\nonumber\\
\Sigma_{11}^{ex}&=& 0,\nonumber\\
\Sigma_{22}^{ex}&=&\half[\sigma_{gg}(x,\bs)-\sigma_{gg,0}(x)]+
\half[\sigma_{gg}(x,\bs+\br-\br')-\sigma_{gg,0}(x)],\nonumber\\ 
\Sigma_{33}^{ex}&=&\half[\sigma_{gg}(x,\bs+\br)-\sigma_{gg,0}(x)]+
\half[\sigma_{gg}(x,\bs-\br')-\sigma_{gg,0}(x)],\nonumber\\ 
\Sigma_{44}^{ex}&=&\half[\sigma_{gg}(x,\bs)+\sigma_{gg}(x,\bs+\br-\br')]
-{C_{27}\over C_A} \sigma_{gg,0}(x),\nonumber\\ 
\Sigma_{55}^{ex}&=&\half[\sigma_{gg}(x,\bs+\br)+\sigma_{gg}(x,\bs-\br')]-
{C_{27}\over C_A} \sigma_{gg,0}(x),
\nonumber\\ 
\hat{\omega}&=& -{1\over \sqrt{2}N_c}\Omega
\{\ket{e_1}\bra{e_2}   + \ket{e_1}\bra{e_3}      + \ket{e_4}\bra{e_2}-\ket{e_5}\bra{e_3}
+ H.c\}.
 \label{eq:AppC.6}
\eea
The final-state summation goes as 
\bea 
\sum_f \bra{f} = \sum_R \sqrt{\textsf{dim[R]}} \bra{R\overline{R}} =
\underbrace{\bra{e_1}}_1 + \underbrace{N_c \sqrt{2} \bra{e_2}}_{8_A
+8_S} + \underbrace{ N_c^2
\bra{e_4}}_{10+\overline{10}+27+R_7}\,. 
\label{eq:AppC.7}
\eea
and the averaging over incoming colors shall introduces a factor $1/N_c$,
\bea 
\ket{\mathrm{in}} = {1 \over  \sqrt{\textsf{dim[8]}}} \ket{8_A 8_A} = {1
\over \sqrt{2}N_c} \Big(\ket{e_2} +\ket{e_3}\Big) \, . 
\label{eq:AppC.8}
\eea
\end{widetext}

\end{document}